\documentclass{jfm}
\usepackage{graphicx}
\usepackage{epstopdf, epsfig}

\usepackage{url}
\usepackage{amsmath}
\usepackage{natbib}
\usepackage{float}
\usepackage{indentfirst}
\usepackage{hyperref}

\usepackage{booktabs}
\usepackage{tabularx}
\usepackage{color}
\usepackage{caption}
\usepackage[position=top, singlelinecheck=false]{subfig}
\usepackage{setspace}
\usepackage{verbatim}
\usepackage{xcolor}
\usepackage{bm}

\shorttitle{Bag Film Breakup of Droplets in Uniform Airflows}
\shortauthor{K. Tang, T.A.A. Adcock and W. Mostert}

\title{Bag Film Breakup of Droplets in Uniform Airflows}

\author{K. Tang\aff{1}
  \corresp{\email{kaitao.tang@eng.ox.ac.uk}},
  T.A.A. Adcock\aff{1}
 \and W. Mostert\aff{1}}

\affiliation{\aff{1}Department of Engineering Science, University of Oxford, Oxford OX1 3PJ, UK}

\begin{document}

\maketitle

\begin{abstract}
We present novel numerical simulations investigating the bag breakup of liquid droplets. We first examine the viscous effect on the early-time drop deformation, comparing with theory and experiment. Next, a bag film forms at late time and is susceptible to spurious mesh-induced breakup in numerical simulations, which has prevented previous studies from reaching grid convergence of fragment statistics. We therefore adopt the manifold death (MD) algorithm which artificially perforates thin films once they reach a prescribed critical thickness independent of the grid size, controlled by a numerical parameter $L_{\rm sig}$. We show grid convergence of fragment statistics when utilising the MD algorithm, and analyse the fragment behaviour and bag film disintegration mechanisms including ligament breakup, node detachment and rim destabilisation. Our choice of the critical thickness parameter $L_{\rm sig}$ is limited by numerical constraints and thus has not been matched to experiment or theory; consequently, the current simulations yield critical bag film perforation thicknesses larger than experimentally observed. The influence of the MD algorithm configuration on the bag breakup phenomena and statistics will be investigated in future work. We also study the effects of moderate liquid Ohnesorge number ($0.005 \leq Oh \leq 0.05$) on the bag breakup process and fragment statistics, where a non-monotonic dependency of the average diameter of bag film fragments on $Oh$ is found. These results highlight the utility of the MD algorithm in multiphase simulations involving topological changes, and pave the way for physics-based numerical investigations into spume generation at the air-sea interface.
\end{abstract}

\section{Introduction}
\label{sec:introduction}

Liquid atomisation refers to the process where a bulk volume of liquid disintegrates into fragments featuring various sizes and shapes \citep{Guildenbecher2009, Pairetti2018}. The fragments generated are described as sprays, which are involved in many natural and industrial processes, including ocean-atmosphere interactions \citep{Veron2015, Erinin2019}, precipitation and rain-drop dynamics \citep{Villermaux2009, Jalaal2012, Veron2015}, combustion of liquid propellant in aerospace applications \citep{Young1995}, pharmaceutical spray generation \citep{Mehta2017} and pathogen transmission \citep{bourouiba2021fluid, kant2022bags}, etc. More specifically, it has recently been found that the atomisation of small-scale sea surface perturbations dominates ocean spume generation under extreme wind conditions, producing large droplets with typical sizes of $10^2 \sim 10^3 \, {\rm \mu m}$ \citep{Troitskaya2017, Troitskaya2018}. In this size range the currently-available sea-spray generation functions (SSGF), crucial for calculations of air-sea momentum and heat exchange in earth-system modelling, show large range of scatter \citep{Veron2015}. However, since the physics governing the fragmentation of bag films have not yet been firmly established, their influences on SSGFs have been difficult to quantify. Improving this understanding is the primary motivation of the present work.
 
Two stages of liquid atomisation have been identified within literature, namely the primary and secondary atomisation. Sheets, ligaments and droplets are stripped from a bulk fluid during primary atomisation, which further decompose until stabilising capillary effects take over during secondary atomisation \citep{Pairetti2018}. Secondary atomisation is typically modelled by the droplet aerobreakup problem characterised by the interaction between an initially spherical droplet with density $\rho_l$, viscosity $\mu_l$ and diameter $d_0$, and an ambient gas flow with density $\rho_a$, viscosity $\mu_a$ and uniform velocity $U_0$ \citep{Guildenbecher2009}. Based on these physical properties, together with surface tension $\sigma$ at the liquid-gas interface, four non-dimensional controlling parameters have been proposed using Buckingham's Pi Theorem (see e.g. Table 1 in \citep{Jalaal2014}):
\begin{equation}
    We \equiv \frac{\rho_a U_0^2 d_0}{\sigma}, \quad Oh \equiv \frac{\mu_l}{\sqrt{\rho_l d_0 \sigma}}, \quad
    \rho^* \equiv \frac{\rho_l}{\rho_a}, \quad
    \mu^* \equiv \frac{\mu_l}{\mu_a}.
    \label{for:non-dimensional-groups}
\end{equation}
Among these, $We$ and $Oh$ are respectively the Weber and Ohnesorge number quantifying the ratio of inertial to capillary and viscous to capillary forces, and $\rho^*$ and $\mu^*$ are respectively the density and viscosity ratios of the liquid and gas phase.

Within literature, various droplet aerobreakup regimes have been observed where the droplet shows different deformation patterns, and the transition thresholds between these regimes have traditionally been delineated using $We$ and $Oh$ \citep{Yang2017, Zotova2019, Marcotte2019}, although some recent works have shown that the density ratio $\rho^*$ may also play an important role \citep{Yang2016, Marcotte2019, Jain2019}. $Oh$ has been reported to influence the transition thresholds only when exceeding a critical value of 0.1 \citep{Hsiang1995}; and as $We$ increases, the breakup becomes more violent and vibrational, bag, multi-mode (bag-stamen), sheet-thinning and catastrophic breakup regimes are observed in succession \citep{Jalaal2014, Kekesi2014}. Alternatively, based on the governing hydrodynamic instability involved in the process, the four breakup regimes mentioned above can be re-grouped into two major categories: Rayleigh-Taylor piercing (RTP) and shear-induced entrainment (SIE) \citep{Theofanous2011}. However, despite the extensive amount of related work, the underlying physics governing the transient drop deformation in each regime are still largely unclear. Furthermore, the empirical transition criteria proposed so far are often contradictory \citep{Theofanous2011, Yang2017}, with the notable exception of a consensus that the critical Weber number beyond which bag breakup initiates is ${We}_c = 11 \pm 2$ when $Oh < 0.1$ \citep{Guildenbecher2009, Yang2017}.

In the bag breakup regime, the initially spherical droplet first flattens and forms a disc, whose centre is then blown downstream and inflates into a hollow bag attached to a toroidal rim. The time it takes for the drop to reach breakup $\Delta t_d$ typically falls within the range of $\tau \leq \Delta t_d \leq 2\tau$, where $\tau \equiv \sqrt{\rho_l / \rho_a} d_0 / U_0$ is the characteristic deformation time proposed by \cite{Nicholls1969}. The swollen bag first ruptures near its centre, triggering expansion of holes on the surface of the bag and eventually bursting into a large number of fragments, which is then followed by the breakup of the remnant toroidal rim into smaller amounts of fragments \citep{Chou1998, Guildenbecher2009, Opfer2014}. Droplet bag breakup and its associated fragment size and velocity distribution functions are of specific interest as they bear a strong resemblance to the previously mentioned bag-mediated fragmentation of small-scale sea-surface perturbations under extreme wind conditions \citep{Troitskaya2017, Troitskaya2018}.

Droplet aerobreakup involves a complex interplay of aerodynamic, capillary and viscous effects that is still poorly understood \citep{Jain2015}. The prevalent theoretical understanding is that hydrodynamic instabilities, particularly Kelvin-Helmholtz (KH) and Rayleigh-Taylor (RT) instability, play an important role in the aerobreakup process \citep{Guildenbecher2009, Theofanous2011, Theofanous2012, Jackiw2021}. KH instability occurs at the interface between two different streams of fluid with different velocities and densities \citep{Kundu2012}. In the context of large-$We$ droplet aerobreakup, it governs the SIE breakup category \citep{Theofanous2011}, and is typically found near the drop periphery where the relative velocity between the liquid and gas phases is the largest \citep{Gorokhovski2008, Jalaal2014}. However, due to strong capillary effects, KH instability is unable to influence droplet deformation in the bag breakup regime \citep{Theofanous2012, Jalaal2014}. RT instability occurs when a corrugated interface separating fluids with different densities undergoes constant acceleration \citep{Zhou2021}, and is hypothesised to cause interfacial perturbation growth on the windward surface of the droplet. The wave number of such perturbations determines whether the droplet undergoes oscillatory deformation, bag breakup or multi-mode breakup \citep{Yang2017}. However, instability theories have difficulty in accounting for the viscous effects \citep{Jalaal2014}, flow dynamics prior to drop flattening, and finite thickness and peripheral boundary of the flattened disc \citep{Jackiw2021}. Alternatively, some works highlight the influence of the internal flow within the droplets on the deformation process \citep{Guildenbecher2009, Villermaux2009, Jackiw2021, Obenauf2021, ling2023detailed}. The internal flow model compensates for the drawback of the RT instability model in predicting early-time drop deformation; however, this approach is somewhat simplified and cannot account for the complex interaction between wake vortices and drop surface \citep{Marcotte2019}. The late-time breakup behaviour, on the other hand, is delineated into a bag-film rupturing event, and the fragmentation of the remnant rim at a later time. The bag film rupture occurs more rapidly and produces much smaller fragments compared with the remnant rim breakup, and is thus more difficult to capture \citep{Guildenbecher2009}. It has only recently been clarified experimentally \citep{jackiw2022prediction} that the major pathways leading to bag fragmentation are the destabilisation and collision of hole rims as they recede over the curved bag and experience centripetal acceleration, which is also observed in the numerical simulations of \cite{ling2023detailed}, where they investigated in detail the morphological changes of the droplet in the moderate $We$ regime, and benchmarked them against existing theoretical and experimental results; based on which they improved the internal flow model of \cite{Jackiw2021} for prediction of drop deformation. Nevertheless, ensemble-averaged size and velocity statistics of aerobreakup fragments are still scarce \citep{Zhao2011}; and given the large span in time and length scales, the understanding of what types of physical mechanisms are involed in the bag film breakup process and how each of them contributes to the statistics of fragments and dictates their subsequent behaviour remains unsatisfactory. Furthermore, the effects of the $Oh$ value on the bag breakup phenomena still remain largely unexplored \citep{jackiw2022prediction}.

The earliest research on droplet aerobreakup are mostly experimental, where the droplet breakup behaviour is recorded and analysed using shadowgraphs, high-speed cameras and particle image velocimetry (PIV) \citep{Hsiang1992, Guildenbecher2009, Jalaal2012, Radhakrishna2021}. Thanks to the recent development of computational power, numerical studies have provided a way to investigate atomisation phenomena and gain insight into fundamental mechanisms that are otherwise difficult to achieve experimentally \citep{Gorokhovski2008, Ling2015}. However, serious challenges are also present for computational studies on droplet aerobreakup, including reaching numerical convergence at large density ratio $\rho^*$ \citep{Zotova2019, Marcotte2019}; high computational cost of fully resolving small-scale fragmentation processes in two-phase turbulence simulations at high $We$ values \citep{Gorokhovski2008, Jalaal2014, Shinjo2018}, where the smallest droplet size may be much less than the Kolmogorov scale \citep{Shinjo2018}. There is also potential need of ensemble averaging when fragments produced from an individual realisation is not sufficient for obtaining statistically meaningful results \citep{Mostert2021}. In particular, as the Navier-Stokes equations do not describe the physical mechanisms that control topological changes at phase boundaries, thin films are subject to uncontrolled numerical perforation when their thickness approaches the minimum grid size \citep{chirco2021manifold}. As a result, the fragment statistics are dependent on grid sizes \citep{jackiw2022prediction}, and numerical convergence with respect to bag fragment statistics has not previously been obtained to our knowledge. It is therefore of paramount importance to improve the grid resolution level and make the onset of breakup independent of the grid size, even though the exact physical mechanism initiating the breakup events remains elusive \citep{kant2022bags}. A few attempts have been made to improve the numerical resolution of fragmentation or coalescence phenomena. Among these, \cite{coyajee2009numerical} first proposed a modified VOF scheme that utilises multiple marker functions for different fluid interfaces to minimise spurious coalescence on coarse meshes. Afterwards, \cite{zhang2019vortex} built a topology-based numerical scheme which automatically refines grid cells containing the liquid film bordered by two adjacent bubble interfaces. Finally, \cite{chirco2021manifold} developed an algorithm that randomly perforates thin films once their thickness reduces to a prescribed critical value independent of the grid size. This algorithm is controllable via a set of tuning parameters and has been shown to improve grid convergence behaviour for various two-phase problems including droplet aerobreakup. 

We present results of novel multiphase direct numerical simulation of droplet bag breakup using both axisymmetric and fully three-dimensional configurations. We conduct axisymmetric simulations to study pre-breakup deformation dynamics, and three-dimensional studies coupled with the MD algorithm of \cite{chirco2021manifold} to shed light on the breakup dynamics of bag films and acquire statistics of bag film fragments for further analysis of their behaviour, while leaving the validation of the MD algorithm with appropriately tuned parameters for the aerobreakup problem to future work. Our study is structured as follows. We present in \S\ref{subsec:description} the configuration of our problem and the parameter space we explore, and then introduce the numerical method in \S\ref{subsec:num-method}. We analyse the axisymmetric simulation results in \S\ref{subsec:pre-breakup} and compare them with previous theoretical predictions, focusing on the early-time deformation period where \cite{Jackiw2021} predicted a constant spanwise growth rate (\S\ref{subsubsec:early}), and the film-thinning period immediately before bag breakup where an exponential decay model of film thickness is available \citep{Villermaux2009} (\S\ref{subsubsec:drainage}). We then investigate the breakup of the bag film based on the three-dimensional simulation results, where we first show grid convergence of fragment statistics using the MD algorithm \citep{chirco2021manifold} (\S\ref{subsec:grid-conv}). We then analyse the size and velocity distributions, and provide an overview for the breakup mechanisms leading to bag disintegration in \S\ref{subsec:frag-tracking}. Afterwards, we track and reconstruct the evolution of individual fragments, and study the dependence of their ejection velocity, lifetime and oscillation patterns in \S\ref{subsec:frag-behaviour}. Finally, we investigate the influence of $Oh$ values on the breakup of bag films (\S\ref{subsec:Oh-effects}). We provide a summary for the numerical convergence of bag fragment statistics in \S\ref{sec:num-summary}, and conclude the study in \S\ref{sec:conclusions} with some remarks on future work.

\section{Formulation and methodology}
\label{sec:formulation}

\subsection{Problem description}
\label{subsec:description}
\begin{figure}
	\centering
	\subfloat[]{
		\label{fig:config-axi}
		\includegraphics[width=.48\textwidth]{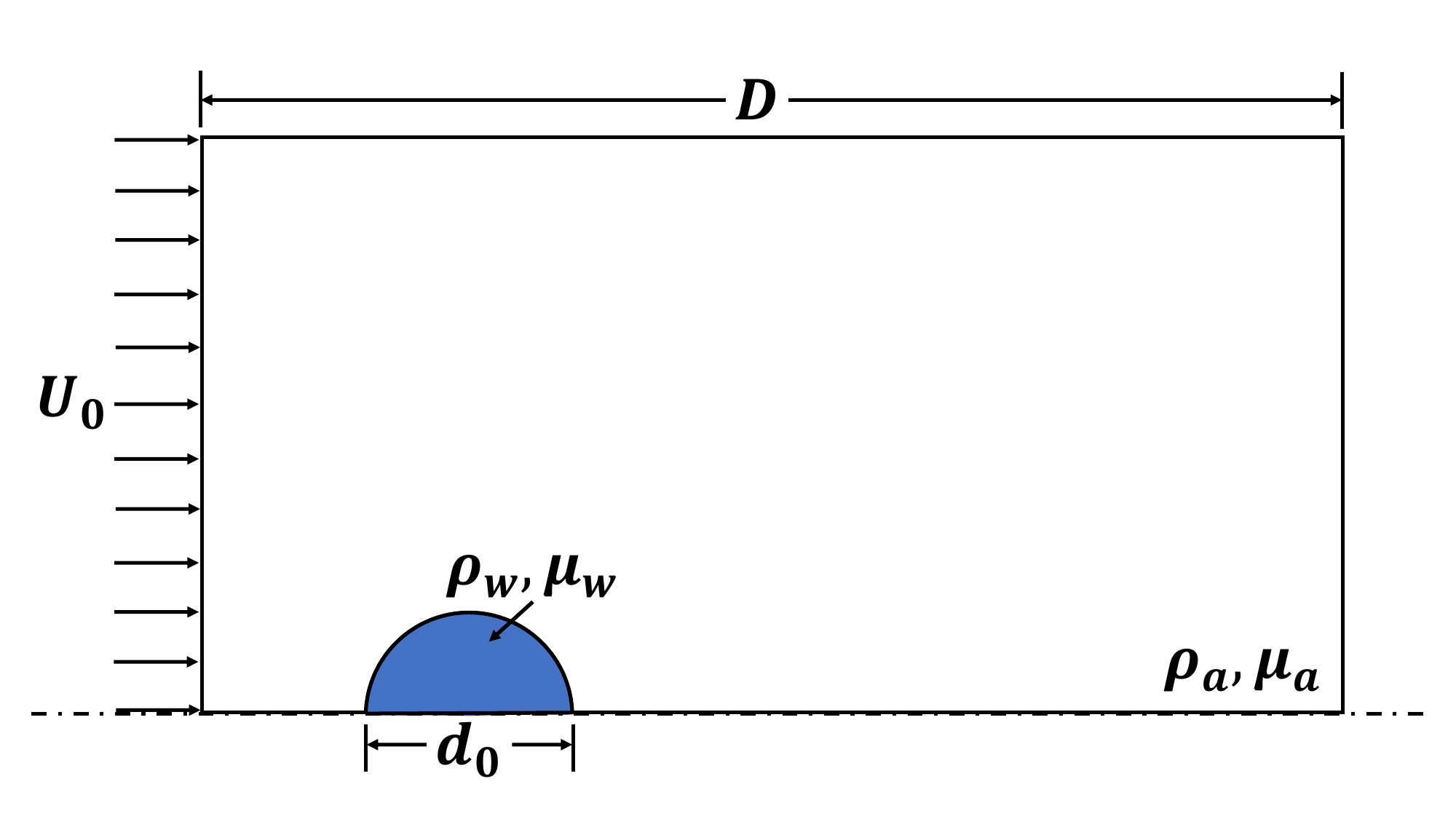}}
	\centering
	\subfloat[]{
		\label{fig:config-3d}
		\includegraphics[width=.48\textwidth]{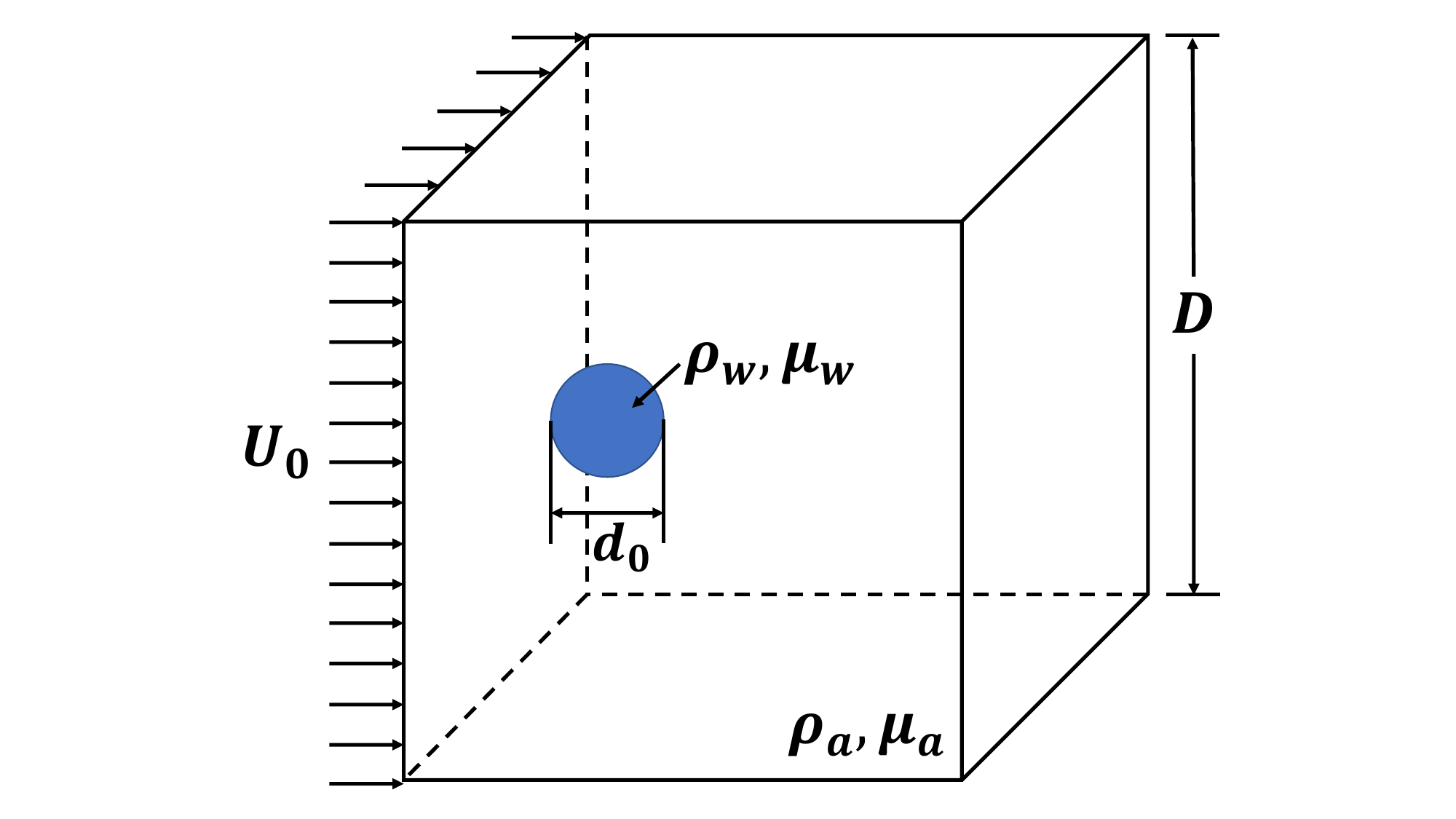}}
	
	\caption{Sketches showing the initial configurations of axisymmetric (a) and three-dimensional (b) droplet aerobreakup simulations. The axis of symmetry is located at the bottom in (a).}
	\label{fig:flow-config}
\end{figure}

The flow configurations for axisymmetric and three-dimensional simulations are shown in figs.~\ref{fig:config-axi} and \ref{fig:config-3d}, respectively. For both axisymmetric and three-dimensional simulations, a stationary liquid droplet with diameter $d_0$, density $\rho_l$ and viscosity $\mu_l$ is placed close to the left boundary, surrounded by an initially quiescent gas phase with density $\rho_a$ and viscosity $\mu_a$. The domain width $D$ is set as $10d_0$ and $15d_0$ for axisymmetric and three-dimensional simulations, respectively, so as to eliminate the influence of finite domain size on the aerobreakup process. A zero-gradient velocity boundary condition is applied at the right boundary and a uniform incoming velocity $U_0$ is imposed on the left boundary, while no-penetration conditions are applied at the other domain boundaries. This velocity initialisation results in an impulsive acceleration of the droplet at the first time step, and induces a flow field satisfying both the incompressible constraint and the conservation of linear momentum \citep{Jalaal2014, Marcotte2019}.

As discussed in \S\ref{sec:introduction}, the problem is defined by four non-dimensional parameters, namely the Weber number $We$, the Ohnesorge number $Oh$, the density ratio $\rho^*$ and the viscosity ratio $\mu^*$. Since we are interested in air-water systems, $\rho^*$ and $\mu^*$ are set as 830 and 55, respectively, following the earlier work of \cite{Pairetti2018}. We vary $We$ between 12 and 25 in our axisymmetric simulations, while in our current 3D simulations we fix it at 15. In the meantime, $Oh$ is varied between $10^{-4}$ and $0.075$, which allows for a comprehensive investigation of viscous effects on bag breakup.

\subsection{Numerical method}
\label{subsec:num-method}
We use the open-source Basilisk numerical library \citep{Popinet2019basilisk} to solve the Navier-Stokes equations for two-phase incompressible, immiscible and isothermal flows, which are written in the following variable-density form,
\begin{gather}
\nabla \cdot \bm{u} =0, \label{for:continuity} \\
\rho \left( \frac{\partial \bm{u}}{\partial t} + \bm{u} \cdot \nabla \bm{u} \right) = -\nabla p + \nabla \cdot \left[ \mu (\nabla \bm{u} + \nabla \bm{u}^T) \right] + \sigma \kappa \delta_s \bm{n}. \label{for:momentum}
\end{gather} 
Equations \eqref{for:continuity} and \eqref{for:momentum} are respectively the continuity and momentum equation, where $p$ is the fluid pressure. Surface tension effects are incorporated in the volumetric form $\sigma \kappa \delta_s \bm{n}$ within Equation \eqref{for:momentum}, where $\sigma$ is the surface-tension coefficient, and $\kappa$ and $\bm{n}$ are respectively the local curvature and normal vector on the interface. The Dirac delta \(\delta_s\) is non-zero only on the interface, indicating the local concentration of surface tension effects \citep{popinet2009accurate, popinet2018numerical}.

The geometric volume-of-fluid (VOF) method is applied in Basilisk to reconstruct the interface and minimize the parasitic currents induced by surface tension \citep{popinet2018numerical}, which solves the following advective equation,
\begin{equation}
    \frac{\partial f}{\partial t} + \bm{u} \cdot \nabla f=0, \label{for:interface-tracking}
\end{equation}
where $f$ is the VOF function that distinguishes the liquid and gaseous phases, taking the value of 1 and 0 in the former and latter respectively. For modelling of surface tension effects, \(\delta_s \bm{n}\) in Equation \eqref{for:momentum} is approximated as the gradient of the VOF function \(\nabla f\) using an adaptation of Brackbill's method \citep{brackbill1992continuum, popinet2009accurate}, and the curvature \(\kappa\) is calculated by taking the finite-difference discretisation of the derivatives of interface height functions \citep{popinet2009accurate}. The quad/octree-based AMR scheme based on the estimation of local discretisation errors of the VOF function gradient $\nabla f$ and flow velocity $\boldsymbol{u}$ is adopted so as to reduce the computational cost at high resolution levels $L$, which is defined using the minimum grid size,
\begin{equation}
    \Delta = \frac{D}{2^L}.
\end{equation}
As $\Delta$ is the smallest length scale at which necks of thinning filaments can be represented, $L$ sets the length scale at which liquid filament breakup occurs.

In the bag breakup regime, the onset of fragmentation is preceded by the inflation of bag structure whose thickness reduces considerably over time. While the mechanism responsible for the puncture of the bag film has been extensively discussed \citep{lhuissier2012bursting, chirco2021manifold, kant2022bags}, in VOF simulations this is initiated when the local thickness of the bag decreases to the finest grid size \citep{ling2017spray}, causing the initiation time of breakup and the size of the finest fragments to be grid-dependent. To circumvent this unphysical and numerically uncontrolled phenomenon, we adopt the manifold death (MD) algorithm recently developed by \cite{chirco2021manifold}, which artificially perforates thin films once their thickness decreases to a prescribed critical value independent of the grid size. This enables grid convergence to be reached in the fragment size distributions and related quantities \citep{chirco2021manifold}. This is realised in the Basilisk framework by first computing quadratic moments of the VOF colour function $f$ on grid cells with a given signature level $L_{\rm sig} \leq L$, which defines the critical thickness $h_c \equiv  3D/2^{L_{\rm sig}}$, the smallest length scale at which liquid films can be presented as below it they will be artificially perforated by the MD algorithm. The signs of the computed quadratic moments indicate the local shape of the interface. If a film with thickness not larger than $h_c$ is detected, the algorithm randomly creates cubic cavities on the ligament by directly setting the value of $f$ to that of the other phase with a probability $p_{\rm perf}$. While the total fluid mass is changed when holes are created on thin films, the MD algorithm is able to minimise this side effect by creating cavities with minimum sizes that allow for Taylor-Culick expansion, and limiting the maximum number of holes perforated at every iteration. Further discussion, and details of the parameters used for the MD algorithm in our study are supplied in \S\ref{subsec:grid-conv}.

Before the formation of thin bag films and their subsequent breakup, the smallest length scale in the aerobreakup problem is the thickness of the viscous air boundary layer $\delta$ around the droplet, through which momentum diffuses from the surrounding airflow into the droplet and drives its deformation. Batchelor's estimation with the defining length scale of the droplet $d_0$ yields $\delta \sim d_0 / \sqrt{Re}$, where $Re \equiv \rho_a U_0 d_0/\mu_a$ is the freestream Reynolds number. For a typical droplet in the bag breakup regime, characterised by Weber and Ohnesorge numbers $We = 15$, $Oh = 10^{-3}$, this corresponds to $\delta \sim 1.2 \times 10^{-2} d_0$. The recommended criteria of $\delta/\Delta \geq 2$ \citep{Mostert2020} then requires the grid resolution level satisfy $L \geq 12$ for simulations with domain size $D = 15 d_0$. The highest grid resolution level we set in our present simulations is $L = 14$, at which the droplet contour in our axisymmetric simulations has reached grid independence. The numerical convergence of fragment statistics will be discussed in detail in \S\ref{subsec:grid-conv}.

Finally, the droplet diameter $d_0$, incoming flow velocity $U_0$, dynamic flow pressure $p_0 \equiv \rho_l U_0^2$ and the characteristic deformation time $\tau$ introduced in \S\ref{sec:introduction} provide the natural reference scales for the length, mass and time quantities that appear in Eqs.~\eqref{for:continuity} and \eqref{for:momentum}, and will be used to non-dimensionalise the numerical results in the remainder of this study unless otherwise specified.

\section{Pre-breakup deformation dynamics}
\label{subsec:pre-breakup}

Before the onset of bag breakup, the shape of the deforming droplet remains largely axisymmetric, although the wake region may have become fully turbulent and three-dimensional. Many previous numerical aerobreakup studies therefore conducted axisymmetric simulations for a parametric study \citep{Yang2017, Marcotte2019, Jain2019}. In this section, we present our axisymmetric results to provide an overview of the pre-breakup deformation characteristics of the droplet, while also verifying our simulation results by comparing with available analytic models and experimental results.

\subsection{Early-time deformation}
\label{subsubsec:early}
\begin{figure}
	\centering
	\subfloat[]{
		\label{fig:contour_Oh_0.001}
		\includegraphics[width=.48\textwidth]{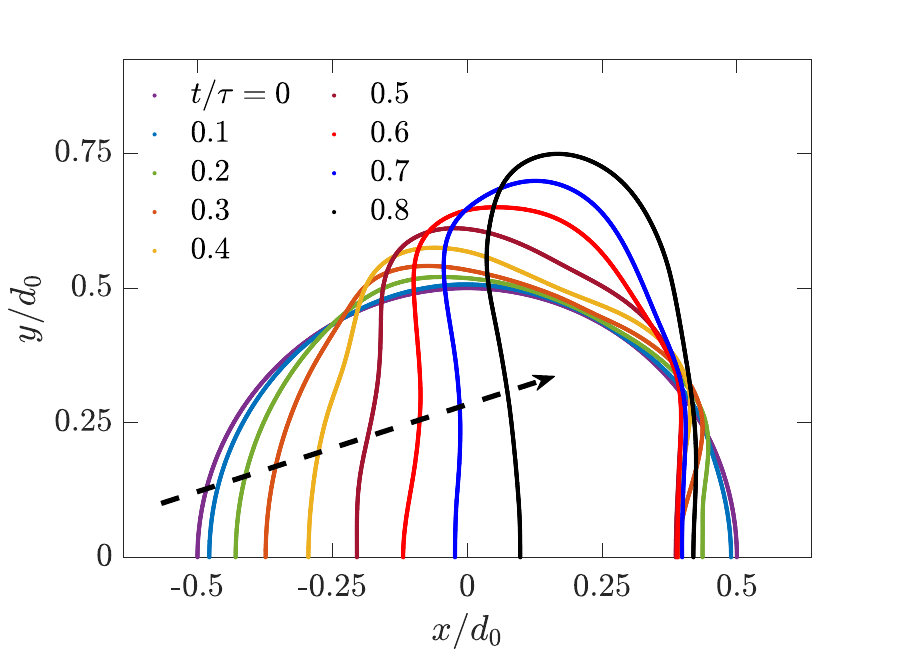}}
	\centering
	\subfloat[]{
		\label{fig:contour_Oh_0.01}
		\includegraphics[width=.48\textwidth]{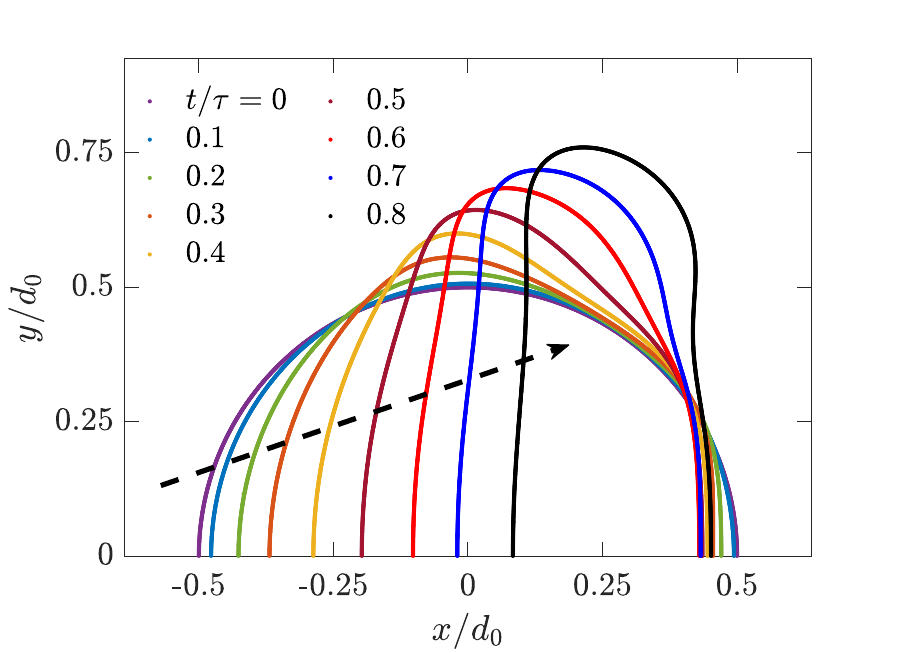}}	
	\caption{Early-time development of droplet contours for axisymmetric simulations with Ohnesorge number $Oh = 10^{-3}$ (a) and $10^{-2}$ (b), while the Weber number $We=15$. The axis of symmetry is at $y=0$.}
	\label{fig:contour_We_15}
\end{figure}

We first discuss the initiation period of aerobreakup, defined by \cite{Jackiw2021} as $0 \leq t \leq T_i$, where $T_i$ is the time when the droplet reaches its minimal streamwise thickness. To provide an overview of the early-time droplet deformation process characterised by spanwise flattening, we first present in fig.~\ref{fig:contour_We_15} the droplet contours extracted from our axisymmetric simulations at various instants within $0 \leq t/\tau \leq 0.8$ for two different Ohnesorge numbers $Oh$ of $10^{-3}$ and $10^{-2}$, while the same Weber number $We = 15$ is set for both cases. The radial profile is shown with $y=0$ as the axis of symmetry.
It is found that during the early deformation stage, the windward surface of the droplet continues moving downstream and pushing liquid to the drop periphery, leading to the gradual spanwise flattening of the droplet. In the meantime, a dimple develops on the windward surface that moves downwards and eventually evolves into a crater on the axis of symmetry for $Oh = 10^{-3}$, as shown in fig.~\ref{fig:contour_Oh_0.001}. The leeward side of the droplet remains relatively stationary after an initial movement to the left. In contrast, fig.~\ref{fig:contour_Oh_0.01} shows that the increase of $Oh$ to $10^{-2}$ postpones the dimple formation on the windward surface significantly, which only begins to appear at $t/\tau = 0.8$. Previous works have attributed the spanwise flattening of the drop to the aerodynamic pressure difference between the frontal stagnation point and the equatorial periphery \citep{Jackiw2021}, which drives the internal flow within the droplet against the restoring effects of surface tension \citep{Marcotte2019}. The airflow quickly separates from the leeward surface, creating a re-circulation region with low pressure which induces little movement at the leeward interface \citep{Jain2015}. Formation of similar dimple structures on the windward surface can also be observed in fig.~1 of \cite{Marcotte2019} within the Weber number range of $11.3 \leq We \leq 24$ corresponding to bag and bag-stamen breakup. 

We briefly examine whether the dimple is a result of RT Instability developing on the windward surface due to wind acceleration. \cite{li2019theoretical} predicted a critical instantaneous Bond number $Bo_c \equiv \rho_l \alpha d_0^2/4\sigma = 11.2$, beyond which the windward surface is destabilised. Here $\alpha$ is the instantaneous acceleration of the liquid droplet. For a droplet with $We = 15, \, Oh = 10^{-3}$, our results show $Bo = 0.57$ at $t/\tau = 0.4$ when the dimple is first observed in fig.~\ref{fig:contour_Oh_0.001}, much smaller than the threshold value of 11.2 predicted by \cite{li2019theoretical}. Taking into account that the liquid is being primarily pushed from the frontal stagnation point to the windward side of the periphery around the time of dimple formation ($t/\tau \sim 0.4$ in fig.~\ref{fig:contour_Oh_0.001}), together with the $We$ range where it is observed in \citep{Marcotte2019}, it is more likely that the dimple formation is caused by the capillary pinching effects against fluid influx, and should therefore be viewed as a precursor of later rim formation.

\begin{figure}
	\centering
	\subfloat[]{
		\label{fig:bag-width-valid}
		\includegraphics[width=.48\textwidth]{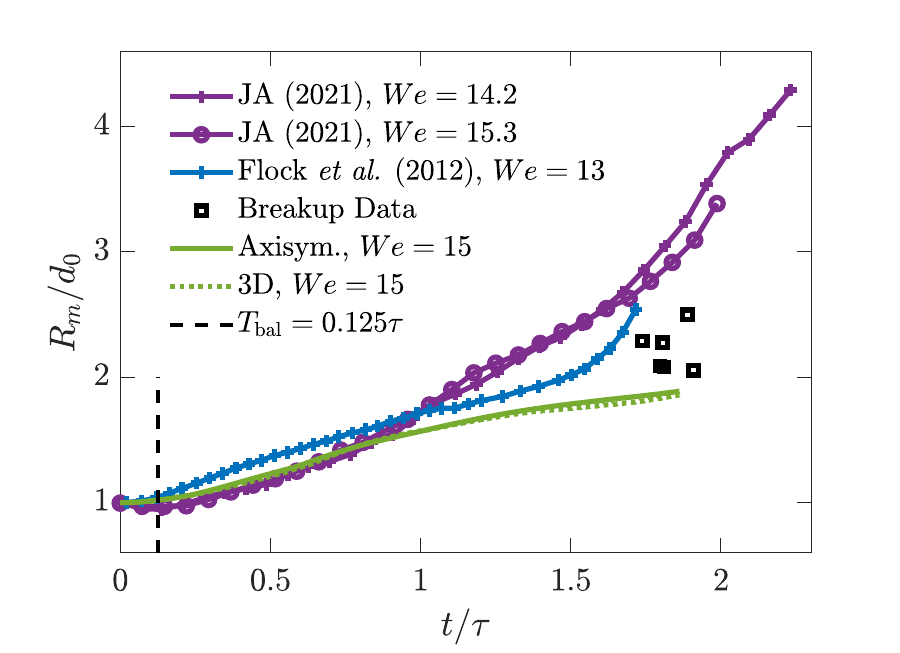}}	
	\centering
	\subfloat[]{
		\label{fig:bag-len-valid}
		\includegraphics[width=.48\textwidth]{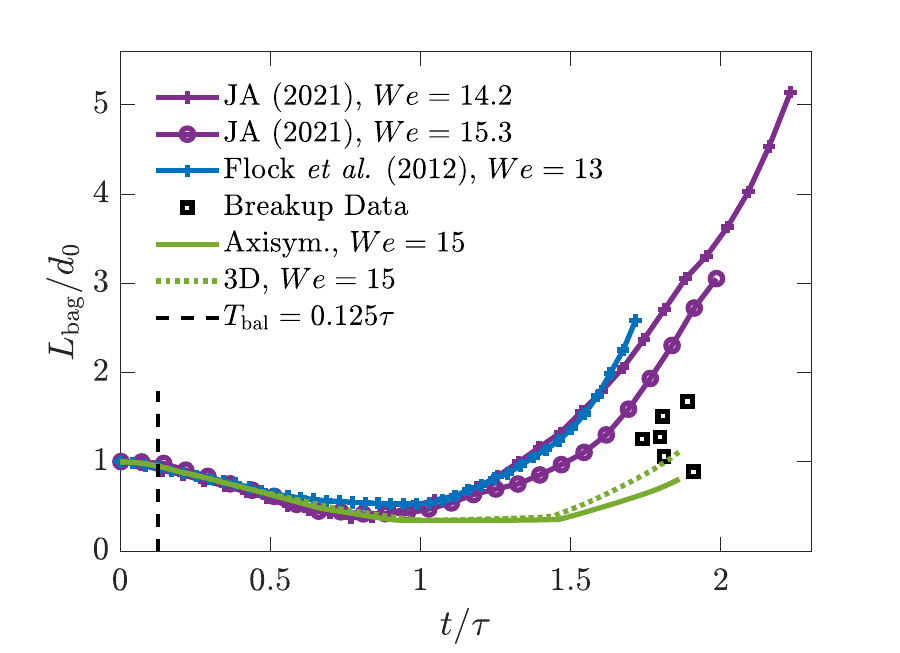}}
  
	\caption{Comparison of our axisymmetric and 3D simulation results for the evolution of bag length (a) and width (b) at $We = 15, \, Oh = 2.5 \times 10^{-3}$ with the experimental data of \cite{Jackiw2021} and \cite{flock2012experimental}. The breakup lengths and widths for various $Oh$ values are included as scattered points, and the balance time $T_{\rm bal} = 0.125\tau$ proposed by \cite{Jackiw2021} is also plotted for reference.}
	\label{fig:rm_We_15_comp}
\end{figure}

For validation of our numerical results, we present in Fig.~\ref{fig:rm_We_15_comp} the evolution of the maximum spanwise radius of the drop $R_m$ and the streamwise length of the bag $L_{\rm bag}$ measured from our axisymmetric and 3D numerical simulations at $We = 15, \, Oh = 2.5 \times 10^{-3}$, and compare them with the available experimental results of \cite{Jackiw2021} and \cite{flock2012experimental}. It can first be seen that the axisymmetric and 3D numerical results agree excellently until $t \approx 1.5\tau$, when the axisymmetric simulation shows a smaller bag length in fig.~\ref{fig:bag-len-valid}. This late-time deviation most likely arises from the lack of 3D flow instability development in axisymmetric simulations \citep{Marcotte2019}, which may break the symmetry of the bag and limits its streamwise growth. Both our axisymmetric and 3D simulation results agree well with the experimental data of \cite{Jackiw2021} up to $t/\tau = 1$, after which the experimental results show faster growth in both $R_m$ and $L_{\rm bag}$. This may be due to the sensitivity of the flattened drop to difference in the ambient flow conditions, as in our numerical simulations the air-phase flow remains laminar, whereas the experimental configuration of \cite{flock2012experimental} and \cite{Jackiw2021} in fact produces air-phase turbulence, which has been shown by \cite{zhao2019effect} to be capable of increasing the height and width of bags at late time (see e.g., their fig.~6). More specifically, \cite{Jackiw2021} used a 5-gauge air needle (whose diameter $D_n$ is only 2.48 times of the droplet diameter $d_0$) to generate air jets with centreline Reynolds number $Re_a$ of $5.2 \times 10^3 \sim 2.5 \times 10^4$, apart from needles for suspending the drop within such air jets. In the case of \cite{flock2012experimental}, air jets are produced through a nozzle with diameter $D_n \approx 11d_0$, but the airflow is also turbulent with $Re_a = 1.8 \times 10^4$. The results of \cite{jackiw2022prediction} are obtained from single experimental runs without being ensemble-averaged, which may lead to larger variations in their results, as also noted in the comparison of numerical results by \cite{ling2023detailed}. Additionally, note that in Fig.~\ref{fig:bag-width-valid}, the experimental results of \cite{Jackiw2021} and \cite{flock2012experimental} show some mutual disagreement in the spanwise radius values within the range of $t/\tau \leq 1.5$.

The bag lengths and widths recorded at various $Oh$ values at the point of breakup are also included as scattered points in fig.~\ref{fig:rm_We_15_comp}, which we will return to in \S\ref{subsec:Oh-effects}. It can be seen that our bags approach breakup within the time range of $1.74 \leq t/\tau \leq 1.91$, earlier than the experimental results of \cite{Jackiw2021} ($t = 2.2\tau$ for $We = 15.3$). On the other hand, \cite{flock2012experimental} did not report the exact time at which bag breakup is initiated. This earlier breakup time is associated with the limit of grid resolution, and hence $L_{\rm sig}$ on the critical thickness at which the bag film is perforated by the MD algorithm, as at $L_{\rm sig} = 13$, the critical thickness is $3D/2^{L_{\rm sig}} = 5.5 \times 10^{-3} d_0$, which is a few times larger than the experimental value of $h/d_0 = 1.2 \times 10^{-3}$ as found by \cite{jackiw2022prediction}, and $5 \times 10^{-5} \leq h/d_0 \leq 5 \times 10^{-4}$ by \cite{Opfer2014}. This is a limitation present in all numerical simulations of droplet aerobreakup, as is also noted in the recent work of \cite{ling2023detailed}. 

Furthermore, \cite{Jackiw2021} found experimentally that there exists an early period featuring constant growth rate of the maximum spanwise radius of the droplet $R_m$, and proposed the following model for its prediction,
\begin{equation}
    \Dot{R} = \frac{d_0 T_{\rm bal}}{8 \tau^2} \left( a^2 - \frac{128}{We} \right),
    \label{for:jackiw-r-dot}
\end{equation}
where $a$ is the axial stretching rate near the frontal stagnation point and approximated as $a \simeq 6$; $\tau$ is the characteristic deformation time introduced in \S\ref{sec:introduction}; and $T_{\rm bal}$ is the time when a constant streamwise deformation rate is reached, taken as $0.125 \tau$ according to the experimental results \citep{Jackiw2021}. We note that this model is derived assuming ellipsoidal or cylindrical droplet shape and a balance between aerodynamic and capillary forces during deformation, which leads to a purely radial internal velocity profile that cancels out the viscous effects. 

\begin{figure}
	\centering
	\subfloat[]{
		\label{fig:r_dot_Oh_0.001_comp}
		\includegraphics[width=.48\textwidth]{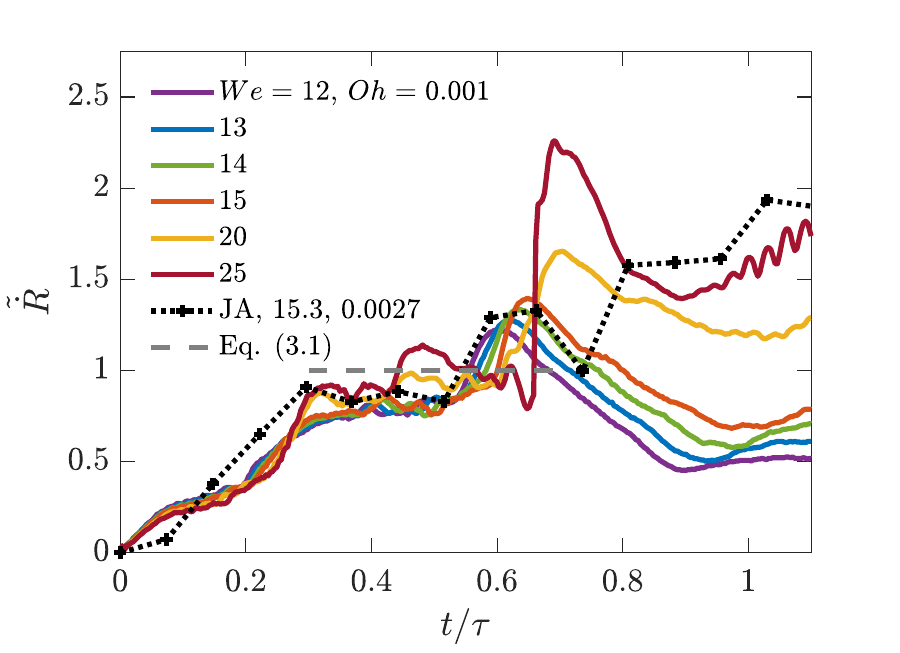}}
	\centering
	\subfloat[]{
		\label{fig:r_dot_We_15_comp}
		\includegraphics[width=.48\textwidth]{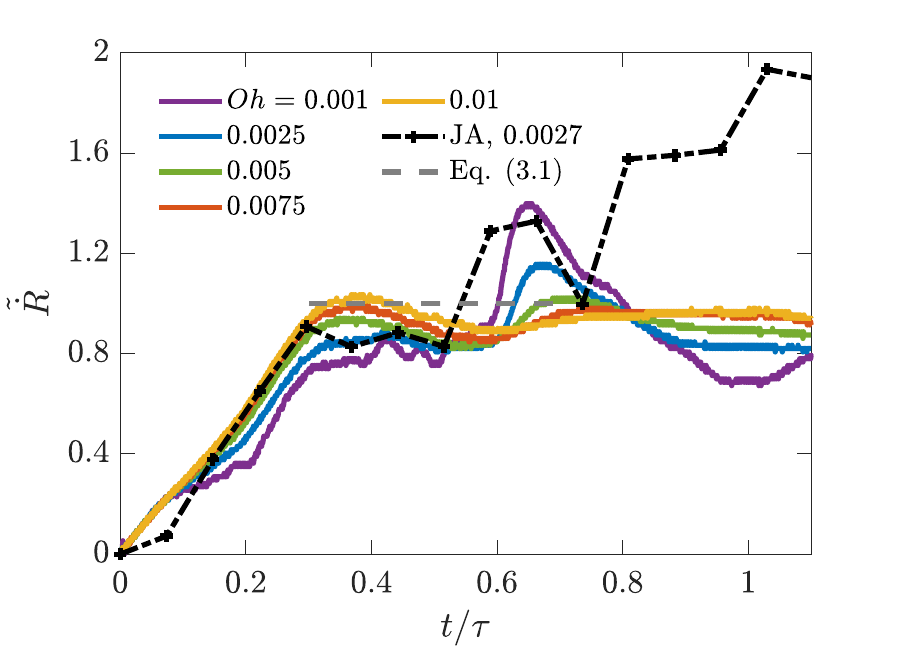}}
	\caption{Measured droplet spanwise growth rate compared with the experimental data of \cite{Jackiw2021}. Evolution of instantaneous spanwise growth rate $\Tilde{\Dot{R}}$ at various $We$ and $Oh = 10^{-3}$ (a) and various $Oh$ with $We = 15$ (b) are plotted; and the results are normalised using Eq.~\eqref{for:jackiw-r-dot}.}
	\label{fig:r_dot_comp_jackiw}
\end{figure}

We now investigate \eqref{for:jackiw-r-dot} using our axisymmetric numerical results. Figs~\ref{fig:r_dot_comp_jackiw}(a) and (b) show respectively the influence of $We$ and $Oh$ on the measured instantaneous spanwise growth rate $\Tilde{\Dot{R}}$, where the tilde indicates normalisation by the theoretical value \eqref{for:jackiw-r-dot}. We also include the growth rate evolution obtained by numerically differentiating the experimental data presented in Fig.~28 of \cite{Jackiw2021} for comparison. For the small $Oh$ value of $10^{-3}$, fig.~\ref{fig:r_dot_Oh_0.001_comp} indicates that the spanwise growth rate $\Tilde{\Dot{R}}$ reaches a plateau with relatively small variations around $t = 0.3 \tau$, where the prediction of \eqref{for:jackiw-r-dot} matches qualitatively with the measured $\Tilde{\Dot{R}}$ values. We note that while \cite{Jackiw2021} set $T_{\rm bal} = 0.125\tau$ as an \emph{a posteriori} estimation based on the evolution of $R_m$ rather than $\Tilde{\Dot{R}}_m$ when analysing their Fig.~17(b), our results agree well with the spanwise growth rate computed from their experimental data up to $t = 0.74\tau$, with their data also reaching a plateau around $t = 0.3 \tau$. The growth rate of \cite{Jackiw2021} becomes much larger than ours for $t > 0.74\tau$, corresponding to the larger $R_m$ values observed in fig.~\ref{fig:bag-width-valid}, which is possibly a result of air-phase turbulence as previously discussed. For cases at $Oh = 10^{-3}$, this period of constant $\Tilde{\Dot{R}}$ ends around $t = 0.55 \tau$, after which $\Tilde{\Dot{R}}$ reaches a peak aound $t = 0.6\tau$ and then decreases, indicating a deviation from model \eqref{for:jackiw-r-dot} absent in the analyses of \cite{Jackiw2021}. On the other hand, fig.~\ref{fig:r_dot_We_15_comp} suggests that as $Oh$ increases beyond $2.5 \times 10^{-3}$, the late-time peaking of $\Tilde{\Dot{R}}$ gradually attenuates, while the match with \eqref{for:jackiw-r-dot} is improved and maintained for longer periods of time, which is particularly interesting as \eqref{for:jackiw-r-dot} is derived based on inviscid flow assumptions and cannot account for viscous influences. Note that \cite{Jackiw2021} tested droplets for which $Oh = 2.7 \times 10^{-3}$; our numerical results are therefore consistent with their experiment.

Returning to fig.~\ref{fig:contour_Oh_0.001} suggests that during the period $0.3\tau \leq t \leq 0.55\tau$ when the constant growth rate $\Tilde{\Dot{R}}$ is observed, the liquid is being pushed from the frontal surface to the windward side of the periphery, where the maximum spanwise radius is reached. However, at $t = 0.6\tau$ when the peaking behaviour is observed, a bulge appears downstream and causes a location shift where the maximum spanwise radius $R$ is reached. This bulging behaviour is also present in the growth rate evolution computed from the experimental data of \cite{Jackiw2021} at $Oh = 2.7 \times 10^{-3}$, but virtually absent when $Oh = 10^{-2}$ in our numerical simulations, as shown in fig.~\ref{fig:contour_Oh_0.01}, where the periphery of the droplet contour only flattens over time.

\begin{figure}
	\centering
	\subfloat[]{
		\label{fig:slice_peak_12_15}
		\includegraphics[width=.48\textwidth]{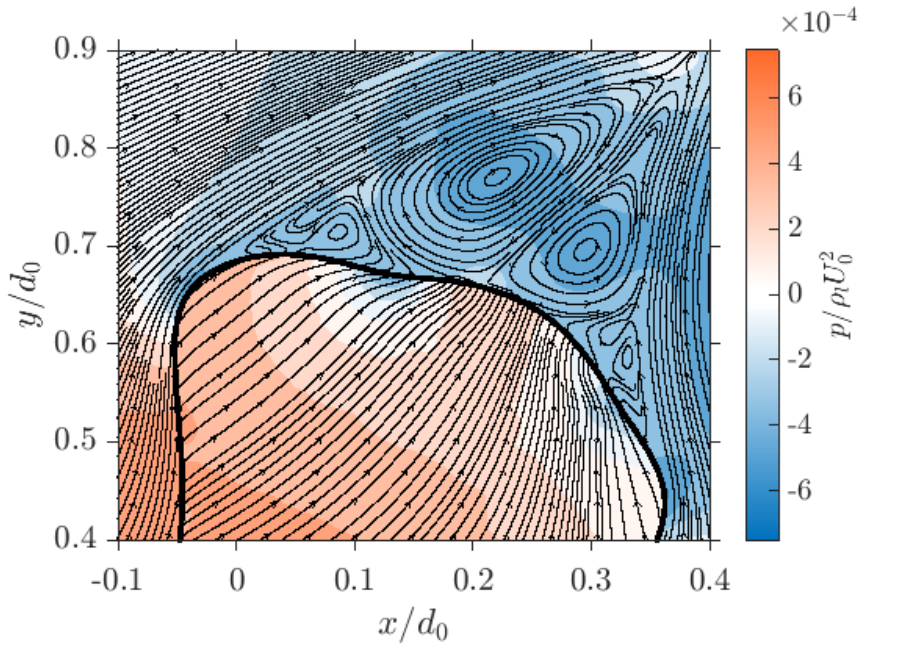}}
	\centering
	\subfloat[]{
		\label{fig:slice_peak_25_15}
		\includegraphics[width=.48\textwidth]{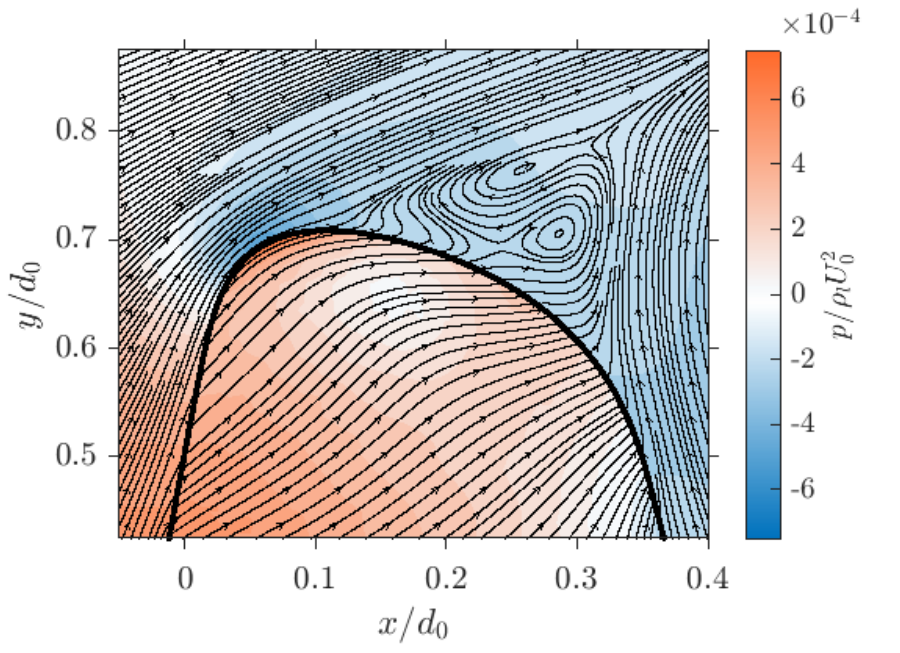}}
	\caption{Flow fields near the tip of a droplet with $We = 20, \, Oh = 10^{-3}$ (a) and $We = 15, \, Oh = 10^{-2}$ (b) when the peaks in $\Dot{R}$ are reached. The non-dimensional times at which (a) and (b) are taken are respectively $t/\tau = 0.62$ and 0.66.}
	\label{fig:slice_peak}
\end{figure}

To provide insights into physical mechanisms governing the peak in the spanwise growth rate observed for low Oh values in fig.~\ref{fig:r_dot_comp_jackiw}, we plot in fig.~\ref{fig:slice_peak} the pressure distribution and streamlines near the drop periphery, when the peaks in the spanwise growth rate $\Tilde{\Dot{R}}$ are reached in fig.~\ref{fig:r_dot_We_15_comp} for $We = 20, \, Oh = 10^{-3}$ and $We = 15, \, Oh = 10^{-2}$. It can be seen that the surrounding gas flow separates from the droplet surface at the windward side of the periphery, creating attached recirculating vortices in its wake with low pressure and slow fluid motion \citep{Marcotte2019, Jain2019}, where the bulges are located. The pressure difference in the surrounding flow between the frontal stagnation point and the recirculating region drives the internal flow within the droplet from the windward surface to the periphery. Furthermore, the peaks in $\Tilde{\Dot{R}}$ observed when $Oh \leq 0.005$ are associated with the formation of a high-pressure region at the bulge on the droplet periphery, as can be seen in fig.~\ref{fig:slice_peak_12_15}, which is caused by surface tension and decelerates the flow into the bulge. Further development of the bulge leads to an increase in the local capillary pressure, which causes the decrease in $\Tilde{\Dot{R}}$ after the peak. Notably, the droplet contour in fig.~\ref{fig:slice_peak_25_15} at $Oh = 10^{-2}$ lacks craters at the axis of symmetry and bulges at the periphery, and therefore more closely resembles the cylindrical shape of the deforming drop assumed in the derivations of \citep{Jackiw2021}, which may explain why the match with the inviscid model \eqref{for:jackiw-r-dot} is improved as $Oh$ is increased. 

\begin{figure}
	\centering
	\subfloat[]{
		\label{fig:pressure_Oh_0.001}
		\includegraphics[width=.48\textwidth]{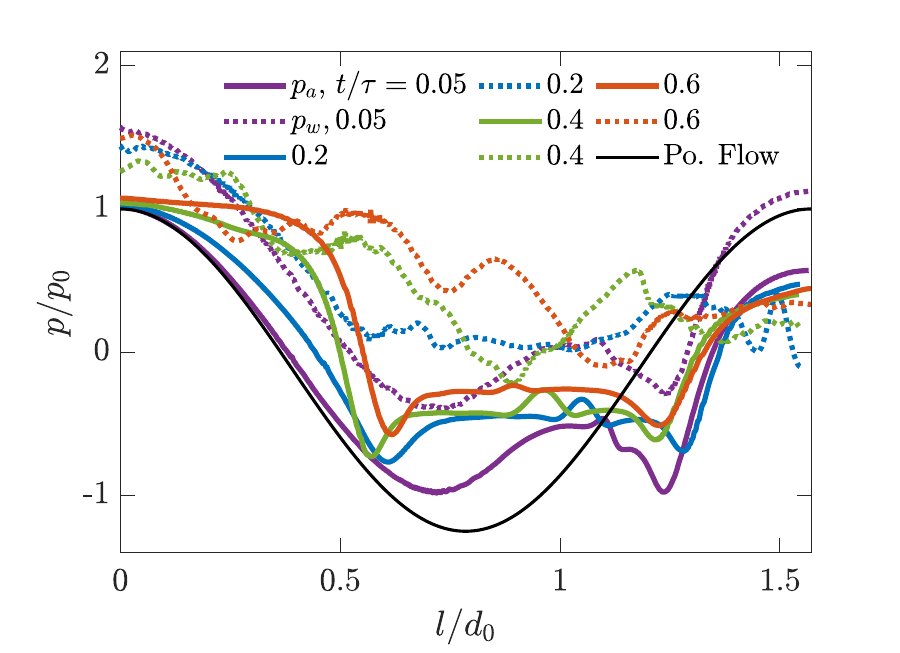}}
	\centering
	\subfloat[]{
		\label{fig:uz_z_Oh_0.001}
		\includegraphics[width=.48\textwidth]{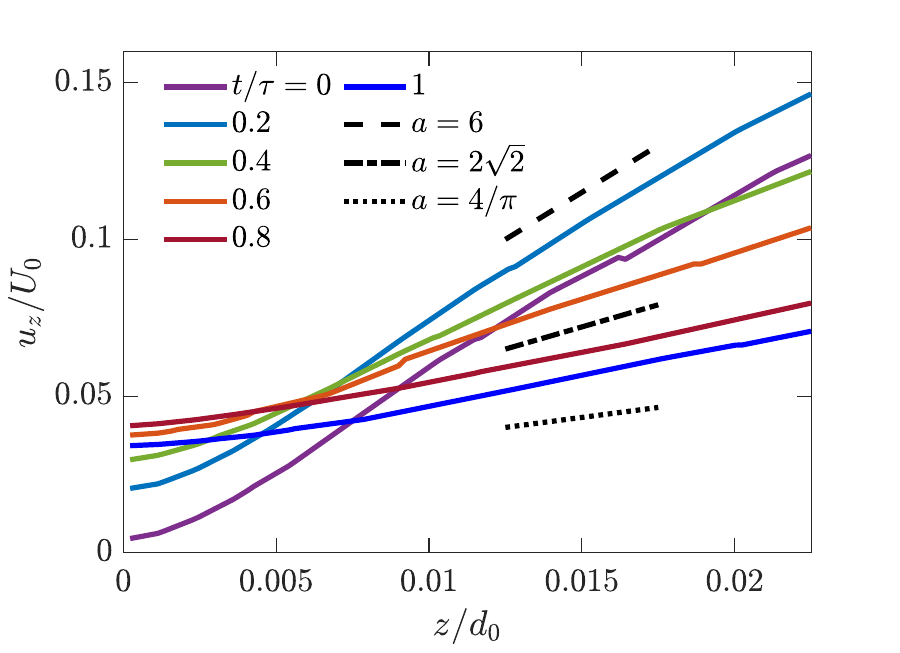}}
		
	\caption{(a): Evolution of air ($p_a$, solid lines) and liquid pressures ($p_w$, dotted lines) on either side of the droplet interface as functions of the interfacial arc length $l$; (b): axial airflow velocity $u_z$ on the axis of symmetry as a function of the distance to the windward stagnation point of the droplet $z$. The values of $We$ and $Oh$ are respectively 15 and $10^{-3}$.}
	\label{fig:p_u_We_15}
\end{figure}

We further investigate the distribution patterns of flow pressure and velocity in the vicinity of the droplet, and their association with prediction~\eqref{for:jackiw-r-dot} of \cite{Jackiw2021}. Figure~\ref{fig:pressure_Oh_0.001} shows the air- and liquid-phase pressure on either side of the drop surface as functions of the arclength $l$ traversing along the axisymmetric droplet contour in the clockwise direction. It can be seen that at very early time $(t/\tau = 5 \times 10^{-2})$, the air-phase pressure profile closely follows the sinusoidal potential-flow solution for $l/d_0 \leq 0.6$, which corresponds to the windward face of the drop; whereas the profile at $l/d_0 > 0.6$ deviates from the potential-flow solution due to flow separation, characterised by a second minimum around $l/d_0 = 1.2$. We also note that at $t/\tau = 5 \times 10^{-2}$ the shape of the liquid-phase pressure profile bears strong resemblance to its airphase counterpart, with a nearly-uniform upshift due to the constant capillary pressure difference $4\sigma/d_0$. As the droplet flattens over time, the air-phase pressure profile on the windward surface increases and the first minimum moves upstream, deviating from the potential-flow solution. In the meantime, the change in liquid-phase pressure for $l/d_0 \leq 0.37$ is relatively small, and the air- and liquid-phase pressure profiles cross over each other at $l/d_0 = 0.37$ and $t/\tau = 0.4$, signalling the dimple formation on the windward surface as the local radius of curvature reaches infinity. It is also noted that the minimum of the liquid-phase pressure profile around $l/d_0 = 0.85$ observed at $t/\tau = 0.4$ becomes a maximum at $t/\tau = 0.6$, which corresponds to the bulge formation observed in fig.~\ref{fig:slice_peak_12_15} that leads to the deviation from \eqref{for:jackiw-r-dot} \citep{Jackiw2021}.

Figure~\ref{fig:uz_z_Oh_0.001} shows the airphase axial velocity $u_x$ measured on the axis of symmetry as a function of the distance to the windward stagnation point of the drop $z/d_0$, where the slope of the curves corresponds to the axial stretching rate $a$ used in model~\eqref{for:jackiw-r-dot}. It is first observed that the axial velocity value at $z/d_0 = 0$ increases gradually over time, which is because the measuring point is located in the airphase boundary layer attached to the accelerating droplet. The axial stretching rate $a$ is found to gradually decrease from 6 and approach $4/\pi$, the extreme values corresponding to spherical and pancake drop shapes as noted in \cite{Jackiw2021}; crossing over the intermediate value of $2\sqrt{2}$ proposed by \cite{kulkarni2014bag}. The decrease of $a$ corresponds to the air-phase pressure increase on the windward surface as observed in fig.~\ref{fig:pressure_Oh_0.001} via the following equation:
\begin{equation}
    p_g(r) - p_g(0) = -\rho_g \frac{a^2U_0^2}{8d_0^2}r^2.
\end{equation}

Consequently, we conclude that our numerical results reproduce the prediction \eqref{for:jackiw-r-dot} of \cite{Jackiw2021} that there exists a period characterised by a constant spanwise radius growth rate $\Tilde{\Dot{R}}$. Furthermore, we find that the later deviation from \eqref{for:jackiw-r-dot} is characterised by a peak in $\Tilde{\Dot{R}}$, which is caused by the capillary deceleration of liquid influx into the drop periphery that causes bulge formation at low $Oh$ values. The increase in $Oh$ eliminates the bulge and the frontal crater on the droplet surface, and the droplet acquires a nearly cylindrical shape which is one of the underlying assumptions by \cite{Jackiw2021} when deriving \eqref{for:jackiw-r-dot}, hence the better match with their model.

\subsection{Film drainage and onset of bag breakup}
\label{subsubsec:drainage}

\begin{figure}
	\centering
	\subfloat[]{
		\label{fig:bag_devel_We_15}
		\includegraphics[width=.48\textwidth]{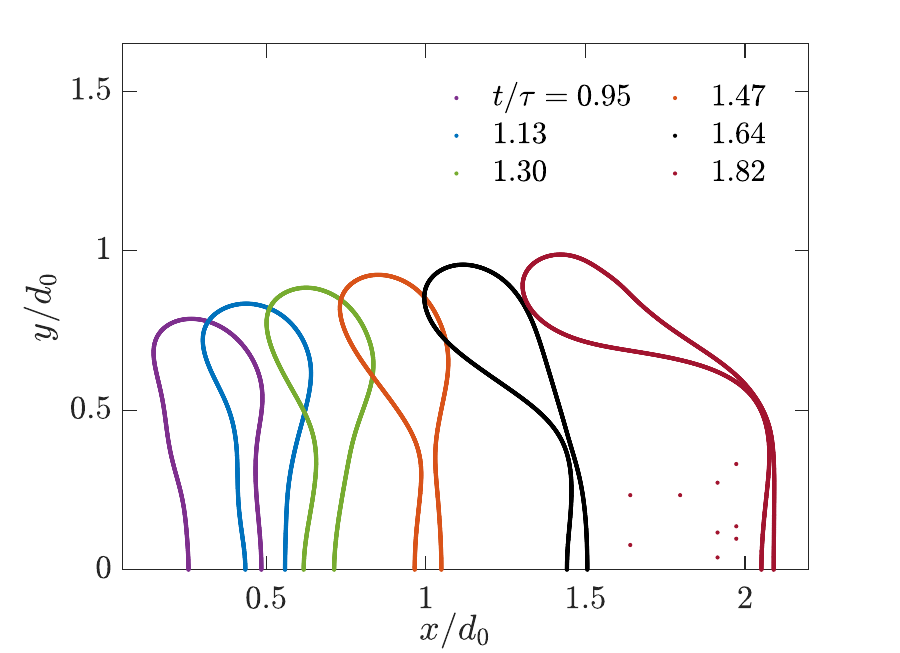}}	
    \centering
	\subfloat[]{
		\label{fig:bag_devel_t_100}
		\includegraphics[width=.48\textwidth]{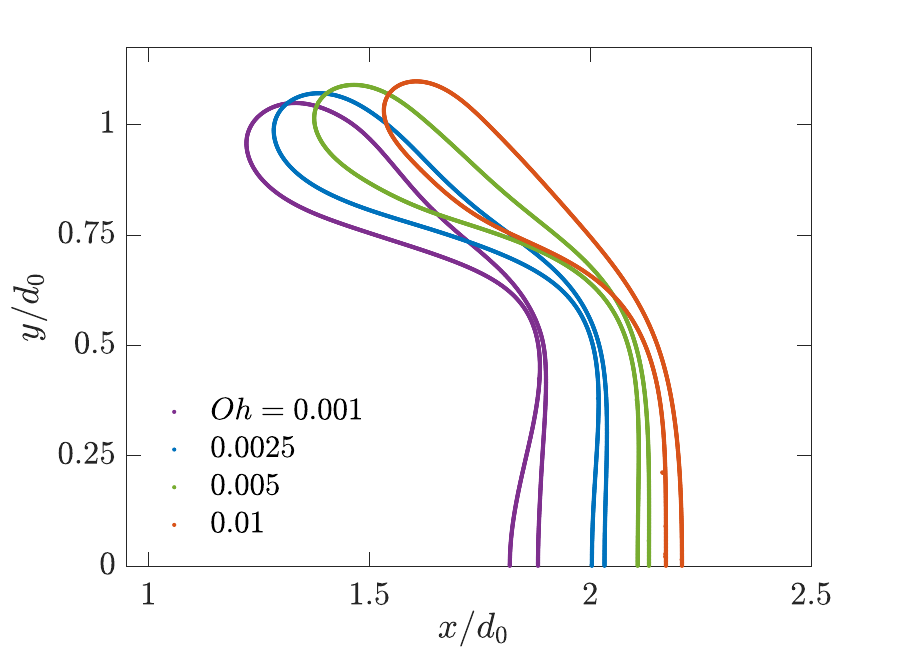}}
		
	\caption{(a): Evolution of axisymmetric droplet contours with $We = 15$ and $Oh=10^{-3}$, where the axis of symmetry is at $y=0$. (b): droplet contours at $t/\tau = 1.73$ with various $Oh$ values and $We = 15$.}
	\label{fig:bag_devel}
\end{figure}

When the droplet deforms into a disc at the end of the early-time deformation period, corrugations develop on their frontal surfaces which have generally been considered as RT perturbation waves \citep{Yang2017}. This appears in fig.~\ref{fig:bag_devel_We_15} at $t/\tau =0.95, \, We=15$, in the form of waves on the windward surface of the droplet. Later on, thick rims are observed to form at the drop periphery due to capillary pinching effects, which extract liquid from the drop centre and contribute to the formation of bag films near the axis of symmetry. Subject to the aerodynamic pressure difference between their frontal and leeward surfaces, these films further bulge out from the rim \citep{Jackiw2021} and cause exponential growth of streamwise bag length before breakup. 

Figure~\ref{fig:bag_devel_We_15} shows non-uniform profiles of the bag thickness $h$, featuring a neck where a local minimum in $h$ is reached and the film breakup eventually occurs. The neck moves outwards radially at $We = 15$, leaving a thickening remnant stamen structure developing at the axis of symmetry \citep{Marcotte2019}. Figure~\ref{fig:bag_devel_t_100}, on the other hand, shows the deformed drop contours at $t/\tau = 1.73$ for various $Oh$ values; and it can be seen that as $Oh$ increases, the neck becomes less obvious as the distribution of bag film thickness becomes more uniform.

\begin{figure}
	\centering
	\subfloat[]{
		\label{fig:neck_Oh_0.001}
		\includegraphics[width=.48\textwidth]{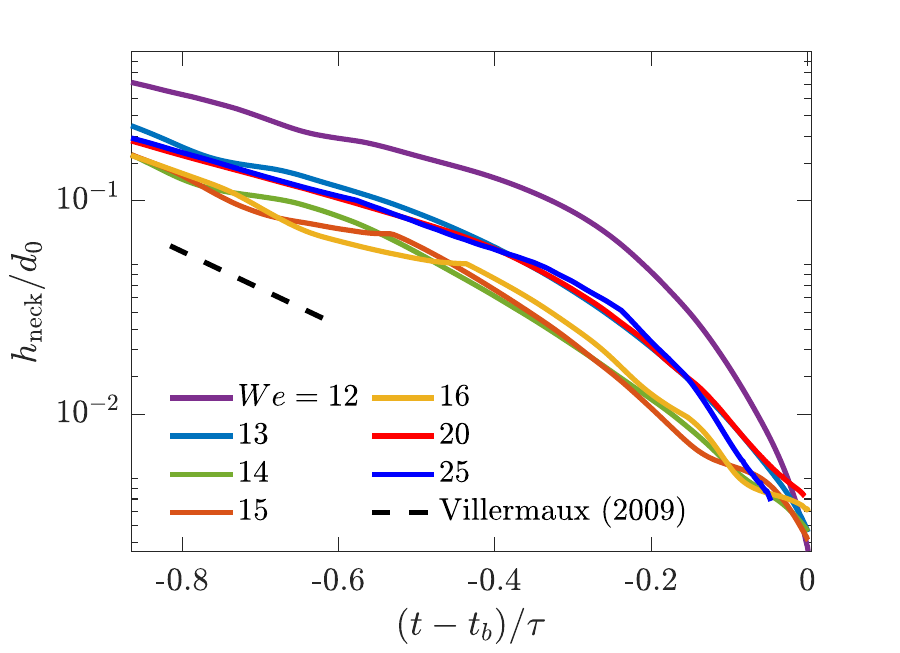}}
	\centering
	\subfloat[]{
		\label{fig:neck_We_12}
		\includegraphics[width=.48\textwidth]{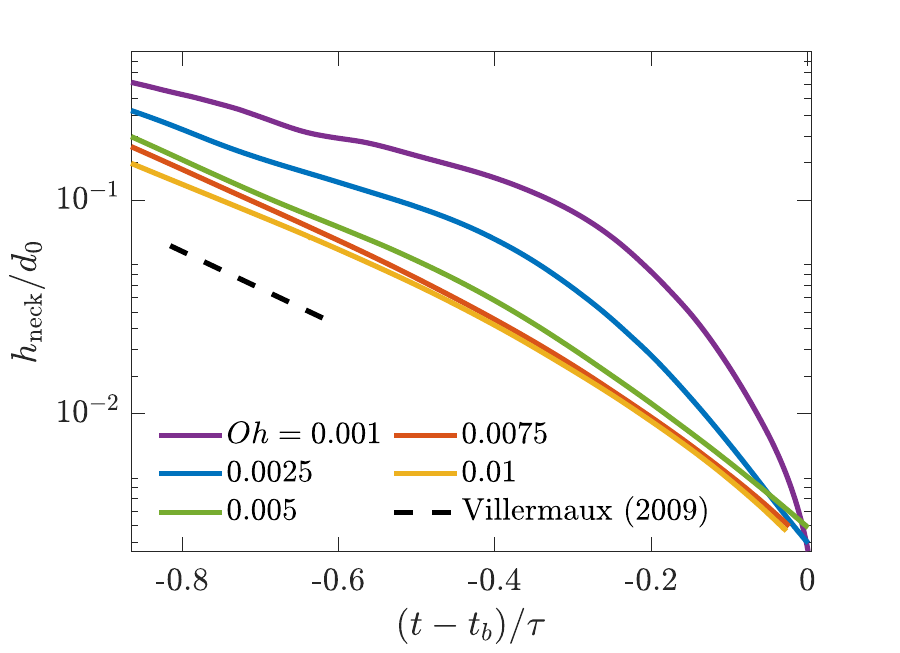}}	

	\caption{The evolution of film thickness $h$ for $t_b - 0.87\tau \leq t \leq t_b$, measured from simulations with various $We$ with $Oh = 10^{-3}$ (a) and various $Oh$ with $We = 15$ (b). For a droplet with $We = 15, \, Oh = 0.001$, the breakup time is $t_b/\tau = 1.84$, and $t_b - 0.87\tau = 0.97\tau$. As fig.~\ref{fig:bag_devel_We_15} shows, over this period a bag is blown out from the centre of the flattened disc. Villermaux and Bossa's prediction \eqref{for:villermaux-neck} is also plotted for comparison.}
	\label{fig:neck_decay}
\end{figure}

It has been argued that the breakup of bag films is due to an RT instability peculiar to thin films rather than a finite-time singularity of the Navier-Stokes equations \citep{Villermaux2009}. Assuming inviscid flow and uniform bag thickness, \cite{Villermaux2009} derived the following exponential decay model for the film thickness $h$,
\begin{equation}
    h(t) \sim d_0 e^{-\lambda t},
    \label{for:villermaux-neck}
\end{equation}
where the exponential decay rate $\lambda$ is given as $4/\tau$. We compare our numerical results with the predictions of model \eqref{for:villermaux-neck} in fig.~\ref{fig:neck_decay}. The film thickness $h$ is calculated by measuring the minimum distance between the windward and leeward surfaces of the deformed drop contour over a time period of $t_b - 0.87\tau \leq t \leq t_b$, where $t_b$ is the time when film breakup is detected. Logarithmic scale is used for the $y$ axis to facilitate comparison of the decay rate $\lambda$.

It can be seen that for the $We$ and $Oh$ range presented in fig.~\ref{fig:neck_decay}, the exponential decay rate $\lambda$ is initially close to the prediction of \eqref{for:villermaux-neck}. This phase, which features a constant thickness decay rate, roughly corresponds to the period of rim development prior to the `bulging' of bag films, which is shown in fig.~\ref{fig:bag_devel}. However, the decay rate increases as the film continues thinning and approaches breakup, similar to the result of \cite{kant2022bags}, which becomes more significant as $We$ and $Oh$ decrease. Most notably, at $We = 12$ the thinning rate continuously increases close to the onset of breakup, which suggests that an exponential decay law in the form of \eqref{for:villermaux-neck} does not fully capture the underlying physics for film drainage with strong surface tension. 

Developing new theoretical models in place of \eqref{for:villermaux-neck} whose predictions match better with the late-time neck drainage behaviour observed within the $We$ and $Oh$ range of interest is out of the scope of the current work. We note briefly that the drainage behaviour of bag films under the influence of aerodynamic pressure difference observed here bears resemblance to the drainage of liquid films between a free air-water surface and a buoyancy-driven air bubble \citep{pigeonneau2011low, kovcarkova2013film, guemas2015slow}, where film drainage models are developed based on lubrication assumptions (see Section 4.2 of \citep{magnaudet2020particles} and references therein for more detailed discussions). More specifically, \cite{pigeonneau2011low} also showed a deviation from exponential decay of bubble film thickness under asymptotically large surface tension, which is ascribed to a finite-time singularity and contrasts with the thin-film RT instability mechanism proposed by \cite{Villermaux2009}. However, the major difference between the bag and the bubble film drainage problem lies in the location of the neck. For the drainage of bag films, the neck can be formed some distance away from the axis of symmetry due to a competition of inertia between the outer rim and the inner stamen \citep{Marcotte2019}, which complicates theoretical modelling due to additional difficulties in predicting time-varying neck locations. In contrast, bubble film drainage always occurs on the axis of symmetry, as the thin bubble film is connected to an infinitely large and quiescent liquid domain.

\section{Breakup of bag films}
\label{sec:film}

In this section, we analyse our three-dimensional simulation results during the period when the bag film undergoes disintegration to form small fragments, and the axisymmetric flow assumption completely breaks down. It is noted that the late stage of aerobreakup is not covered when the receding remnant bag collides with the surrounding main rim and triggers the fragmentation of the latter \citep{jackiw2022prediction}, which will be reserved for future work.

\subsection{Grid convergence for fragment statistics}
\label{subsec:grid-conv}

While the physical mechanisms responsible for the onset of liquid film breakup in general remain an active research topic, with various candidates proposed including chemical or thermal inhomogeneities \citep{kant2022bags}, Marangoni effects \citep{lhuissier2012bursting} or presence of surface contamination \citep{neel2018spontaneous}, it has been argued that for bag films under normal acceleration, thickness modulations arise across the film due to the RT instability, resulting in perforation when the perturbation amplitude becomes comparable to the film thickness $h$ \citep{Villermaux2009, Jackiw2021}. 

\begin{figure}
	\centering
	\subfloat[]{
		\label{fig:snap_L12_no_chirco}
		\includegraphics[width=.32\textwidth]{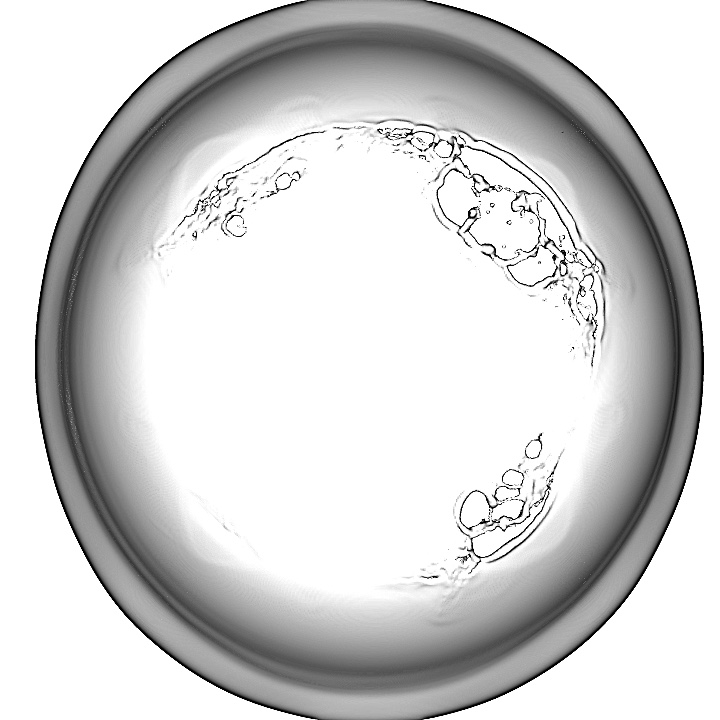}}
	\centering
	\subfloat[]{
		\label{fig:snap_L12}
		\includegraphics[width=.32\textwidth]{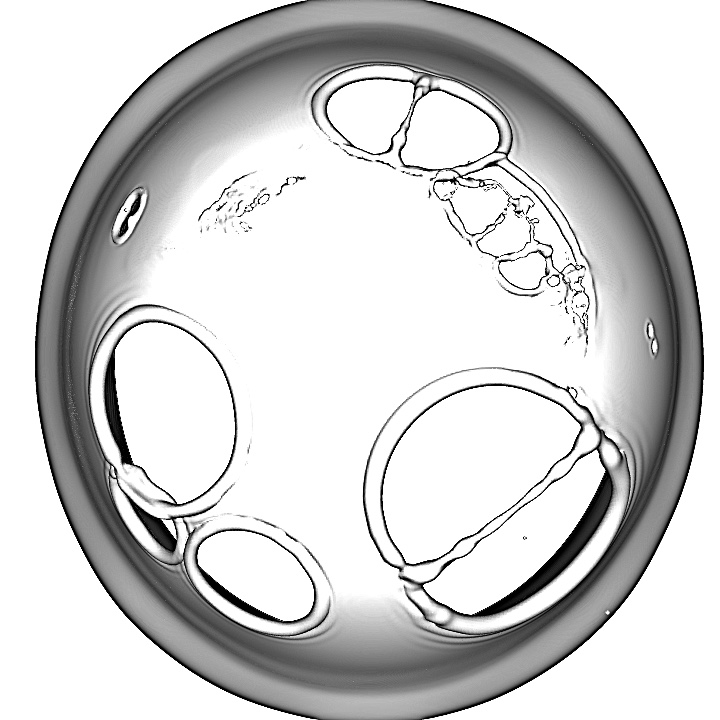}}
	\centering	
	\subfloat[]{
		\label{fig:snap_L13}
		\includegraphics[width=.32\textwidth]{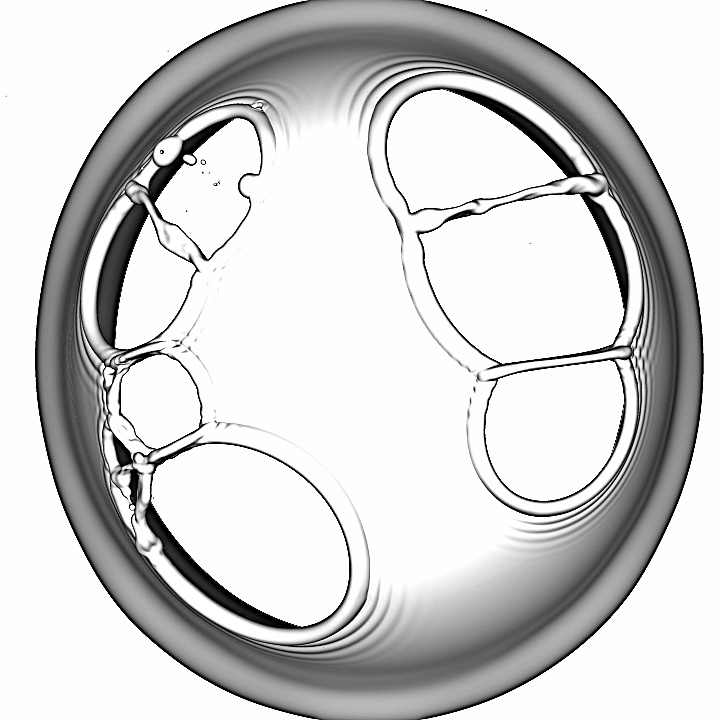}}

	\caption{Effect of the MD algorithm on the bag breakup behaviour at grid level $L=12$ and 13 for $We = 15, \, Oh = 10^{-3}$. (a)-(c): Simulation snapshots showing fragmenting bag films at $t/\tau = 1.909$ without (a) and with artificial perforation (b,c). The grid resolution level is $L=12$ for (a,b) and 13 for (c), while the MD signature level for (b,c) is $L_{\rm sig} = 12$.}
	\label{fig:3D_convergence}
\end{figure}

Numerically, the perforation and subsequent fragmentation of thin films in droplet breakup problems has historically been challenging to represent in a physically consistent manner. Given that perforation involves a topological change in the air-water interface, numerical studies have usually employed interface-capturing techniques such as the geometric VOF approach employed here \citep{tang2021effects, Jain2015, Jain2019}; such methods can represent interfacial topological change without a need for extensive special treatment. However, such techniques also tend to suffer from an unphysical and numerically uncontrolled perforation and fragmentation mode in thin films, which moreover compromises numerical convergence of the statistics of the resulting fragment populations \citep{chirco2021manifold}. In this phenomenon, when the film thickness approaches the local mesh size, it begins to destabilise, generating a large number of small fragments without a well-defined fragmentation mechanism; fig.~\ref{fig:snap_L12_no_chirco} shows a qualitative illustration of this phenomenon in our own simulations. This phenomenon also appears in images of droplet breakup in \cite{Jain2015}; unfortunately, information on numerical convergence of fragment statistics is not supplied in that study.

The manifold death (MD) algorithm constructed and implemented by \cite{chirco2021manifold} into Basilisk aims to bypass this spurious mode of fragmentation. This algorithm detects and artificially perforates thin films periodically by removing liquid mass once their thickness decreases to a prescribed critical value. The key point is that VOF breakup is circumvented because the film is perforated at a thickness greater than what is required for VOF breakup to occur. A limitation of the MD method is that it removes mass from the droplet in order to generate the hole, which disturbs the momentum and mass conservation properties of the VOF scheme as implemented in Basilisk. We find in practice that the parameters of the MD algorithm can be adjusted so that sufficiently few holes are formed in these simulations and this mass loss becomes insignificant (see below).

The holes created by the MD perforation mechanism resemble those appearing in experiments (such as \cite{lhuissier2013effervescent, jackiw2022prediction, kant2022bags}) - see fig.~\ref{fig:snap_L12}. The algorithm also affords considerable user control over the frequency and location, for example, of the perforations. In the present study, we control perforation frequency using a probabilistic approach which is scaled to be independent of parallelisation, resolution, and the calling interval of the algorithm, in order to minimise the number of free parameters governing the perforation problem. The appropriate choice of probability to match the efficiency of hole generation in experimental studies, such as those seen in \cite{lhuissier2012bursting, vledouts2016explosive}, is a complicated problem which is left for future work. Our aim in this study is instead to establish numerical convergence and verify various aspects of the resulting fragmentation process with experiment and theory, which in our knowledge has not been established in previous numerical studies.

All grid convergence test cases we conduct in this section are reloaded from a single three-dimensional simulation snapshot with an intact bag at $We = 15, \, Oh = 10^{-3}$, and the perforation probability and calling interval for the MD algorithm are set as $p_{\rm perf} = 5.7 \times 10^{-5}$ and $\Delta t_c = 0.5d_0/U_0$, with grid level $L = 12, \, 13, \, 14$ and signature level $L_{\rm sig} = 11, \, 12, \, 13$ (see \S\ref{sec:formulation} for their definition). The fragment statistics are collected and output at fixed time intervals until the bags have fully disintegrated, and then post-processed to obtain time- and/or ensemble-averaged data. It is noted that as the test cases run without using the MD algorithm are deterministic in the sense that VOF breakup appears repeatably in the same locations of the thin film, while producing a large number of very small fragments, only one realisation is completed at each $L$ value; whereas the MD algorithm introduces randomness in the perforation location and subsequent fragment formation, while also reducing the number of fragments, and therefore multiple realisations are completed for a given set of $L$ and $L_{\rm sig}$ to generate sufficient total number of fragments for ensemble-averaging. A full list for the configurations of the three-dimensional numerical simulations is available in Table~\ref{tab:ensembles}.

\begin{table}
\setlength{\tabcolsep}{1.25em}
\centering
\begin{tabular}{c|c|c|c|c|c}
    $We$ & $Oh$ & $L$ & $L_{\rm sig}$ & Realisation No. & Category \\ [0.5ex] 
    \midrule
    15 & $10^{-3}$ & 12 & N/A & 1 & Convergence - VOF \\
    15 & $10^{-3}$ & 13 & N/A & 1 & Convergence - VOF \\
    15 & $10^{-3}$ & 14 & N/A & 1 & Convergence - VOF \\
    15 & $10^{-3}$ & 13 & 11 & 10 & Convergence - MD \\
    15 & $10^{-3}$ & 13 & 12 & 5 & Convergence - MD \\
    15 & $10^{-3}$ & 13 & 13 & 3 & Convergence - MD \\
    15 & $10^{-3}$ & 14 & 13 & 7 & Convergence - MD \\ [0.5ex] 
    15 & $10^{-3}$ & 14 & 13 & 3 & $Oh$ Study \\
    15 & $10^{-4}$ & 14 & 13 & 2 & $Oh$ Study \\
    15 & $5 \times 10^{-4}$ & 14 & 13 & 1 & $Oh$ Study \\
    15 & $10^{-3}$ & 14 & 13 & 3 & $Oh$ Study \\
    15 & $5 \times 10^{-3}$ & 14 & 13 & 3 & $Oh$ Study \\
    15 & $10^{-2}$ & 14 & 13 & 3 & $Oh$ Study \\
    15 & $2.5 \times 10^{-2}$ & 14 & 13 & 3 & $Oh$ Study \\
    15 & $5 \times 10^{-2}$ & 14 & 13 & 3 & $Oh$ Study \\
    15 & $7.5 \times 10^{-2}$ & 14 & 13 & 1 & $Oh$ Study \\
\end{tabular}
\caption{List of ensemble realisations for three-dimensional numerical simulations carried out in this work, where the drop Weber and Ohnesorge numbers $We$ and $Oh$, the grid and signature levels $L$ and $L_{\rm sig}$, the number of individual realisations, and the purpose for using the ensemble data (the grid convergence study for \S\ref{subsec:grid-conv} and \S\ref{subsec:frag-tracking}, or the $Oh$ effect study in \S\ref{subsec:Oh-effects}) are indicated.}
\label{tab:ensembles}
\end{table}

We first demonstrate the effects of the MD algorithm on the breakup behaviour of the bag film in fig.~\ref{fig:3D_convergence}. The film rupture behaviour is qualitatively different with and without application of the MD algorithm. Figure~\ref{fig:snap_L12_no_chirco} is a snapshot for a simulation case run without using the MD algorithm, featuring numerous small-scale irregular corrugations and ligament breaking on the bag, which reflect the uncontrolled nature of VOF breakup. Figures.~\ref{fig:snap_L12} and \ref{fig:snap_L13} show that the MD algorithm is able to create holes on the bag film in a controlled manner, and reduce the influence of VOF breakup on the bag dynamics. These holes created by the MD algorithm feature well-defined bordering rims that recede over the bag and create surface capillary waves ahead of them, and these may collide with one another and form a few long stretching liquid bridges \citep{agbaglah2021breakup}, which are distinct from the numerous short and irregular bridges observed with VOF breakup. Note that fig.~\ref{fig:snap_L12} still shows some VOF breakup behaviour which is absent in fig.~\ref{fig:snap_L13}. This reflects the fact that, even though the film is perforated when the film thickness reaches the order of $3\Delta_{\rm sig}$, the perforation probability and the rate of hole expansion are sufficiently low such that there are regions of the film that continue to thin down to the order of $\Delta$, where VOF breakup begins. In fig.~\ref{fig:snap_L12}, $\Delta = \Delta_{\rm sig}$, so that VOF-induced fragmentation still appears, but this can be further minimized by choosing $L > L_{\rm sig}$, such as in fig.~\ref{fig:snap_L13}. 

\begin{figure}
    \centering
	\subfloat[]{
		\label{fig:size_pdf_vof}
		\includegraphics[width=.48\textwidth]{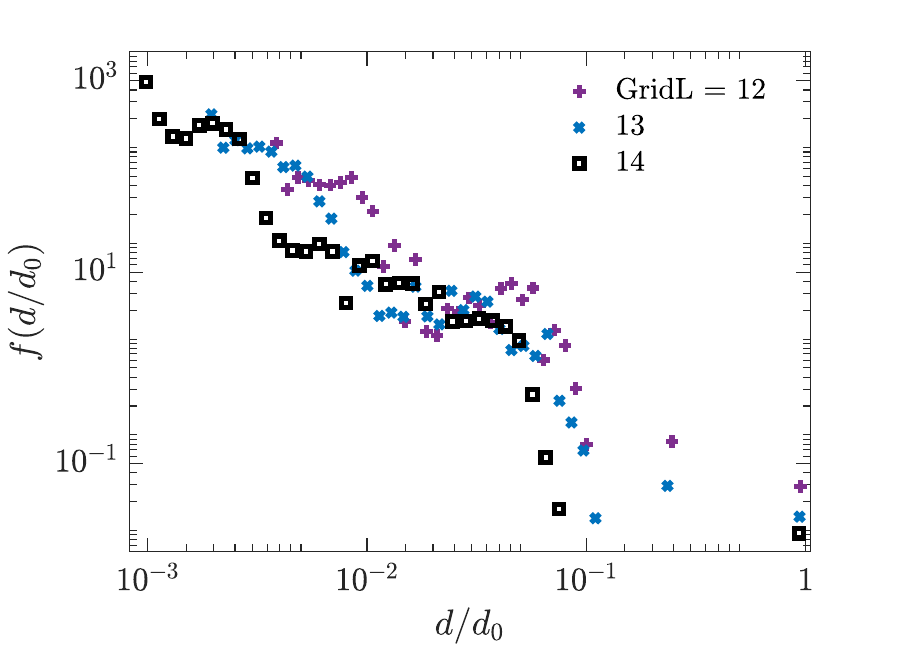}}
	\centering
	\subfloat[]{
		\label{fig:vbar_pdf_vof}
		\includegraphics[width=.48\textwidth]{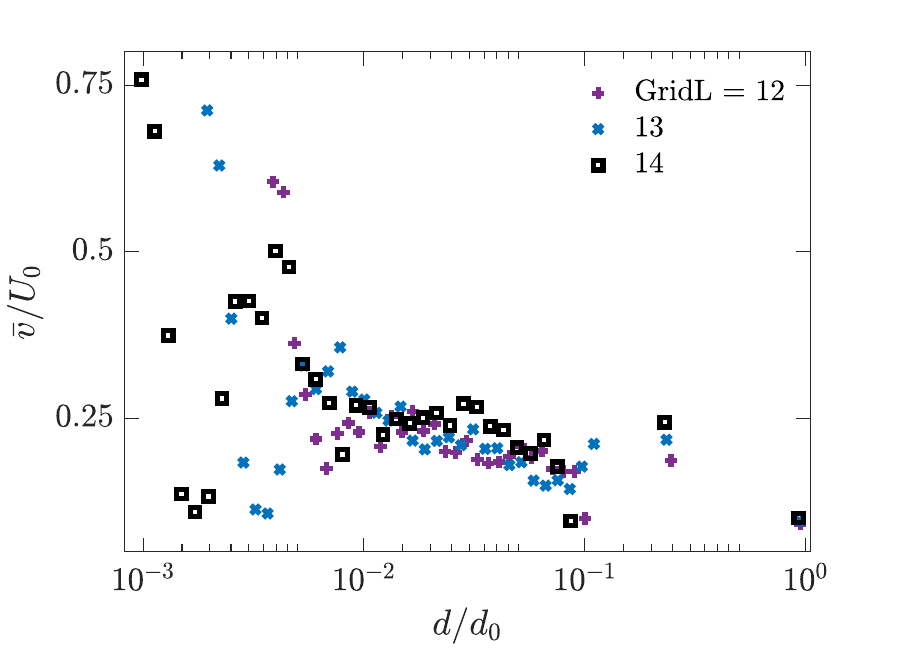}}

    \centering
	\subfloat[]{
		\label{fig:size_pdf_single}
		\includegraphics[width=.48\textwidth]{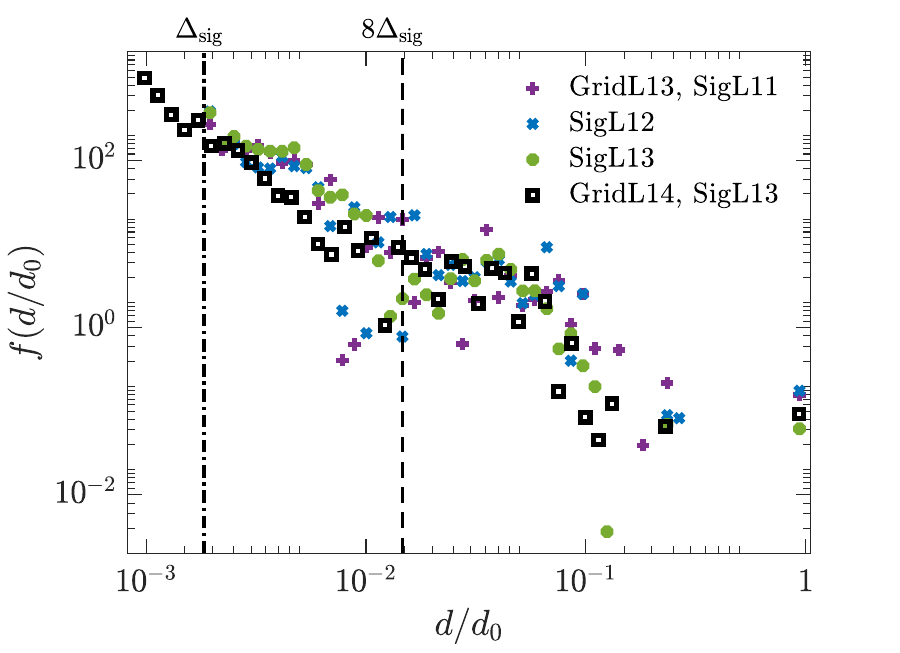}}
	\centering
	\subfloat[]{
		\label{fig:vbar_pdf_single}
		\includegraphics[width=.48\textwidth]{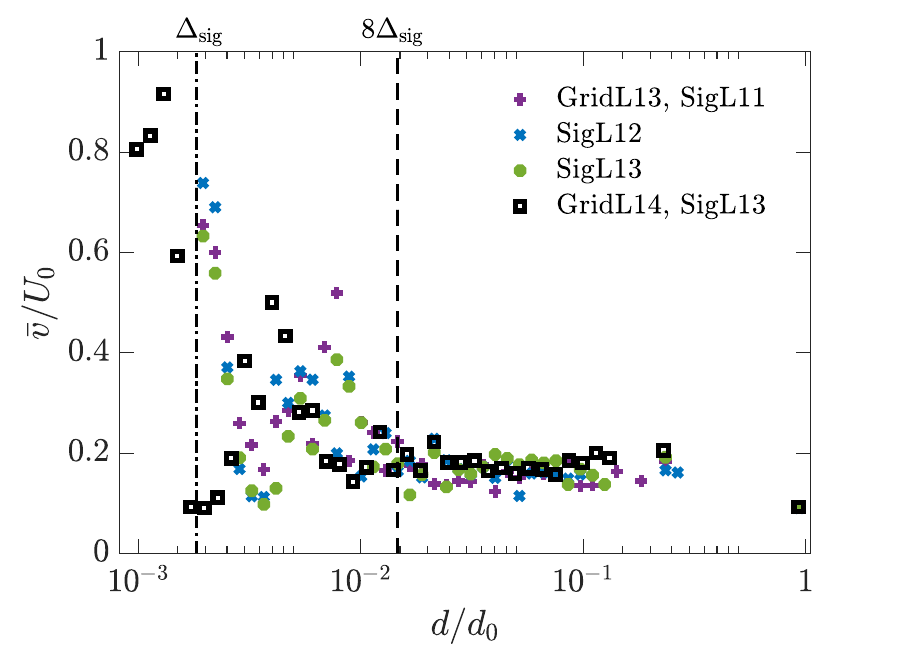}}	

	\centering
	\subfloat[]{
		\label{fig:size_pdf}
		\includegraphics[width=.48\textwidth]{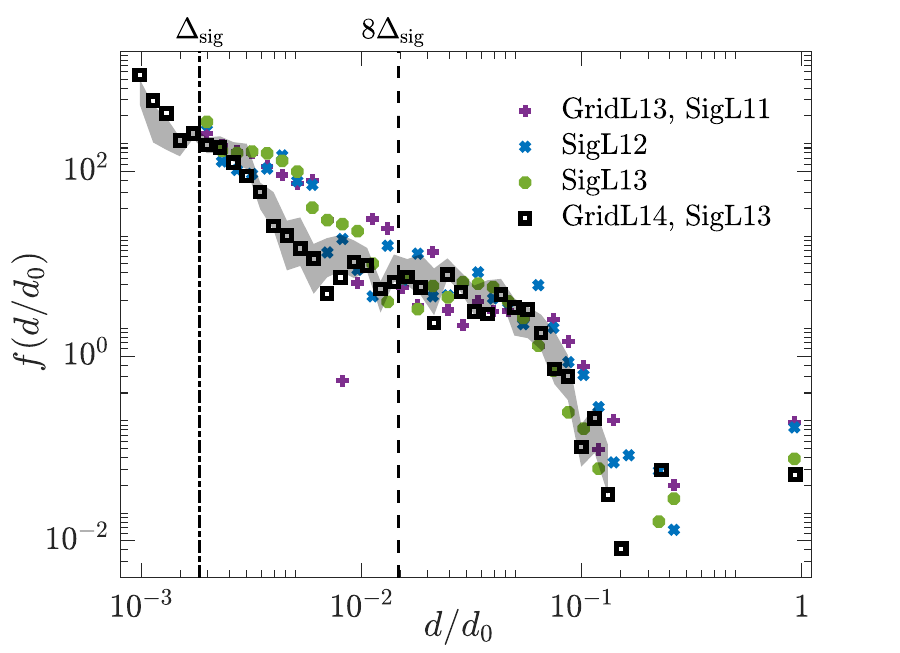}}
	\centering
	\subfloat[]{
		\label{fig:vbar_pdf}
		\includegraphics[width=.48\textwidth]{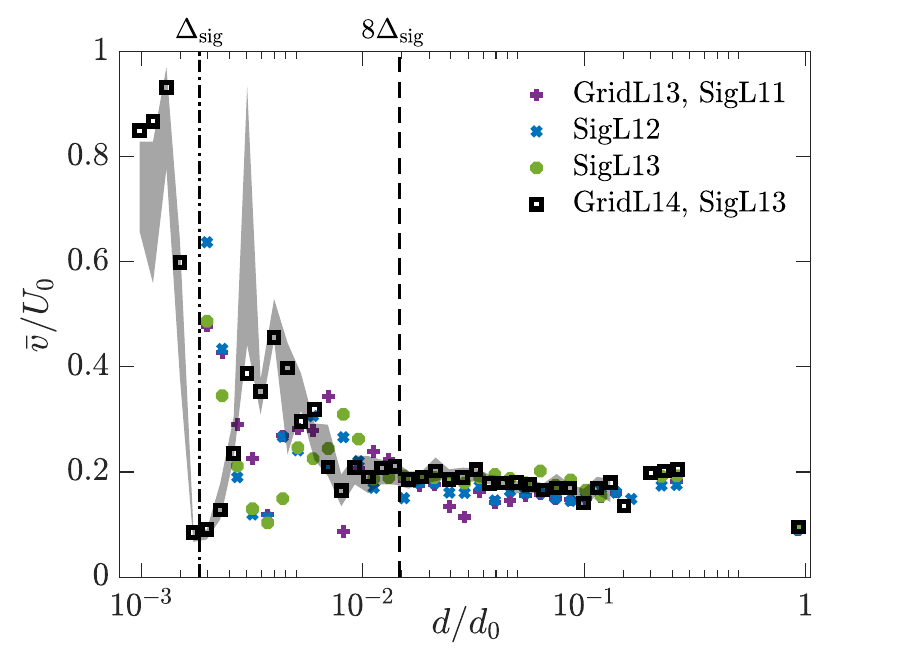}}	
	\caption{Time- and ensemble-averaged size (left column) and speed (right column) probability distribution functions of aerobreakup fragments obtained from simulations without using the MD algorithm (upper row), from an individual realisation (middle row) and from ensemble-averaged data across various realisations with the MD algorithm applied (lower row) at various grid resolution and signature levels. Confidence bounds for each bin are computed across different ensemble realisations at $L=14, \, L_{\rm sig} = 13$ using the bootstrapping method, and plotted in (e) and (f) using shaded area. For all test cases, $We = 15$ and $Oh = 10^{-3}$.}
	\label{fig:pdf-converge-tests}
\end{figure}

Figure~\ref{fig:pdf-converge-tests} further compares the grid convergence behaviour for the size and velocity distribution of bag fragments without and with the MD algorithm applied. The fragment data are sampled at different times throughout the bag film breakup period, and then collected and binned based on the equivalent fragment diameter $d$ to produce the size and velocity distribution functions. While the distribution functions presented in figs.~\ref{fig:size_pdf_vof}-\ref{fig:vbar_pdf_single} are for single realisations, those in figs.~\ref{fig:size_pdf}-\ref{fig:vbar_pdf} are ensemble-averaged for each bin over different realisations; and we have verified that the total number of bins does not significantly influence the shape of size and velocity distributions. Figure~\ref{fig:size_pdf_vof} shows the fragment size distribution functions obtained from individual simulations without application of the MD algorithm. It can be seen that while the distributions have similar shapes at various grid levels $L$, i.e. featuring large number densities of small fragments near the minimum grid size, followed by a fall-off at large fragment sizes, there is no clear indication of the distribution functions reaching grid convergence. In particular, it is observed that as $L$ increases, the entire size distribution shifts to smaller sizes. In contrast, fig.~\ref{fig:size_pdf_single} presents the fragment size distribution functions obtained from individual realisations within the range of $13 \leq L \leq 14$ and $11 \leq L_{\rm sig} \leq 13$ when the MD algorithm is used. While more scatters in the size distribution functions are seen when compared with fig.~\ref{fig:size_pdf_vof} due to smaller amounts of fragments produced, we no longer observe the shift to small fragment sizes for the distribution tail with $d \geq 8\Delta_{\rm sig}$, which is the range for well-resolved fragments as observed by \cite{chirco2021manifold}; and size distributions at different grid and signature levels appear to overlap for $d \geq 8\Delta_{\rm sig}$ despite these scatters. Figure~\ref{fig:size_pdf} further presents the ensemble-averaged size distribution functions obtained with the MD algorithm applied, which features much smaller range of scatter, as indicated by the confidence bounds represented by the grey shade at $L=14, \, L_{\rm sig} = 13$, showing clearly that the distributions of fragment statistics overlap for $d \geq 8\Delta_{\rm sig}$ at different values of $L$ and $L_{\rm sig}$. From this we conclude that the ensemble-averaged data are grid-converged for $d \geq 8\Delta_{\rm sig}$ and $13 \leq L \leq 14, L > L_{\rm sig}$. Moreover, together figs.~\ref{fig:size_pdf_vof}, ~\ref{fig:size_pdf_single} and ~\ref{fig:size_pdf} establish that the lack of grid convergence in the no-MD case (fig.~\ref{fig:size_pdf_vof}) is attributable not to the scatter of individual realisations, but specifically to numerically uncontrolled VOF-breakup. 

\begin{figure}
	\centering
	\includegraphics[width=.6\textwidth]{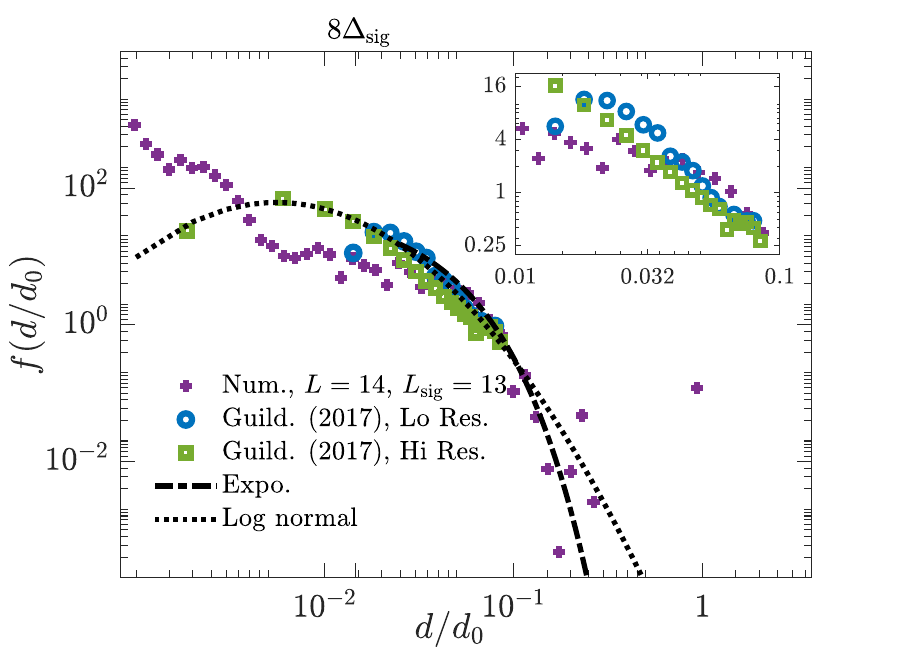}  
	\caption{Fragment size distribution function measured from our $L = 14, \, L_{\rm sig} = 13$ simulations, compared with the experimental data of \cite{guildenbecher2017characterization} measured at two different apparatus resolutions. A zoom-in view is provided as an inset to facilitate comparison of different size distribution functions within the size range of $0.01 \leq d/d_0 \leq 0.1$. Exponential and log-normal functions fitted to the experimental size distribution function are also included.}
	\label{fig:size_pdf_comp_exp}
\end{figure}

We also include the experimental data of \cite{guildenbecher2017characterization} obtained with $We = 13.8$ and $Oh = 5.43\times 10^{-3}$ in fig.~\ref{fig:size_pdf_comp_exp}, together with exponential and log-normal models fitted to their data. \cite{guildenbecher2017characterization} noted the difference between the size distributions obtained using two experimental techniques with different resolution levels, and expressed most confidence in the upper tail of the distributions satisfying $d \geq 0.01 d_0$. Within this size range, the tails of our size distribution and those of \cite{guildenbecher2017characterization} show excellent agreement, which further validates our numerical results within the size range of $d \geq 8\Delta_{\rm sig}$. While we leave for future work the detailed investigation of possible differences in fragmentation mechanisms between our present results and the experiments of \cite{guildenbecher2017characterization} and \cite{Jackiw2021, jackiw2022prediction}, the present remarkable agreement with experimental data at larger fragment sizes suggests that the upper tails of the size distribution do not depend on whatever these differences may be. Both the exponential and the log-normal model are found to match well with the current size distribution functions for $d \geq 8\Delta$, while both differ from the current results within the range of $d < 8\Delta$, which may suggest that no single function can represent the complete spectrum of the current size distribution of bag fragments.

It is noted that in fig.~\ref{fig:size_pdf}, the fragment statistics are not fully converged for $d \leq 8\Delta_{\rm sig}$, where compared with its counterparts at $L=13$, the size distribution at $L=14$ shows more fragments satisfying $\Delta \leq d \leq \Delta_{\rm sig}$, and fewer fragments with $\Delta_{\rm sig} \leq d \leq 8\Delta_{\rm sig}$. This is probably because the fragments within this range are primarily formed due to the breakup of liquid ligaments, especially the smallest fragments near the grid size, which are most likely the satellite drops produced from the capillary breakup of corrugated slender ligaments \citep{pal2021statistics}. These are controlled by the grid level $L$ rather than the signature level $L_{\rm sig}$, as the geometry-specific MD algorithm only targets thin liquid films in 3D simulations and do not act in the stretch-induced breakup of liquid ligaments. Our numerical results for $d \leq 8\Delta_{\rm sig}$ also deviate from the log-normal function fit of \cite{guildenbecher2017characterization}, which may be due to multiple factors including the difference in $We$, the presence of additional flow perturbations in experiments, and possibly resolution limits in experimental equipment, as exemplified by a comparison performed in fig.~9 of \cite{guildenbecher2017characterization}. On the other hand, the size distribution function of very large fragments satisfying $d \geq 0.1 d_0$ show relatively larger range of scatter compared with their smaller counterparts around $8\Delta_{\rm sig}$, which likely arises from the smaller number of these fragments produced in each ensemble realisation and can be further reduced by increasing the ensemble size. Finally, we remark that the influence of MD on mass conservation is minimal as the loss of liquid mass incurred by the MD algorithm does not exceed 0.023\% for $t/\tau \leq 2.18$ at $L=13$ and 14.

Figures~\ref{fig:vbar_pdf_vof}, \ref{fig:vbar_pdf_single} and \ref{fig:vbar_pdf} show the average speed $\Bar{v}$ of fragments as functions of the fragment diameter $d$ obtained from simulations without and with the MD algorithm applied. Similar to the size distribution functions, the shapes of the distribution of $\Bar{v}$ clearly indicate grid convergence for the tail constituted by the well-resolved fragments with $d \geq 8\Delta_{\rm sig}$ in fig.~\ref{fig:vbar_pdf}, which is not observed in fig.~\ref{fig:vbar_pdf_vof} where large scatters across various grid levels are present. Interestingly, in fig.~\ref{fig:vbar_pdf}, fragments with diameter $d \geq 8\Delta_{\rm sig}$ show little variation in the average speed, which appears to be a constant independent of the values of $L$ and $L_{\rm sig}$; whereas a peak can be observed within the range of $\Delta_{\rm sig} \leq d \leq 8\Delta_{\rm sig}$ which is grid-dependent, along with the large increase of the speed of tiny fragments close to $\Delta_{\rm sig}$ which approaches the freestream velocity $U_0$. While it is clear therefore that the production mechanisms of droplets satisfying $d \leq 8\Delta_{\rm sig}$ may not be grid-converged, the resulting dynamics of these small droplets turn out to be well-resolved; more detailed analysis establishing this will follow in \S\ref{subsec:frag-tracking}.

In summary, our results in this section demonstrate that the application of the MD algorithm helps to establish grid convergence of fragment size and speed statistics for well-resolved fragments with diameter $d \geq 8\Delta_{\rm sig}$, which is not achieved when VOF breakup is dominant. Based on these results, all following three-dimensional studies of bag film breakup are conducted at 
$L = 14$ and $L_{\rm sig} = 13$.

\subsection{Mechanisms leading to bag fragmentation}
\label{subsec:frag-tracking}

In this section, we further analyse the fragment statistics obtained from our grid convergence tests run at $L = 14$ and $L_{\rm sig} = 13$, to provide insight into the shapes of the size and distribution functions observed in \S\ref{subsec:grid-conv}, and the physical mechanisms governing the formation of fragments and their subsequent evolution patterns. These choices of $L$ and $L_{\rm sig}$, together with the MD parameters specified in \S\ref{subsec:grid-conv}, allow the creation of only a few holes on the bag film, which are not only enough to avoid the onset of VOF breakup, but also preserves abundant film breakup phenomena including rim recession, collision and destabilisation behaviour that would otherwise be hard to recover with more holes created, where rim collision would dominate \citep{vledouts2016explosive}. As is noted in \S\ref{subsubsec:early}, our film is thicker and breaks up earlier compared with experimental results due to the limit of grid resolution. This leads to smaller Taylor-Culick velocity values, which reduces the probability of destabilisation of receding liquid rims \citep{jackiw2022prediction} and production of fine drops \citep{neel2018spontaneous}; but our results show that many interesting breakup mechanisms can already be captured with this choice of $L_{\rm sig}$, which we will present further below.

\begin{figure}
    \centering
	\subfloat[]{
		\label{fig:size_pdf_evolution}
		\includegraphics[width=.6\textwidth]{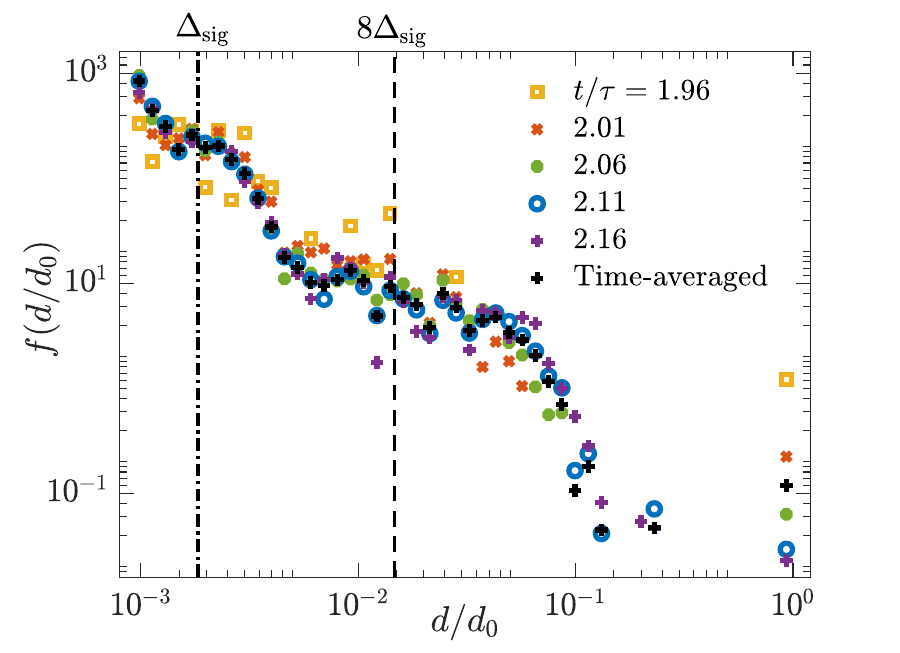}}
		
	\centering
	\subfloat[]{
		\label{fig:ux_pdf_evolution}
		\includegraphics[width=.48\textwidth]{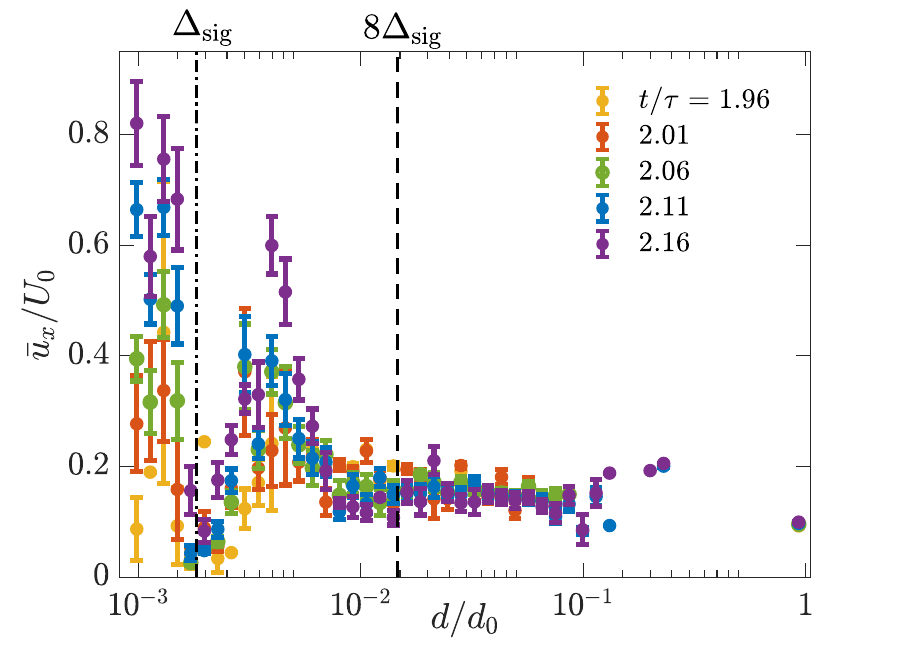}}
	\centering
	\subfloat[]{
		\label{fig:ur_pdf_evolution}
		\includegraphics[width=.48\textwidth]{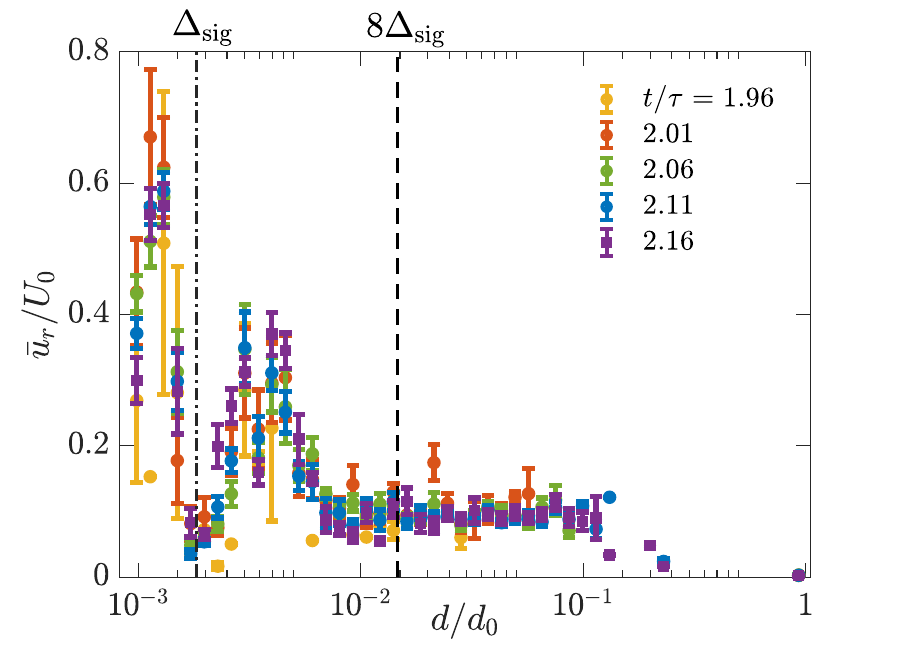}}	
	\caption{Ensemble-averaged instantaneous size distribution functions (a), and probability distribution functions of axial (b) and radial (c) speed of aerobreakup fragments calculated at $L = 14$ and $L_{\rm sig} = 13$. Ensemble- and time-averaged fragment size distribution function is also plotted in (a) for reference.}
	\label{fig:ur-ux-pdfs}
\end{figure}

We first show in fig.~\ref{fig:ur-ux-pdfs} the time evolution of the instantaneous distributions of the size, axial and radial speed distributions of the fragments produced from bag breakup. Figure~\ref{fig:size_pdf_evolution} indicates that immediately after the onset of bag breakup ($t/\tau = 1.96$) only small fragments close to the minimum grid size are produced, and well-resolved larger fragments satisfying $d \geq 8\Delta_{\rm sig}$ only come into existence as time elapses, and are always fewer compared with small fragments near the grid size. The shape of the size distribution function gradually stabilises, and reaches a steady state by $t/\tau = 2.11$ that is very close to the ensemble- and time-averaged size distribution function. These findings suggest that the smaller and larger fragments are produced through different physical mechanisms that arise at different stages of bag breakup, and eventually these fragmentation mechanisms die out as the bag approaches full disintegration and the fragment size distribution is well-represented by time-averaged results. The remaining rim will then disintegrate at still later times, whose investigation we leave for future work.

Figures~\ref{fig:ux_pdf_evolution} and \ref{fig:ur_pdf_evolution} show the instantaneous distribution of fragment axial and radial speeds $u_x$ and $u_r \equiv \sqrt{u_y^2 + u_z^2}$ as functions of their sizes, with the ensemble-wide variations of velocity components in each averaging bin shown in error bars. It can be seen that the speed of well-resolved fragments satisfying $d \geq 8\Delta_{\rm sig}$ remains close to a constant value without significant variations. While the statistics of smaller fragments with $d \leq 8\Delta_{\rm sig}$ are not fully numerically converged, they do show considerably larger variation around the binned average value in qualitative agreement with the experimental results of \cite{guildenbecher2017characterization}. Interestingly, we observe peaks around $d/d_0 = 5 \times 10^{-3}$ in the distributions of both $u_x$ and $u_r$, whose location does not appear to vary with time. Moreover, despite the presence of velocity variations, fig.~\ref{fig:ux_pdf_evolution} suggests that the average axial speed $u_x$ of smaller fragments with $d \leq 8\Delta_{\rm sig}$ increases over time, whereas the radial speed $u_r$ does not show similar increasing trend in fig.~\ref{fig:ur_pdf_evolution}. This is most likely because the smaller fragments are generated earlier and therefore are exposed to the airflow for much longer periods of time compared with larger fragments; together with their smaller mass, this means that they are much more easily accelerated by the axial velocity component of the airflow, hence the continuous increase in their $u_x$ values. On the other hand, $u_r$ does not increase significantly over time, likely because the airflow in the wake region does not have a large radial velocity component that can accelerate bag fragments as they migrate downstream.

\begin{figure}
	\centering
	\subfloat[]{
		\label{fig:lig_breakup_fim_610}
		\includegraphics[width=.24\textwidth]{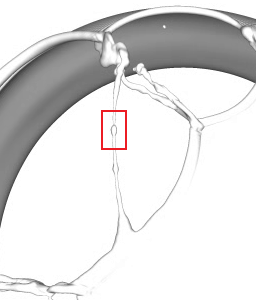}}
	\centering
	\subfloat[]{
		\label{fig:lig_breakup_fim_620}
		\includegraphics[width=.24\textwidth]{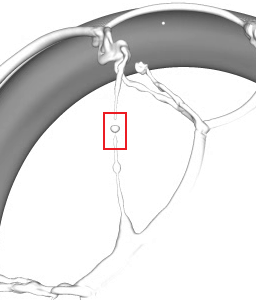}}
	\centering
	\subfloat[]{
		\label{fig:lig_breakup_fim_630}
		\includegraphics[width=.24\textwidth]{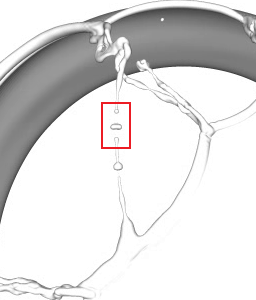}}
	\centering
	\subfloat[]{
		\label{fig:lig_breakup_fim_640}
		\includegraphics[width=.24\textwidth]{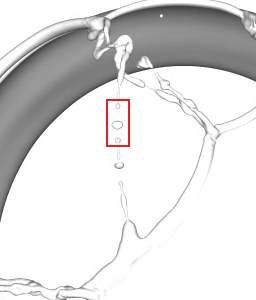}}
		
	\centering
	\subfloat[]{
		\label{fig:lig_breakup_fim_650}
		\includegraphics[width=.24\textwidth]{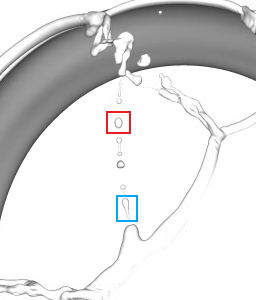}}
	\centering
	\subfloat[]{
		\label{fig:lig_breakup_fim_660}
		\includegraphics[width=.24\textwidth]{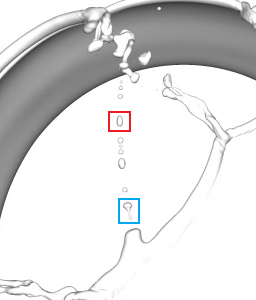}}
	\centering
	\subfloat[]{
		\label{fig:lig_breakup_fim_670}
		\includegraphics[width=.24\textwidth]{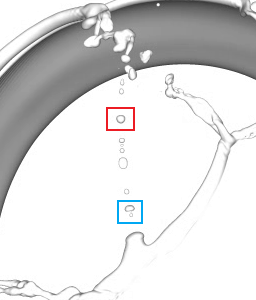}}
	\centering
	\subfloat[]{
		\label{fig:lig_breakup_fim_680}
		\includegraphics[width=.24\textwidth]{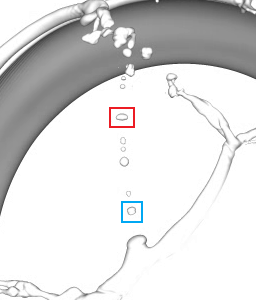}}		
	\caption{Snapshots showing the non-local breakup of a long ligament into multiple fragments during bag film fragmentation with $We = 15$ and $Oh = 10^{-3}$. The red boxes shows formation of a single fragment through non-local end-pinching and its subsequent oscillation, and the blue boxes show the formation of two fragments through a local breakup event and their subsequent coalescence.}
	\label{fig:lig-breakup-snapshots}
\end{figure}

We will hereafter discuss qualitatively several mechanisms through which the bag film undergoes fragmentation and form small droplets, which can be identified by inspecting typical simulation snapshots taken from our $L = 14, \, L_{\rm sig} = 13$ simulations. Firstly, fig.~\ref{fig:lig-breakup-snapshots} shows the breakup of a stretched long ligament neighbouring two enlarging holes into a series of small drops. As the ligament is itself connected to the main drop, there is a significant size difference between the parent and child drops produced from its breakup, which is an example of \emph{non-local} breakup events (see Eq.~\eqref{for:locality} in \S\ref{subsec:frag-behaviour} for a definition of non-local breakup). It can be seen from fig.~\ref{fig:lig_breakup_fim_610} that significant cross-sectional diameter variations have developed on the ligament before the onset of its breakup, which can be viewed as the result of the nonlinear development of the RP instability \citep{pal2021statistics}. Afterwards, the ligament shrinks to form sharp tips and then breaks up on multiple sites, as shown in fig.~\ref{fig:lig_breakup_fim_620}, and forms a primary drop which continues to undergo periodic prolate-oblate shape oscillations resembling droplets produced by breaking Rayleigh jets \citep{hu2021deformation}, as highlighted in the red boxes in figs.~\ref{fig:lig_breakup_fim_620}-\ref{fig:lig_breakup_fim_680}. This is because the pinch-off of the stretching ligament induces an inner velocity field within the detaching droplet that drives it in the oblate direction (see e.g. fig.~7b in \citep{hu2021deformation}), matching the perturbation shape of the second Rayleigh mode, which then excites oscillation modulated by capillary effects. In the meantime, the other parts of the ligament do not pinch off to form a series of fragments at once, but first break up into several elongated debris, and then split into large primary and small satellite drops via the well-known end-pinching mechanism \citep{castrejon2012breakup, pal2021statistics}, which is an example of \emph{local} breakup events as the parent (the elongated debris) and child (satellite drops) do not differ significantly in their sizes. Under certain circumstances, the primary and satellite drops might coalesce and form a larger fragment as highlighted in the blue boxes, resembling the `immediate satellite merge' mechanism discussed by \cite{vassallo1991satellite}.

\begin{figure}
	\centering
	\subfloat[]{
		\label{fig:node_breakup_fim_550}
		\includegraphics[width=.24\textwidth]{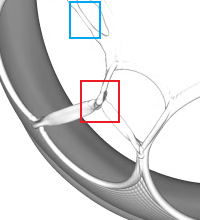}}
	\centering
	\subfloat[]{
		\label{fig:node_breakup_fim_570}
		\includegraphics[width=.24\textwidth]{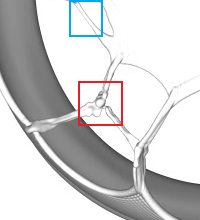}}
	\centering
	\subfloat[]{
		\label{fig:node_breakup_fim_590}
		\includegraphics[width=.24\textwidth]{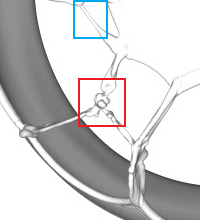}}
	\centering
	\subfloat[]{
		\label{fig:node_breakup_fim_610}
		\includegraphics[width=.24\textwidth]{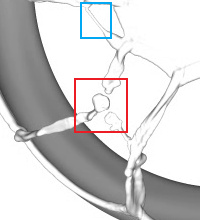}}
		
	\centering
	\subfloat[]{
		\label{fig:node_breakup_fim_630}
		\includegraphics[width=.24\textwidth]{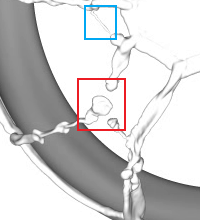}}
	\centering
	\subfloat[]{
		\label{fig:node_breakup_fim_650}
		\includegraphics[width=.24\textwidth]{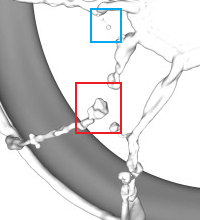}}
	\centering
	\subfloat[]{
		\label{fig:node_breakup_fim_670}
		\includegraphics[width=.24\textwidth]{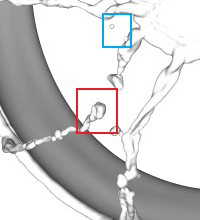}}
	\centering
	\subfloat[]{
		\label{fig:node_breakup_fim_690}
		\includegraphics[width=.24\textwidth]{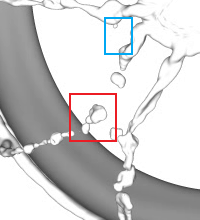}}		
	\caption{Snapshots showing the detachment of a liquid node from ligament webs (highlighted in the red box) and the evolution of a short ligament into a single drop (highlighted in the blue box) during bag film fragmentation with $We = 15$ and $Oh = 10^{-3}$.}
	\label{fig:node-breakup-snapshots}
\end{figure}

Figure~\ref{fig:node-breakup-snapshots} first shows an example of short ligament breakup and its eventual contraction into a single droplet, as highlighted in the blue boxes. Compared with the breakup of long ligaments demonstrated in fig.~\ref{fig:lig-breakup-snapshots}, this type of short ligament breakup bears stronger resemblance to the breakup phenomena of liquid bridges studied by \cite{agbaglah2021breakup}, where fragments produced from the same liquid bridge do not feature significant size variations. This is most likely because the initial holes are placed very close to each other in \cite{agbaglah2021breakup}, and the receding rim does not have enough time to grow in size and momentum before their impact. 

Another type of fragmentation mechanism can also be identified in fig.~\ref{fig:node-breakup-snapshots}; as highlighted in the red boxes, three adjacent holes have merged with each other, and their three bordering rims converge on a common `node' as they are stretched, which is also observed in the breakup of ligament webs formed on Savart sheets by \cite{lhuissier2013effervescent}. Compared with the ligament pinch-off mechanism discussed earlier, the surface evolution of this `node' shows much more complicated corrugation patterns as the rims it was connected to are gradually detached, and therefore is not dominated by the second Rayleigh mode alone. The `node' drop that eventually forms in this case also has a much larger size compared with its counterparts formed from ligament pinch-off events. However, different from an earlier study by \cite{vledouts2016explosive}, our choice of $L = 14$ and $L_{\rm sig} = 13$ does not produce holes on the bag film with as high a number density, and therefore we are not able to directly measure the wavelength of the RT instability responsible for the film fragmentation from the average distance between the centres of adjacent holes. The relatively smaller number density of holes formed also means that the larger `node' drops formed due to the merging of three or more adjacent and similarly sized holes are relatively rare compared with generally smaller fragments formed from ligament breakup, which require the collision of only two adjacent liquid rims. Furthermore, no less than three holes should fully expand and arrive at the same region on the bag film where the node is located, and each of the connecting rims need to break off successively before the node can be treated as a seperate fragment by the fragment counting algorithm. These factors help explain the tail of the size distribution function of aerobreakup fragments taking shape at much later time as we observed in fig.~\ref{fig:size_pdf_evolution}; namely, that larger droplets are relatively few, and produced at generally later times during the fragmentation process. 

\begin{figure}
	\centering
	\subfloat[]{
		\label{fig:lam_destab_fim_600}
		\includegraphics[width=.24\textwidth]{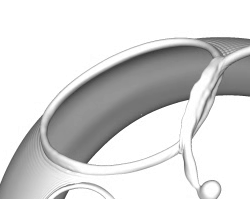}}
	\centering
	\subfloat[]{
		\label{fig:lam_destab_fim_625}
		\includegraphics[width=.24\textwidth]{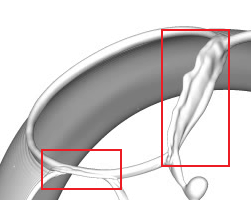}}
	\centering
	\subfloat[]{
		\label{fig:lam_destab_fim_650}
		\includegraphics[width=.24\textwidth]{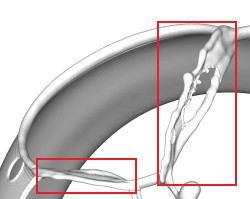}}
	\centering
	\subfloat[]{
		\label{fig:lam_destab_fim_675}
		\includegraphics[width=.24\textwidth]{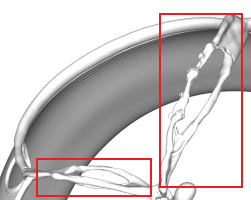}}
		
	\caption{Snapshots showing the evolution of `fingering' liquid lamellae during bag film fragmentation with $We = 15$ and $Oh = 10^{-3}$.}
	\label{fig:lam-destab-snapshots}
\end{figure}

We also note that while we often observe the formation, oscillation and subsequent corrugation development of liquid ligaments after the impact of receding rims, as shown on the ligament to the left of the red box in fig.\ref{fig:node-breakup-snapshots}, destabilisation of such structure due to the RT instability and its subsequent evolution into fully developed transverse `fingers' and `fine drops'  are only occasionally observed in our current simulations. According to \cite{neel2020fines}, these two regimes are separated by a critical local Weber number for rim collision $We_c \equiv \rho_l (2v_{\rm TC})^2 d_l / \sigma = 66$, where $d_l$ is the rim diameter. Neglecting the curved geometry of the bag film, liquid mass conservation further yields $We_c = 8 \sqrt{D_c/\pi h}$, where $D_c$ and h are respectively the distance between the centre of two neighbouring holes and the film thickness. Taking $h = 3D/2^{L_{\rm sig}} = 3D/2^{13}$, we find that $We_c = 66$ corresponds to $D_c = 1.2 d_0$, which we expect might be reached for some pairs of sufficiently separated holes, as the bag diameter $d_f$ before the onset of fragmentation typically approaches $2d_0$, as shown in both our numerical results and the experimental data of \cite{Jackiw2021} (see e.g. their fig.~30). Figure~\ref{fig:lam-destab-snapshots} highlights two such examples in red boxes, where we observe the transverse growth of the lamella and the growth of finger-shaped corrugations on its edges; however, before the fingers fully develop and detach as `fine' drops, holes are observed to form on the thinning lamella, which then expand and collide with the fingering lamella edges, turning them into isolated breaking ligaments. Similar phenomena of lamellae rupture and their edges forming corrugated ligaments can also be observed in fig.~14 of \cite{vledouts2016explosive}, although in that case the lamellae appear to remain within the plane of the film surface, and do not experience transverse growth; and VOF breakup may play a role in the present examples. Nevertheless, the liquid ligaments found in the current simulations have already displayed a variety of well-documented physical phenomena that collectively contribute to the large span of fragment sizes found in our fragment distribution functions.

\begin{figure}
	\centering
	\subfloat[]{
		\label{fig:rim_destab_fim_700}
		\includegraphics[width=.24\textwidth]{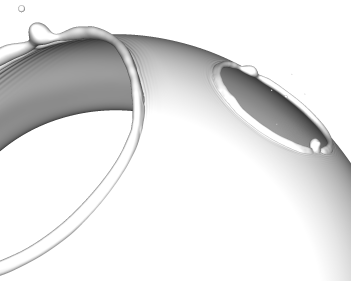}}
	\centering
	\subfloat[]{
		\label{fig:rim_destab_fim_725}
		\includegraphics[width=.24\textwidth]{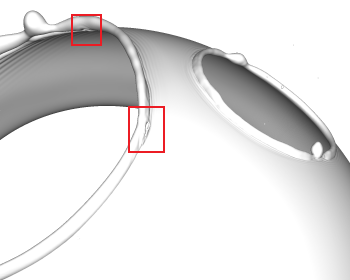}}
	\centering
	\subfloat[]{
		\label{fig:rim_destab_fim_750}
		\includegraphics[width=.24\textwidth]{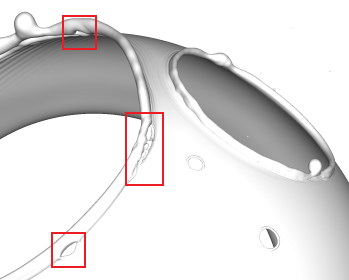}}
	\centering
	\subfloat[]{
		\label{fig:rim_destab_fim_775}
		\includegraphics[width=.24\textwidth]{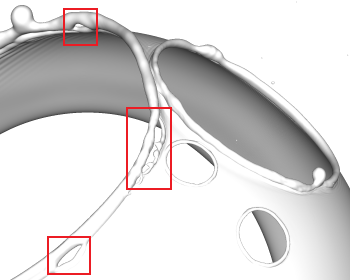}}
		
	\caption{Snapshots showing the receding liquid rim destabilisation during bag film fragmentation with $We = 15$ and $Oh = 10^{-3}$. The sites where the rim is detached from its base is highlighted in red boxes.}
	\label{fig:rim-destab-snapshots}
\end{figure}

Lastly, we observed a few examples showing the destabilisation of receding liquid rims, as demonstrated in fig.~\ref{fig:rim-destab-snapshots}. These destabilisation phenomena are absent in the recent work of \cite{agbaglah2021breakup} where holes expand over a flat liquid film, and are therefore most likely linked with the influence of centrifugal acceleration caused by the curved bag film \citep{lhuissier2012bursting, jackiw2022prediction}. While we are not yet able to measure the wavelength or the linear growth rate of the instability, and therefore have not identified the type of hydrodynamic instability involved here; we observe in fig.~\ref{fig:rim_destab_fim_775} regular-spaced holes highlighted in the red boxes forming at the foot of the rim bordering the hole on the left. These are not attributable to VOF breakup because they are not observed to appear elsewhere on the bag at the same time; while the rim bordering the larger hole on the right is seen to develop regular corrugation patterns, which might be an indication that the receding liquid rim is experiencing the RP instability. Any further development of the instability is interrupted by the eventual collision between adjacent rims (not shown in fig.~\ref{fig:rim-destab-snapshots}). The readers are referred to \cite{jackiw2022prediction} for a more comprehensive discussion on rim destabilisation. There a few candidates are proposed, including the RT instability mechanism governing the `fingering' behaviour on bursting surface bubbles \citep{lhuissier2012bursting}; and it is concluded that the centrifugal acceleration does not govern the instability of the rim directly, but instead regulates the thickness of the rim via a local-Bond-number criterion \citep{wang2018universal}, with the rim in turn susceptible to the RP instability. With the present methodology, these rim instabilities can be investigated in more detail with higher signature levels $L_{\rm sig}$, and concomitantly higher $L$. Given the large computational expense of such simulations, it is not feasible to include such an analysis in the present study.

\subsection{Behaviour of bag fragments}
\label{subsec:frag-behaviour}

In the following, we move away from considering the dynamics and numerical characteristics of the production mechanisms of fragments to examine instead those of the fragments themselves. To provide further insight into the evolution of individual fragments rather than their collective behaviour, we utilise the droplet tracking algorithm proposed by \cite{chan2021identifying} in post-processing to reconstruct their breakup lineage. This toolbox assumes breakup and coalescing events to be binary (i.e. at most two parent droplets may collide and form one large child droplet, or two child droplets may be produced from the breakup of one parent droplet in a single breakup/coalescing event), and is capable of identifying all coalesce/breakup events and differentiate between the new drops produced from these events and those which do not undergo such changes. It requires only the instantaneous fragment size, location and velocity output from the simulation at given time intervals, instead of knowledge of the entire flow field at successive simulation time steps, and therefore incur only limited computational cost \citep{chan2021identifying}.

\begin{figure}
	\centering
	\subfloat[]{
		\label{fig:lifespan-r}
		\includegraphics[width=.48\textwidth]{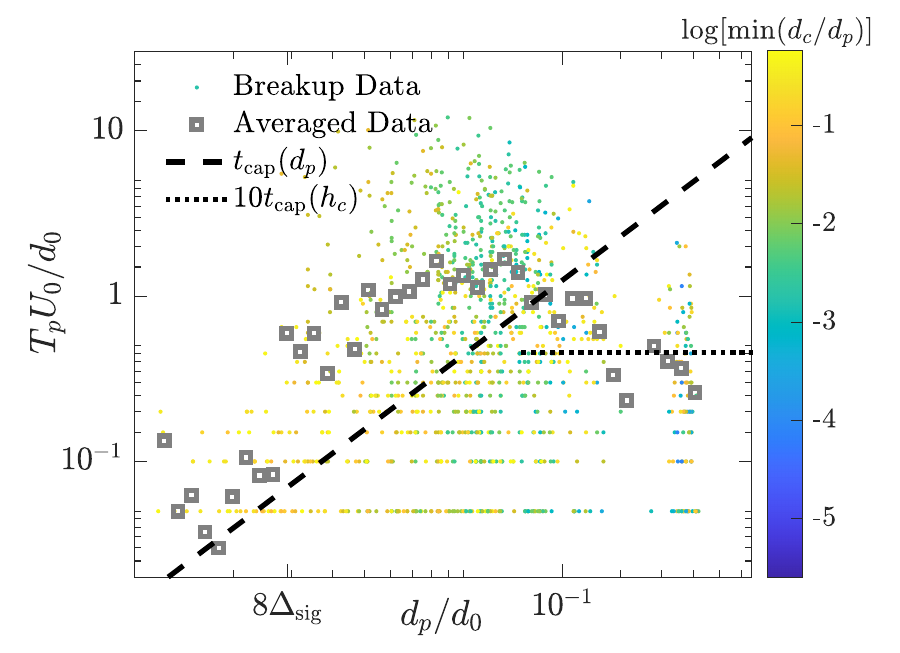}}
	\centering
	\subfloat[]{
		\label{fig:vel-diff-pcratio}
		\includegraphics[width=.48\textwidth]{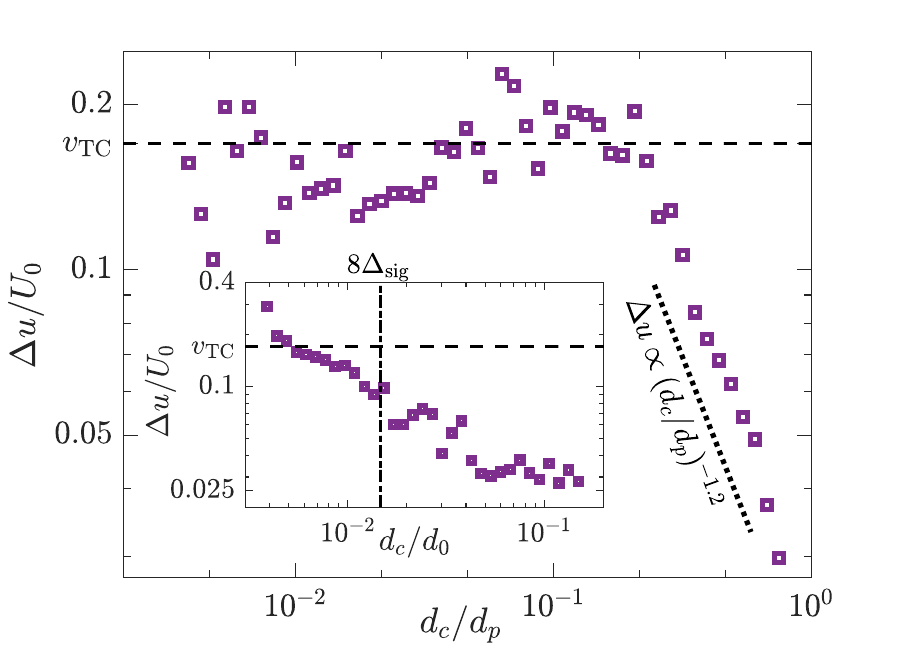}}
	\caption{(a): The lifetime of the parent fragments $T_p$ as a function of their diameter $d_p$, where $We = 15$ and $Oh = 10^{-3}$. The bin-averaged results are shown in grey hollow squares, and the original data are shown as solid dots, whose colour represents the value of the child/parent diameter ratio. It is noted that this plot does not include the main drop as a parent which features $d_p/d_0 \approx 1$. (b): Velocity difference between parent and child fragments $\Delta u$ as a function of the child/parent diameter ratio $d_c/d_p$ is shown in the main plot, whereas the inset plots $\Delta u$ as a function of the diameter of child fragments $d_c$.}
	\label{fig:vel-diff}
\end{figure}

As the fragments produced from bag breakup are much smaller compared with the parent drop, and therefore have a much smaller Weber number, it is highly unlikely that they will undergo another bag breakup event. However, they may still experience secondary breakup to form smaller fragments as they evolve over time. It is therefore of interest to determine the lifetime of breaking parent fragments $T_p$ using the toolbox of \cite{chan2021identifying}, defined as the interval between their birth and death in two successive breakup events \citep{riviere2022capillary}. Figure \ref{fig:lifespan-r} shows $T_p$ as a function of the diameter $d_p$ of parent fragments, with the bin-averaged values of $T_p$ shown in grey squares. The solid dots plotted in the background represent recorded individual breakup events, and are colour-coded by the logarithm value of the minimum child/parent diameter ratio $d_c/d_p$, highlighting a broad distribution of parent fragment lifetime. For comparison, the characteristic capillary time of fragments $t_{\rm cap}$ is also plotted in fig.~\ref{fig:lifespan-r} as a function of $d_p$, which is defined as follows:
\begin{equation}
    t_{\rm cap}(d_p) = \sqrt{\frac{\rho_l d_p^3}{8\sigma}}.
    \label{for:cap-time}
\end{equation}
It can first be seen from the scattered original data that most of the bag film fragments that undergo a secondary breakup fall within the range of $d_p \geq 8\Delta_{\rm sig}$, which correspond to the `well resolved' fragments discussed in \S\ref{subsec:grid-conv}. 
Furthermore, the lifetime of fragments $T_p$ satisfying $d_p \leq 0.05d_0$ shows a dependency on $r_p$ that roughly scales with the characteristic capillary time, but this trend breaks down for even larger fragments with $d_p \geq 0.05d_0$. It is noted that the capillary time defined in Eq.~\eqref{for:cap-time} is proportional to the oscillation period of droplet spherical harmonic modes (Eq.~\eqref{for:rayleigh-freq}). Therefore, the scaling of $T_p$ with $t_{\rm cap}$ for $d_p \leq 0.05d_0$ may suggest that the fragmentation of these fragments is primarily due to large-amplitude nonlinear oscillations which can trigger a capillary breakup \citep{lalanne2019model}. This also explains the large scatters we observe in the lifetime of parent fragments within this size range, as when nonlinearity becomes dominant, the surface oscillations cannot be represented by a single mode, and different modes of perturbation with different oscillation periods might trigger breakup depending on specific fragments. As for even larger fragments with $d_p \geq 0.05d_0$, their lifetime $T_p$ appears to scatter around an average value of $10t_{\rm cap}(h_c)$, where $t_{\rm cap}(h_c)$ is the characteristic capillary time based on the critical film thickness $h_c = 3D/2^{L_{\rm sig}}$. Here $10t_{\rm cap}(h_c)$ is an estimation of the inertial timescale leading to the capillary breakup of stretching liquid ligaments, which are formed due to hole collision \citep{agbaglah2021breakup}. This suggests that this type of non-local breakup is only dependent on the topological evolution of the stretching liquid ligament, rather than that of the entire parent drop from which the child fragments are torn off; but this remains to be verified in future work.

We were also able to compute the magnitude of the velocity differences $\Delta u$ between fragment parents and their children at two successive instants when the fragment statistics are collected, and plot them in fig.~\ref{fig:vel-diff-pcratio} as a function of the ratio between the child and parent diameter $d_c/d_p$. It is found that for breakup events where a small child/parent size ratio $(d_c/d_p \leq 0.22)$ are detected, the velocity difference $\Delta u$ appear to show little dependence on $d_c/d_p$, despite significant scatter. Based on our findings in fig.~\ref{fig:lifespan-r}, these breakup events with $d_c/d_p \leq 0.22$ mostly feature small children with large parents. On the other hand, breakup events satisfying $d_c/d_p \geq 0.22$ are dominated by small parents and children, and their $\Delta u$ decreases with increasing values of $d_c/d_p$, roughly following a power-law scaling $\Delta u \propto (d_c/d_p)^{-1.2}$. We further note that the $\Delta u$ values of breakup events satisfying $d_c/d_p \leq 0.22$ roughly coincides with the inviscid Taylor-Culick velocity $v_{\rm TC}$, with an estimation for the bag film thickness $h$ based on the signature level $L_{\rm sig}$:
\begin{equation}
    v_{\rm TC} \equiv \sqrt{\frac{2\sigma}{\rho_l h}} = \sqrt{\frac{2^{L_{\rm sig}+1} \sigma}{3\rho_l D}} \approx 0.17U_0.
\end{equation}
This agrees with the recent confirmation by \cite{neel2022velocity} that the speed of film drops produced from bubble-bursting can be estimated by $v_{\rm TC}$ to an order of magnitude. An explanation for this approximate agreement is as follows. Prior and up to collision between adjacent hole rims, each rim travels at the Taylor-Culick velocity. The colliding rims of these holes then form liquid ligaments, which exhibit an axial stretching rate comparable to the pre-collision rim speed (i.e. the Taylor-Culick velocity). This stretching rate in turn sets the relative speed of sufficiently small fragments ejected from the parent ligament.
The parent ligaments may constitute part of the parent drop, or may themselves have separated from it. Therefore, for a given child droplet size produced by this mechanism, the ratio $d_c/d_p$ may see considerable variation. Consequently, we observe the large range of $d_c/d_p \leq 10^{-2}$ where the parent/child velocity difference remains close to $v_{\rm TC}$. The inset of fig.~\ref{fig:vel-diff-pcratio} further shows the velocity difference in breakup events as a function of the child diameter $d_c$ alone; and this time we find that it is the small fragments satisfying $d_c \leq 8 \Delta_{\rm sig}$ that appear to approach $v_{\rm TC}$, agreeing with our analysis that it is the fragments produced from liquid ligament breakup that are represented within the range of $d_c/d_p \leq 0.22$. Note however that in the inset of fig.~\ref{fig:vel-diff-pcratio}, $\Delta u$ continues to increase beyond $v_{\rm TC}$ with decreasing $d_c$, and does not scatter around $v_{\rm TC}$ as in the main plot. This is most likely because the droplet tracking algorithm has a finite fragment detection frequency \citep{chan2021identifying}, and the smallest fragments may be generated and accelerated by the ambient airflow between two successive instants when this algorithm is called for their detection. The nominal velocity difference $\Delta u$ for the smallest fragments therefore includes contributions from both the initial ejection velocity (primarily in the radial direction) and the increment due to airflow acceleration (primarily in the axial direction), which causes $\Delta u$ to increase steadily beyond $u_{\rm TC}$. As these smallest fragments are scattered in different averaging bins according to the child/parent diameter ratio of the breakup events, the contribution of airflow acceleration to $\Delta u$ becomes less obvious.

Based on our findings in fig.~\ref{fig:vel-diff-pcratio}, we introduce the following criteria for determining `non-local' breakup and coalescing events, respectively:
\begin{equation}
    \max \left( \frac{d_{c,i}}{d_p} \right) \leq 0.22, \quad \max \left( \frac{d_c}{d_{p,i}} \right) \leq 4.64, \quad i = 1,2.
    \label{for:locality}
\end{equation}
Otherwise we term the breakup or coalescing events `local'. Here the critical child/parent diameter ratio of 0.22 for fragment breakup (and analogously 4.64 for coalescence) separates the two breakup regimes found in fig.~\ref{fig:vel-diff-pcratio}, where the velocity difference $\Delta u$ either scatters around $v_{\rm TC}$ or scales with $d_c/d_p$. Note that the term `local' here does not mean the parent and child fragments are close to each other in terms of their locations in the physical space, which all such events satisfy; but rather in the sense of the parent and its two children being close in their respective sizes \citep{chan2021turbulent}.

\begin{figure}
	\centering
	\subfloat[]{
		\label{fig:breakup_num_hist}
		\includegraphics[width=.48\textwidth]{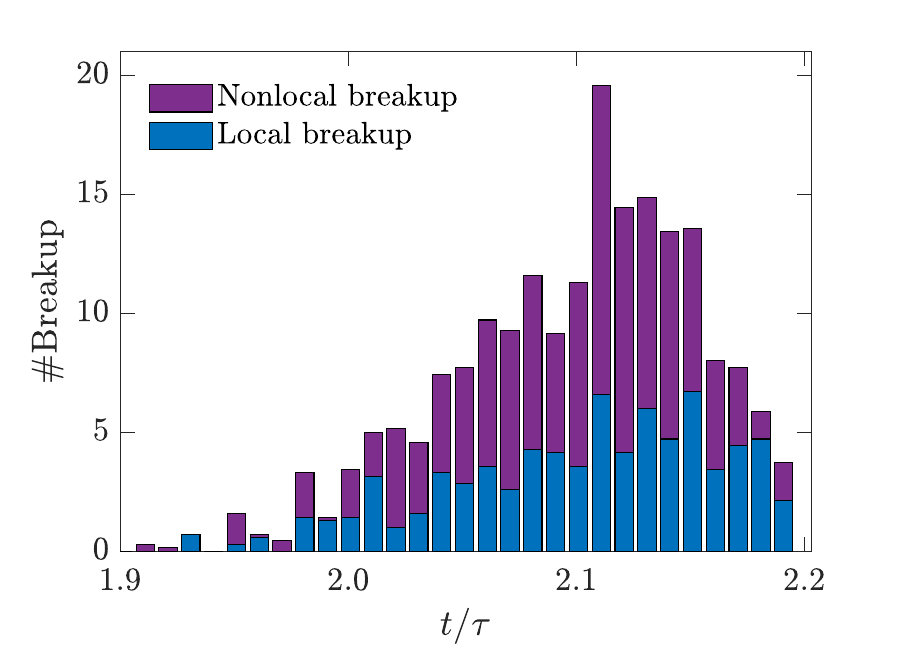}}
	\centering
	\subfloat[]{
		\label{fig:coalesce_num_hist}
		\includegraphics[width=.48\textwidth]{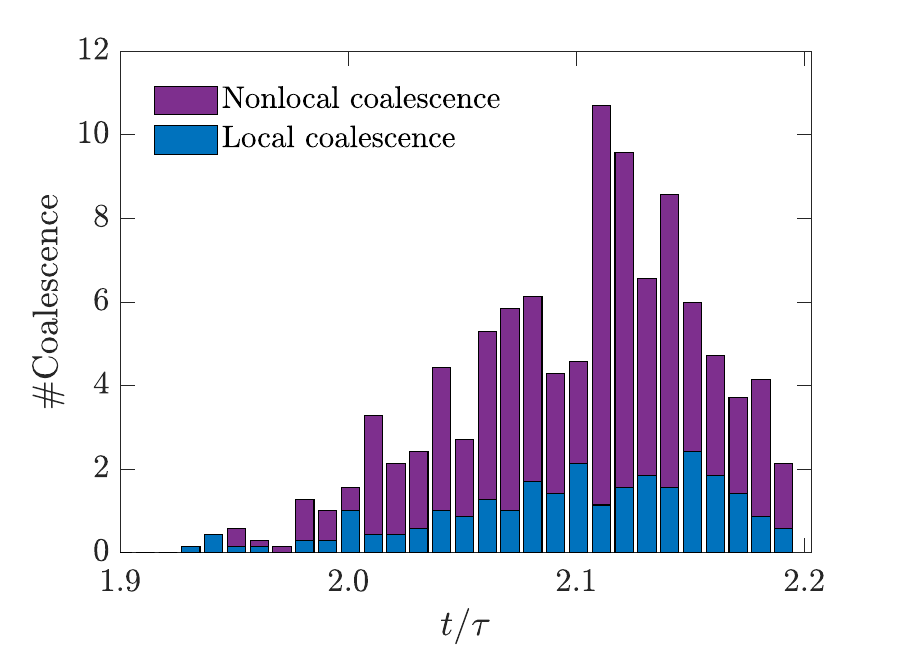}}	
	\caption{Ensemble-averaged evolution of number of breakup (a) and coalesce (b) events during the breakup of bag films produced from an initial droplet with $We = 15$ and $Oh = 10^{-3}$. All ensemble realisations are run at $L = 14, \, L_{\rm sig} = 13$.}
	\label{fig:frag-breakup-track}
\end{figure}

We further plot in fig.~\ref{fig:frag-breakup-track} the ensemble-averaged number density of breakup and coalescing events detected during the bag fragmentation period. It is found that both breakup and coalescing events occur most frequently around $t/\tau = 2.10$. This is most likely when the receding rims fully absorb the bag film and collide with each other, which then triggers a series of corrugated ligament breakup and fragment coalescing events. After this the breakup and coalescing behaviour become less frequent, as the corrugated ligaments gradually disintegrate without liquid mass input from the bag film. While there exist `multistep' breakup events in our aerobreakup simulation outputs, our results in fig.~\ref{fig:frag-breakup-track} suggest that the fragmentation process involved in the aerobreakup problem cannot be well described by a breakup cascade model \citep{garrett2000connection, chan2021turbulent}, as non-local breakup and coalescing events producing children with sizes drastically different from their parents are found to dominate this problem, which is different from the entrained air bubble breakup scenario in breaking wave studies \citep{garrett2000connection, deane2002scale, deike2016air, chan2021turbulent, Mostert2021} where the prevalence of local breakup events leads to a well-defined bubble-mass flux supporting breakup cascade models. It is also noted that breakup events occur much more frequently than coalescing events for bag films, which is expected as the latter requires two adjacent fragments to cross paths at the same time, which  only happen for a small portion of neighbouring fragments with specific initial position and velocity configurations.

\begin{figure}
	\centering
	\subfloat[]{
		\label{fig:es_frags}
		\includegraphics[width=.48\textwidth]{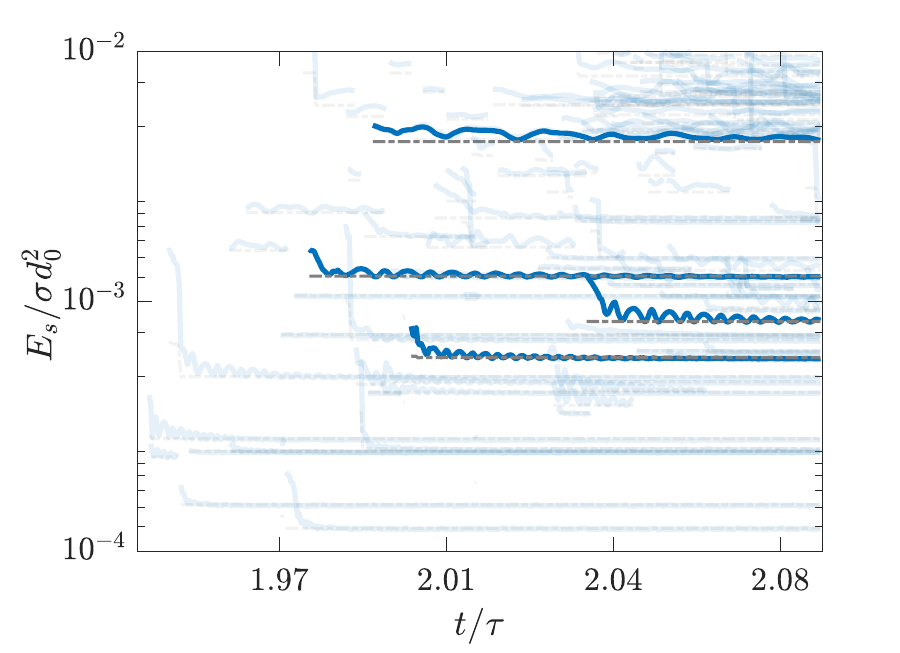}}
	\centering
	\subfloat[]{
		\label{fig:freq_frags}
		\includegraphics[width=.48\textwidth]{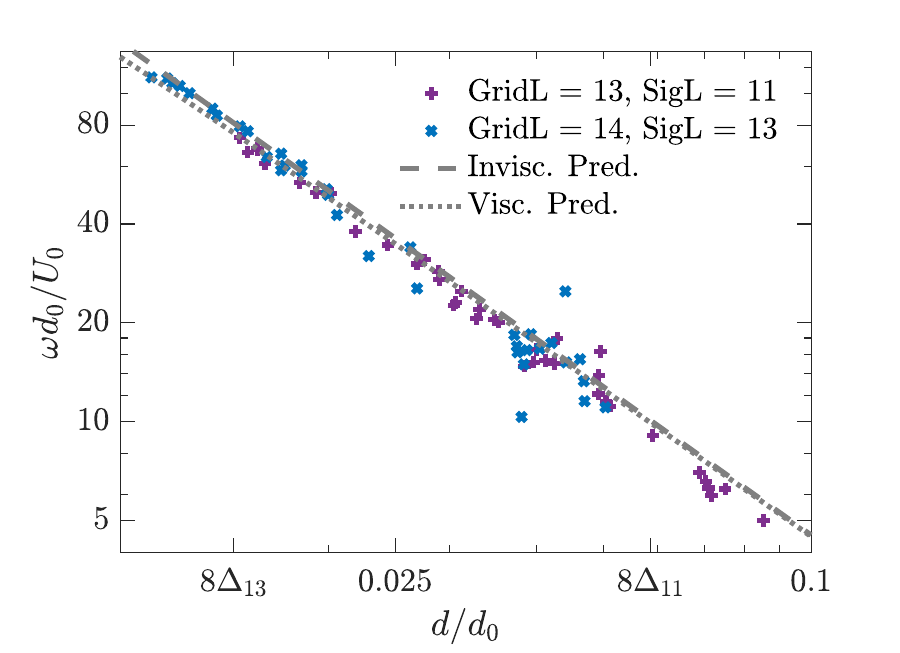}}	
	\caption{Oscillatory behaviour of very small aerobreakup fragments produced from an initial droplet with $We = 15$ and $Oh = 10^{-3}$, with the simulation run at $L=14$ and $L_{\rm sig} = 13$. Left: surface energy evolution of individual fragments (blue curves) with their steady-state surface energy values plotted (dashed lines) for reference, with the records of only a few representative fragments highlighted for clarity; right: frequency of the dominant fragment oscillation mode as a function of the fragment radius.}
	\label{fig:frag-oscillations}
\end{figure}

Apart from identifying all breakup and coalescing events in the spray formed due to aerobreakup, the toolbox developed by \cite{chan2021identifying} also enables us to track the evolution of properties of individual fragments during their lifetime. For example, fig.~\ref{fig:es_frags} shows the evolution of surface energy $E_s$ of individual small fragments recorded from simulations run at $L=14$ and $L_{\rm sig} = 13$, with the records of only a few representative fragments highlighted for clarity. The steady-state values of surface energy is also computed based on the volume of corresponding fragments, and plotted in grey dashed lines for reference. It is seen that the oscillation frequency and amplitude vary for each fragment, but all of them clearly demonstrate decaying oscillation behaviour, with their oscillation frequency generally increasing with decreasing fragment radius $r$ (hence decreasing steady-state surface energy values). We further extract the frequency of the dominant oscillation mode of these small fragments at two different grid resolution configurations ($L = 14, \, L_{\rm sig} = 13$ and $L = 13, \, L_{\rm sig} = 11$), and plot them against the fragment radius in fig.~\ref{fig:freq_frags}, where an excellent agreement is found for small fragments within the diameter range of $0.01d_0 \leq d \leq 0.1d_0$ with the theoretical predictions of \cite{prosperetti1980free} for the second Rayleigh mode, which is given as follows for an inviscid droplet with density $\rho_l$ and radius $R^*$ at equilibrium:
\begin{equation}
    \omega_{n,0} = \sqrt{(n-1)n(n+2) \frac{\sigma}{\rho_l R^*}},
    \label{for:rayleigh-freq}
\end{equation}
where $n$ is the spherical harmonic mode number at which the interface of the droplet is perturbed. The second Rayleigh mode corresponds with $n=2$, and this mode number is associated with the oblate-prolate shape perturbations which we observed in fig.~\ref{fig:lig-breakup-snapshots}. Both the viscous and inviscid theoretical model of \cite{prosperetti1980free} are plotted in fig.~\ref{fig:frag-oscillations}, which almost completely overlap except for small fragments below $8\Delta_{13}$, where the viscous model shows a slightly better match with the numerical results. This is because the $Oh$ value of $10^{-3}$ at which simulations discussed in this section are run is very low, and viscous effects become non-trivial only for very small fragments. Furthermore, for results at both resolution levels shown in fig.~\ref{fig:freq_frags}, the agreement between numerical and theoretical results reaches into their corresponding range of small fragments with $d \leq 8\Delta_{\rm sig}$, although the fragment size and velocity distributions within this range have not reached grid convergence. Overall, these results demonstrate that fragments satisfying $d \leq 8\Delta_{\rm sig}$ are governed by well-documented physical mechanisms, e.g. rim retraction at the Taylor-Culick velocity and the Rayleigh oscillation theory \citep{prosperetti1980free}, even though full grid independence in terms of fragment size and velocity statistics is still to be established. This suggests that it is the droplet production mechanism through ligament fragmentation which remains somewhat grid-dependent, while the droplets resulting from these production events are themselves well-resolved. Nevertheless, we highlight that the film fragmentation process is well-resolved in our numerical simulations with the aid of the MD algorithm \citep{chirco2021manifold}.

\subsection{Viscous effects on bag breakup}
\label{subsec:Oh-effects}

Having provided an overview of the physical mechanisms governing bag breakup and the subsequent evolution patterns of the fragments at a specific low $Oh$ value of $10^{-3}$ in \S\ref{subsec:frag-tracking}, in this section we increase $Oh$ up to 0.05, and examine its influence on the bag breakup phenomena. This has not been examined in depth in currently available aerobreakup studies as most research efforts have been carried out in the limit of very low $Oh$ values \citep{guildenbecher2017characterization, jackiw2022prediction, kant2022bags}.

\begin{figure}
    \centering
	\subfloat[]{
		\label{fig:Oh_0_0001_snap_1}
		\includegraphics[width=.3\textwidth]{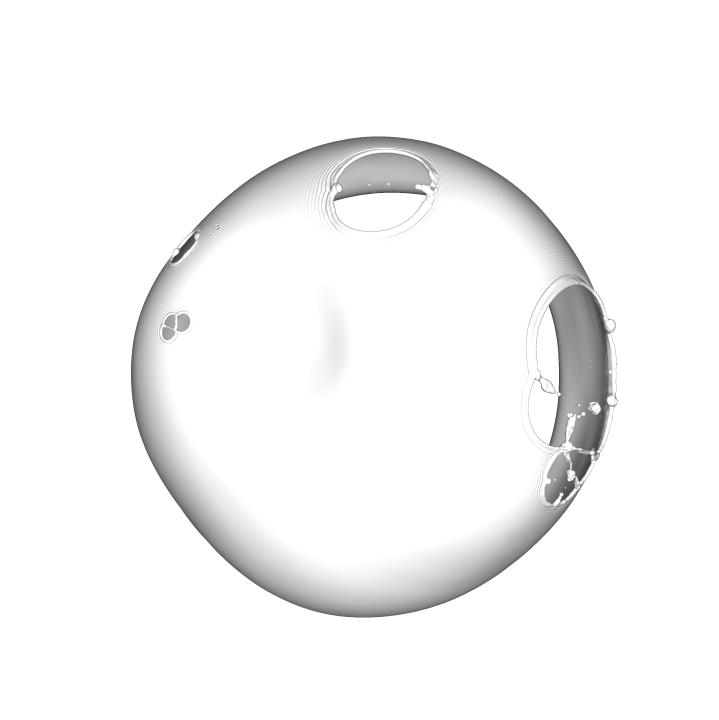}}
	\centering
	\subfloat[]{
		\label{fig:Oh_0_0001_snap_2}
		\includegraphics[width=.3\textwidth]{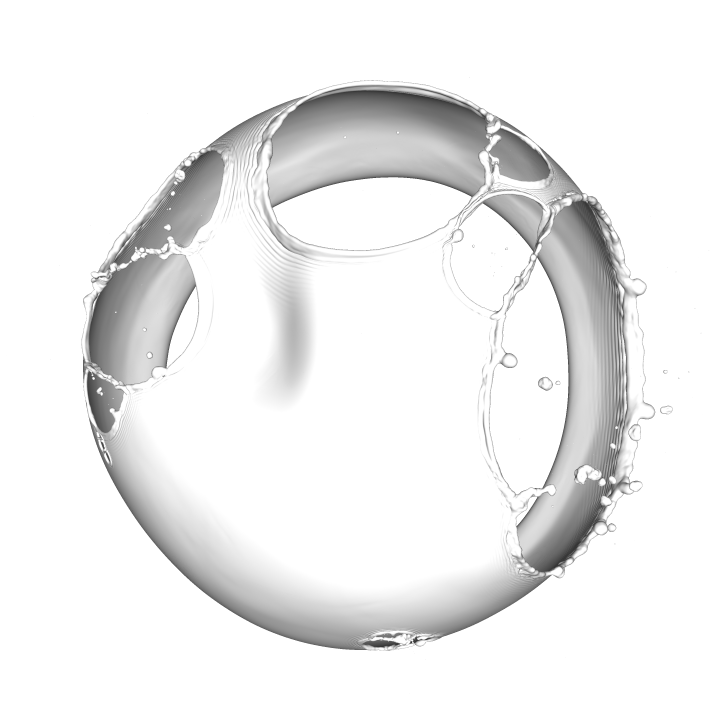}}
	\centering	
	\subfloat[]{
		\label{fig:Oh_0_0001_snap_3}
		\includegraphics[width=.3\textwidth]{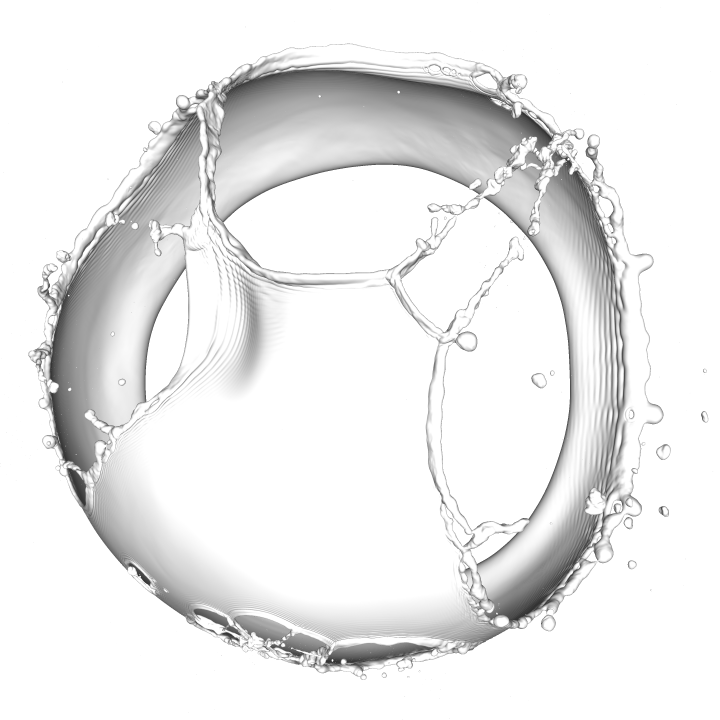}}

	\centering
	\subfloat[]{
		\label{fig:Oh_0_001_snap_1}
		\includegraphics[width=.3\textwidth]{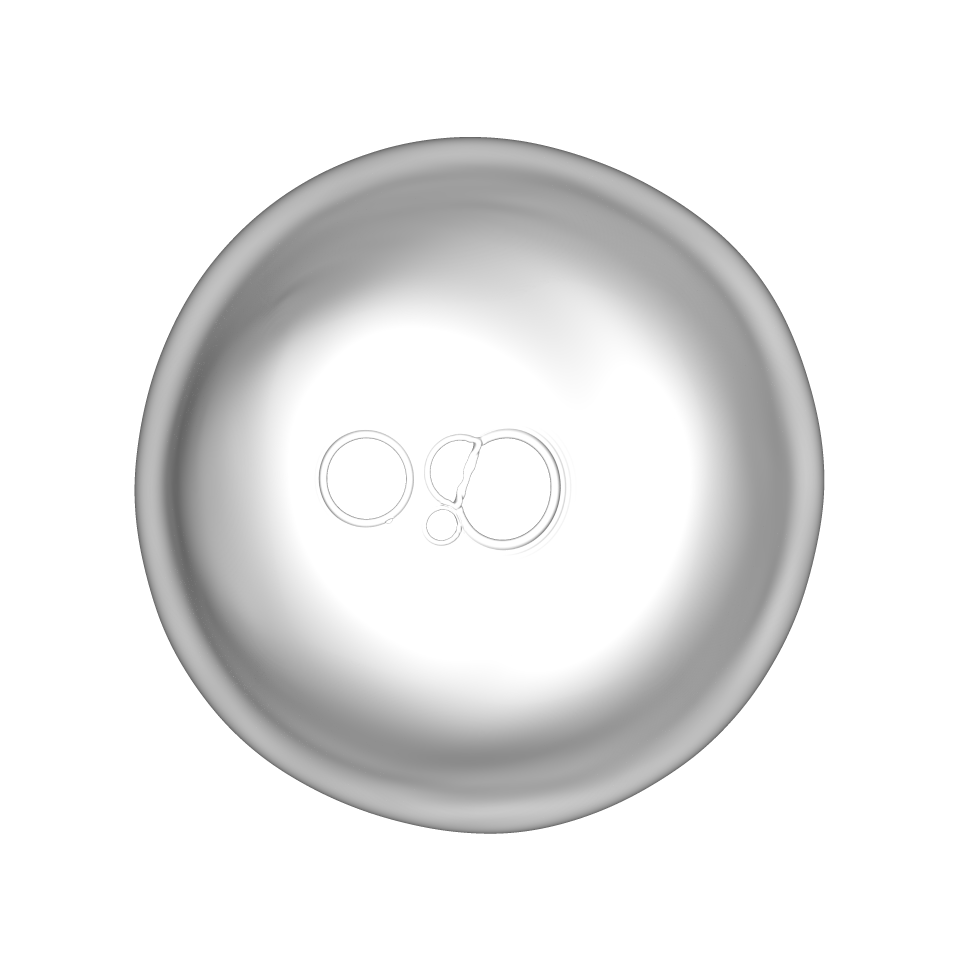}}
	\centering
	\subfloat[]{
		\label{fig:Oh_0_001_snap_2}
		\includegraphics[width=.3\textwidth]{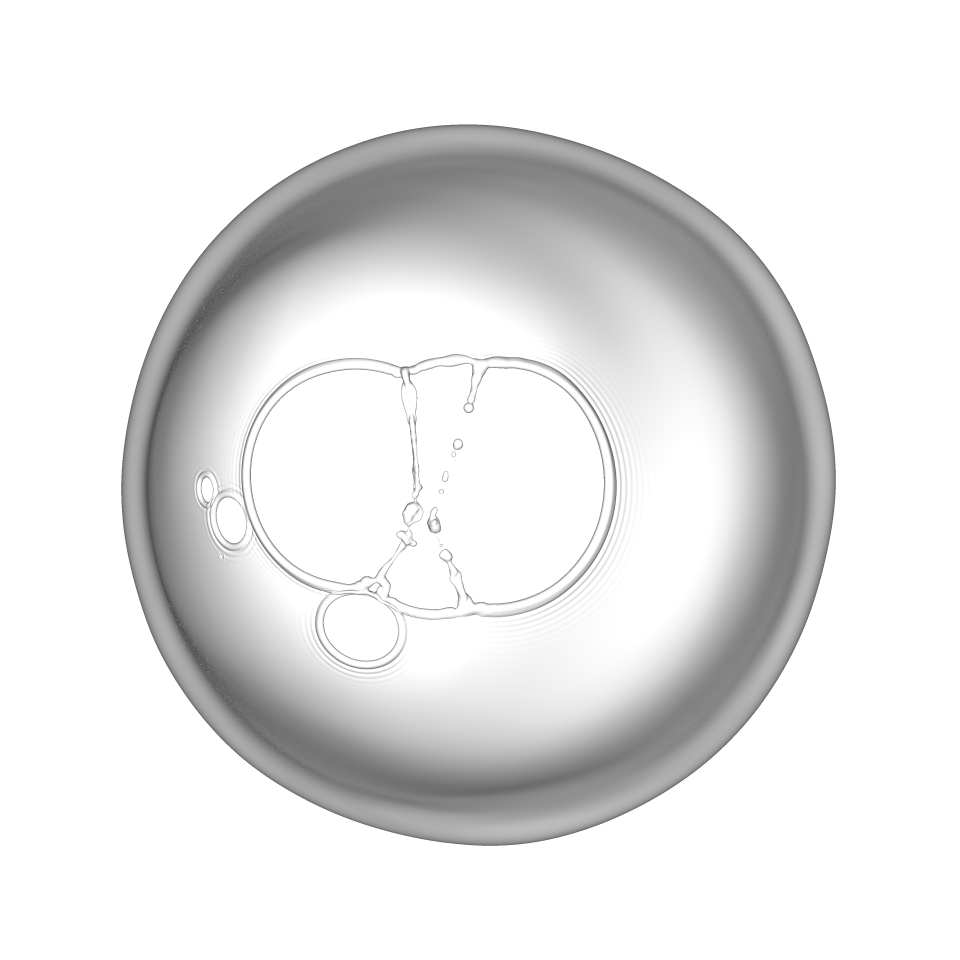}}
	\centering	
	\subfloat[]{
		\label{fig:Oh_0_001_snap_3}
		\includegraphics[width=.3\textwidth]{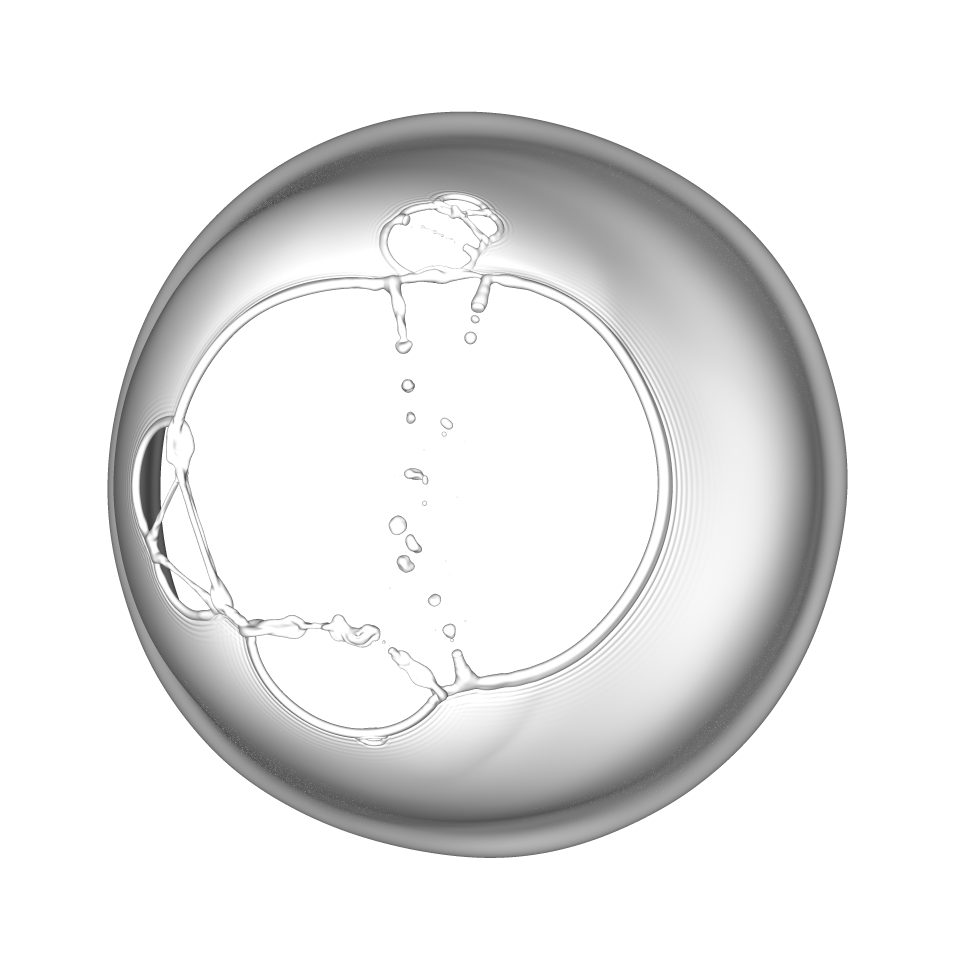}}
		
    \centering
	\subfloat[]{
		\label{fig:Oh_0_01_snap_1}
		\includegraphics[width=.3\textwidth]{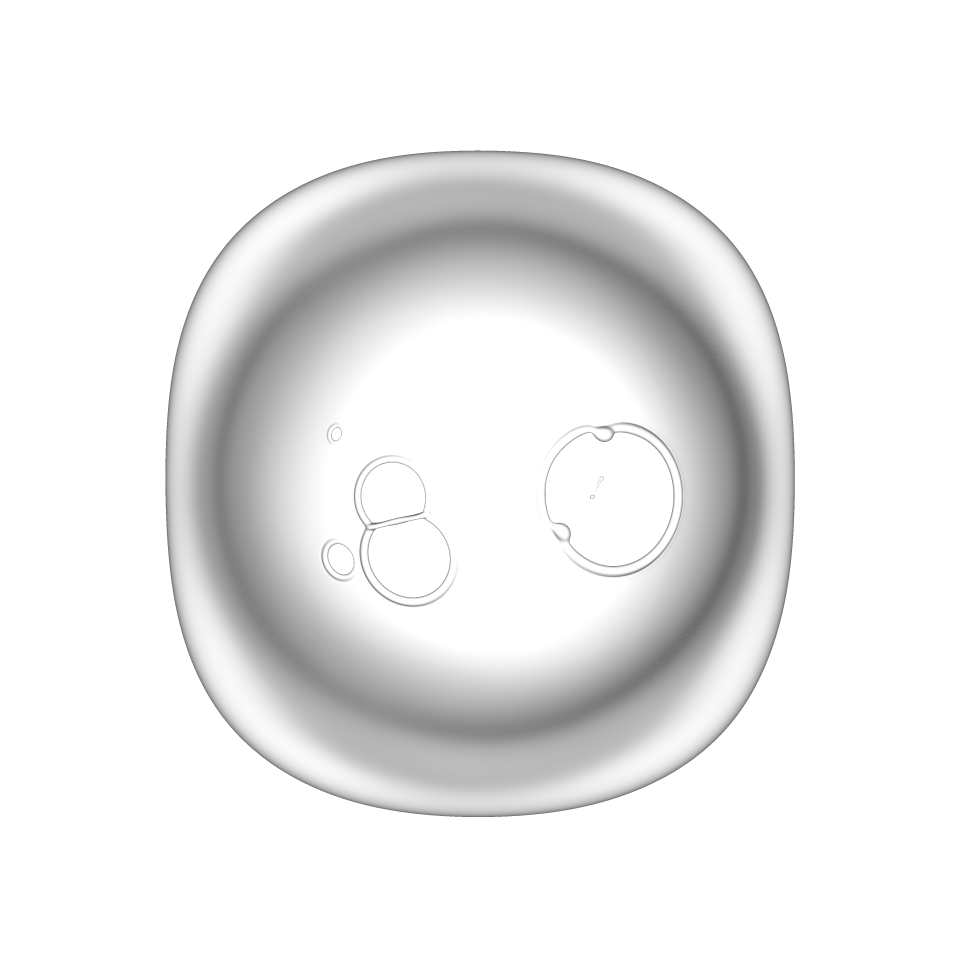}}	
	\centering
	\subfloat[]{
		\label{fig:Oh_0_01_snap_2}
		\includegraphics[width=.3\textwidth]{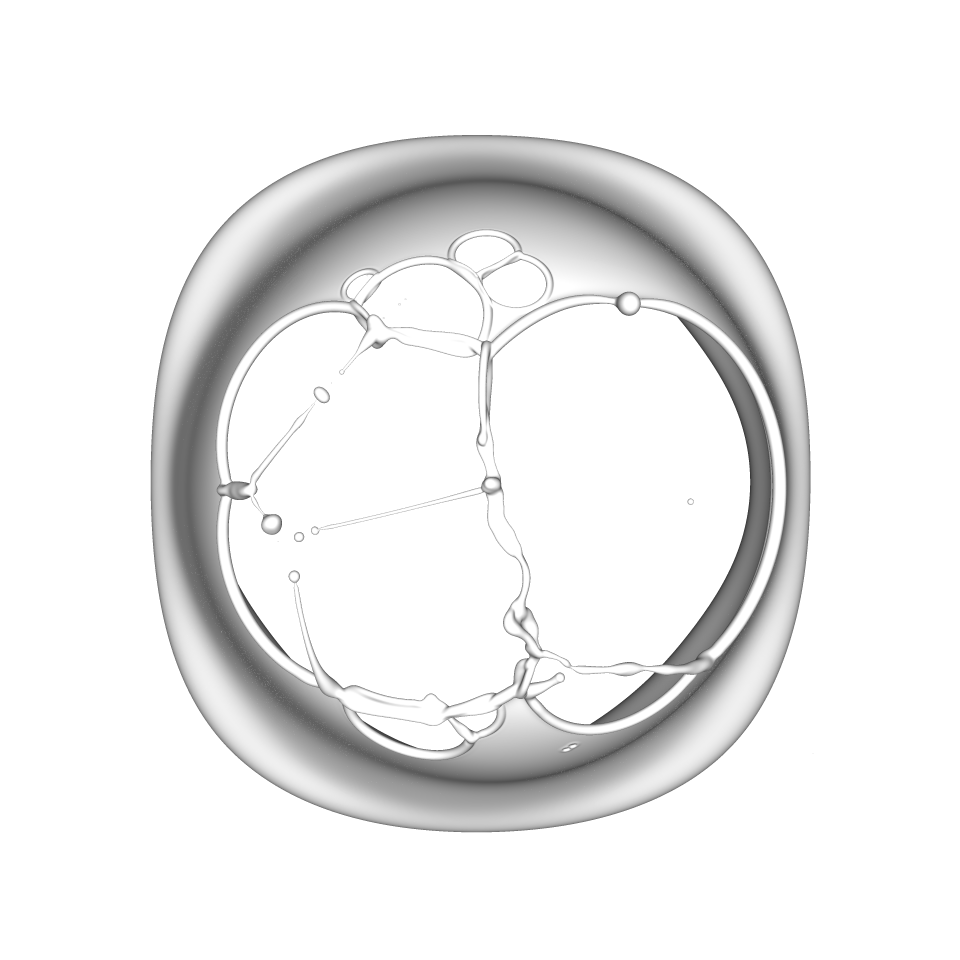}}	
	\centering	
	\subfloat[]{
		\label{fig:Oh_0_01_snap_3}
		\includegraphics[width=.3\textwidth]{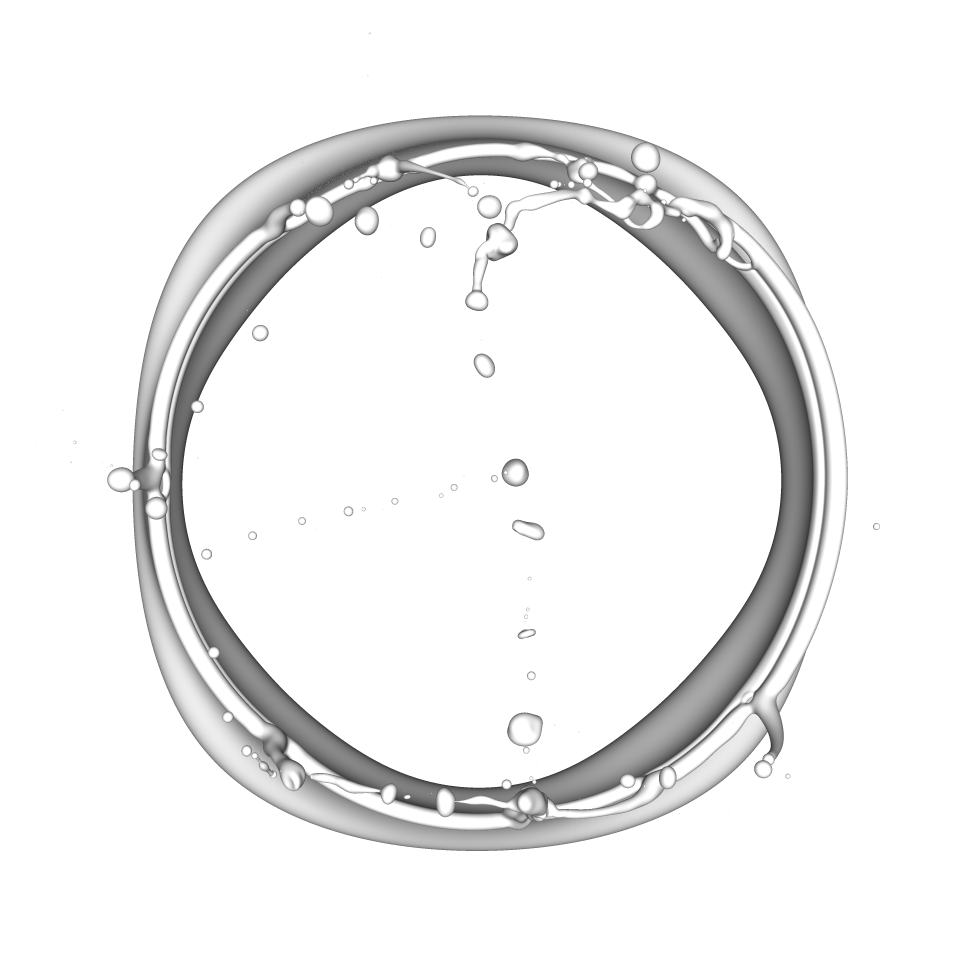}}	
		
	\centering
	\subfloat[]{
		\label{fig:Oh_0_05_snap_1}
		\includegraphics[width=.3\textwidth]{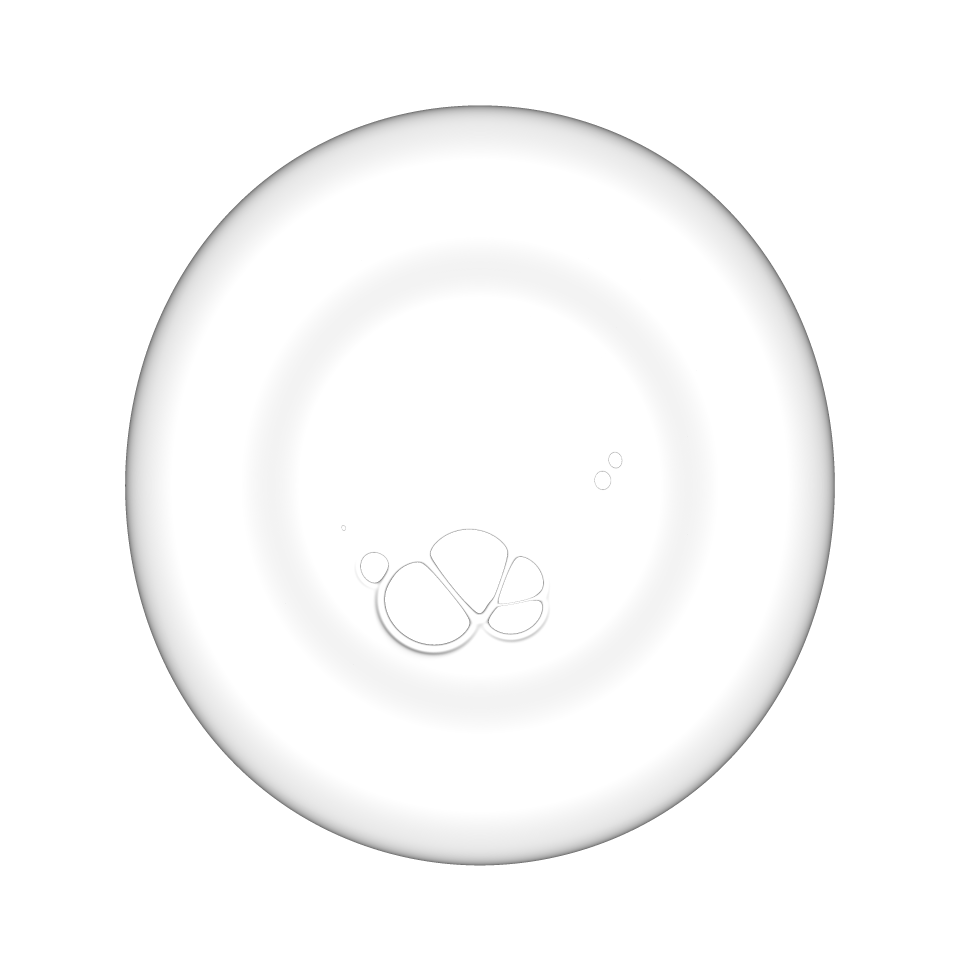}}	
	\centering
	\subfloat[]{
		\label{fig:Oh_0_05_snap_2}
		\includegraphics[width=.3\textwidth]{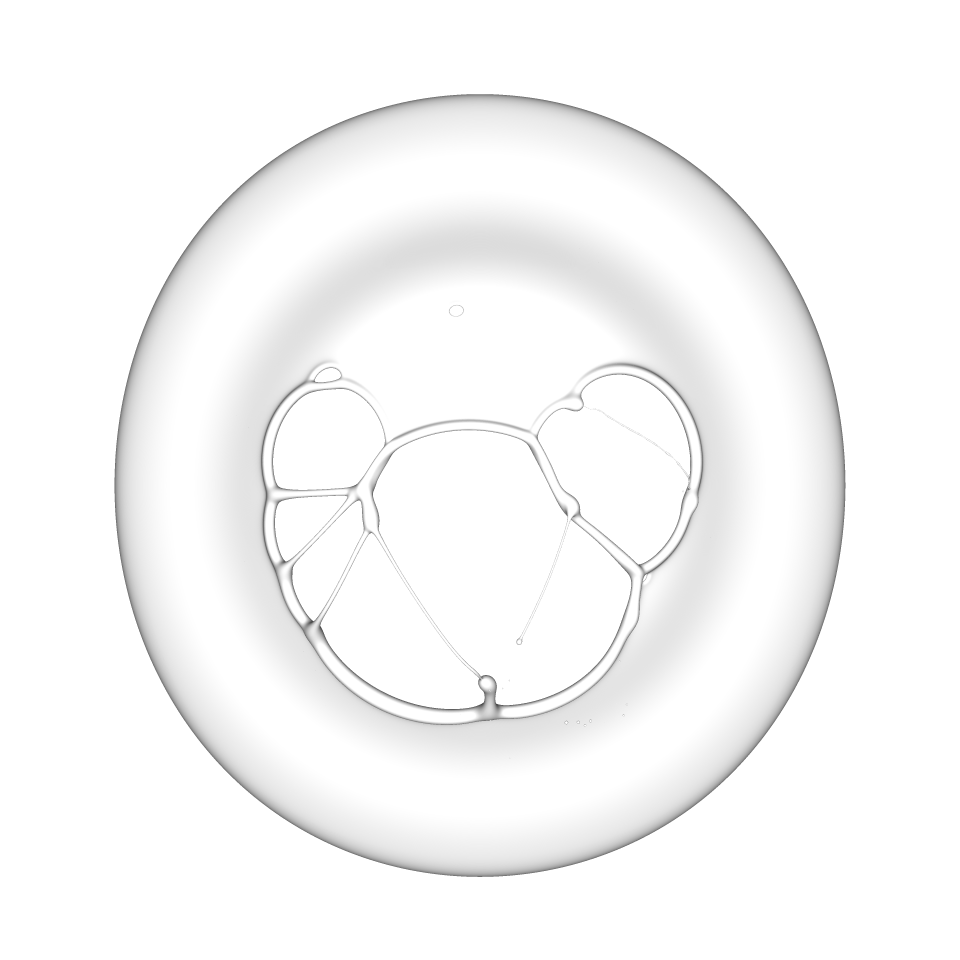}}	
	\centering	
	\subfloat[]{
		\label{fig:Oh_0_05_snap_3}
		\includegraphics[width=.3\textwidth]{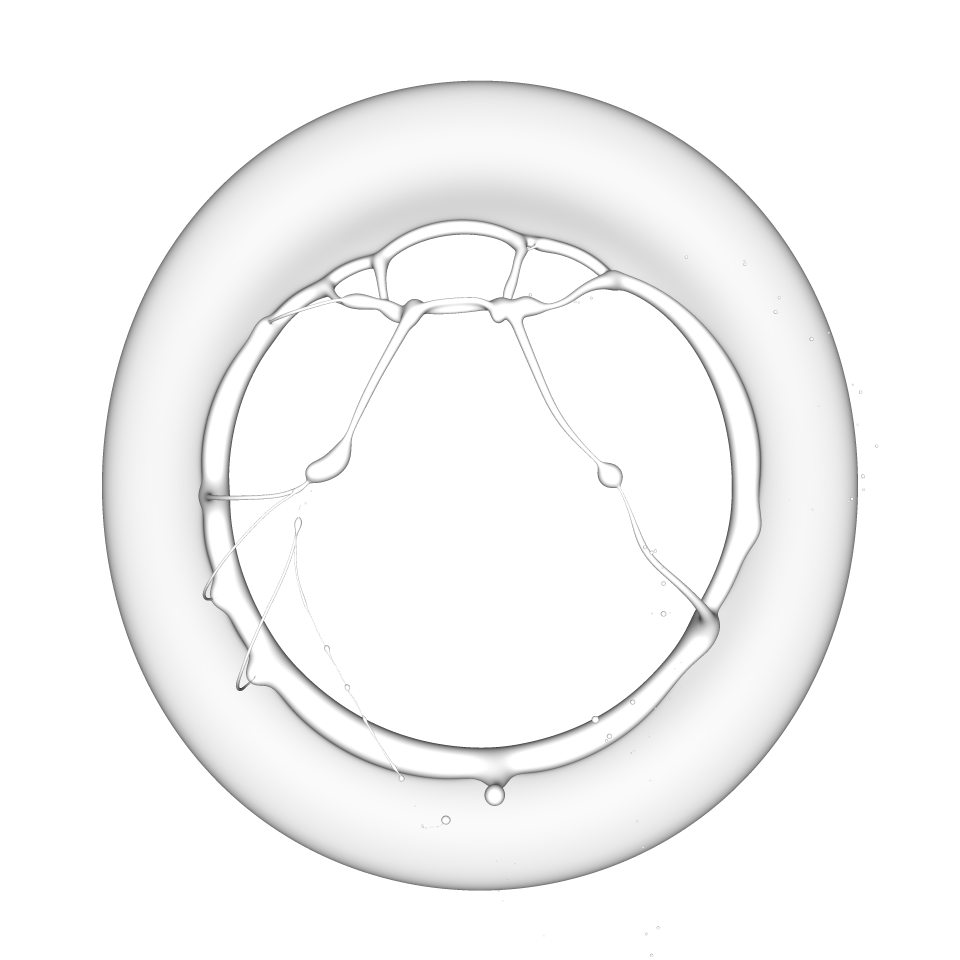}}	
	\caption{Simulation snapshots showing the bag breakup process at different $Oh$ values ($10^{-4}$, $10^{-3}$, $10^{-2}$ and $5 \times 10^{-2}$ from the top to the bottom row), where $We$ is fixed as 15. For all cases, $L = 14$ and $L_{\rm sig} = 13$.}
	\label{fig:Oh_sweep_snapshots}
\end{figure}

We first provide in fig.~\ref{fig:Oh_sweep_snapshots} simulation snapshots showing the bag breakup process for $Oh = 10^{-4}$, $10^{-3}$, $10^{-2}$ and $5 \times 10^{-2}$ for a qualitative analysis, with $We = 15$ for all cases. It is first seen that as $Oh$ increases from very low ($Oh \leq 0.005$) to moderate ($0.005 \leq Oh \leq 0.05$) values, the bag becomes more `flattened' and its surface area becomes smaller, and correspondingly the surrounding rim around the bag becomes more prominent. This implies that the inviscid model proposed by \cite{Jackiw2021} predicting the volume of the bag film and rim may need to be extended for a generalisation to the moderate-$Oh$ regime. Furthermore, it is observed that the ligament breakup behaviour changes significantly as the $Oh$ value increases. While at $Oh = 10^{-4}$ the receding liquid rims generate capillary waves propagating through the entire bag \citep{savva2009viscous}, and undergo destabilisation patterns similar to what we observed in fig.~\ref{fig:rim-destab-snapshots}, these are not found at higher $Oh$ values, which suggests that these phenomena are highly sensitive to viscous damping effects, and their contribution to fragment statistics becomes negligible with increasing $Oh$. At $Oh = 10^{-3}$, liquid ligaments typically show long periods of radial oscillation after their formation out of colliding hole rims, and then break up into `primary' and `satellite' drops that differ significantly in their sizes. When $Oh$ increases to the moderate value of $10^{-2}$, it is found that far fewer satellite drops are produced from ligament breakup, and the `end-pinching' breakup mechanism comes to dominate as the ligaments now tend to break up on one end repeatedly and form small drops. This may be because the ligaments can be stretched longer and thinner with a higher $Oh$, and the smaller ligament radius impedes the formation of satellite drops. According to \cite{vassallo1991satellite}, smaller radius of ligaments induces a larger pressure difference that pushes their free ends back in the axial direction much quicker, hence preventing capillary pinch-off in the radial direction that produces the satellite drops. Further increasing $Oh$ to $5 \times 10^{-2}$ causes the ligaments to be stretched even thinner and produce smaller fragments once they break up, which is because increased viscosity smooths out the variation of the axial velocity along the ligament that  drive the pinch-off events \citep{hu2021deformation}. Another side effect appearing at $Oh = 5 \times 10^{-2}$ is that fewer node fragments are observed, which is because the decreased bag area leaves smaller room for generation and mutual collision of more than three holes which  produce node fragments. \cite{savva2009viscous} suggest that at even higher $Oh$ values the liquid rims will disappear, and the thickness of the entire bag film will correspondingly increase as the holes enlarge, although we do not reach this limit in our current numerical simulations.

\begin{figure}
    \centering
	\subfloat[]{
	    \label{fig:N-frag-Oh-sweep}
	    \includegraphics[width=.48\textwidth]{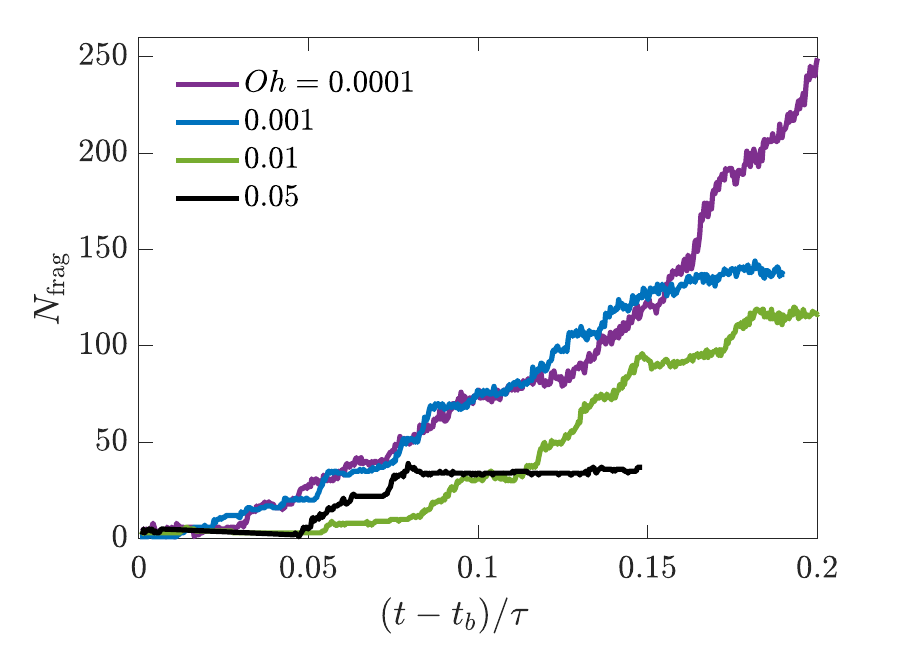}}
    \subfloat[]{
	  \label{fig:Lbag-dbag-Oh-sweep}
        \includegraphics[width=.48\textwidth]{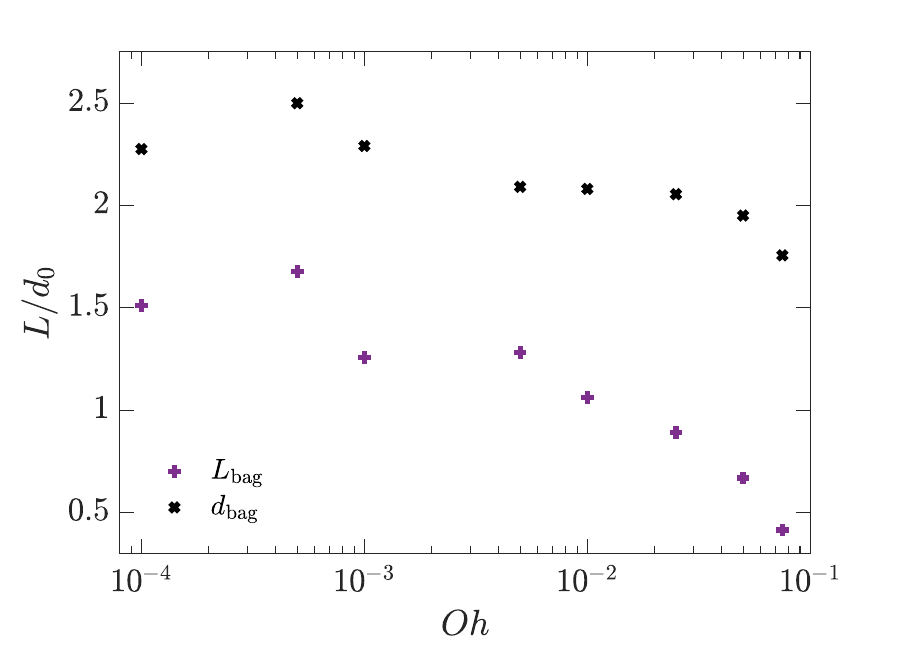}}
	    
	\centering
	\subfloat[]{
	    \label{fig:size-pdf-Oh-sweep}
	    \includegraphics[width=.48\textwidth]{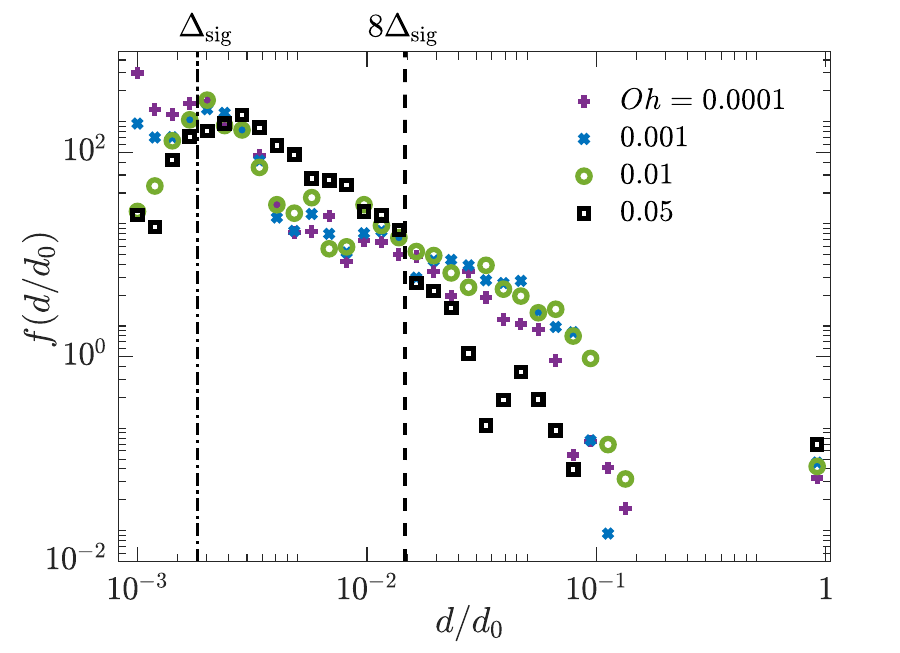}}
	\centering
	\subfloat[]{
	    \label{fig:vel-pdf-Oh-sweep}
	    \includegraphics[width=.48\textwidth]{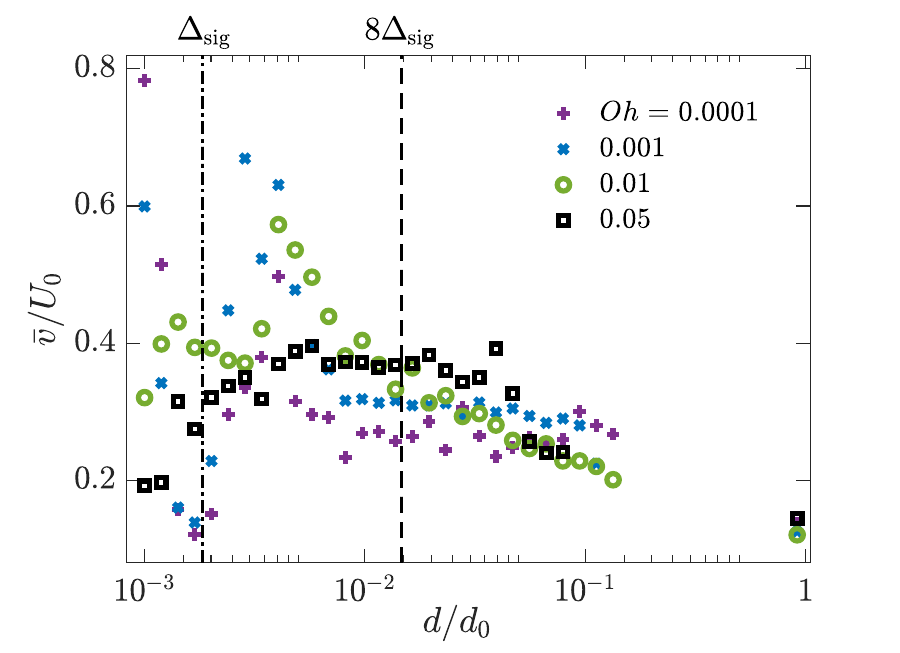}}
	    
	\caption{(a): Time evolution of the total number of film fragments after the onset of bag breakup for one ensemble realisation with different $Oh$ values. (b): The bag length $L_{\rm bag}$ and width $d_{\rm bag}$ just before the breakup of bag films as functions of the $Oh$ values. (c)(d): Time- and ensemble-averaged size (c) and speed (d) probability distribution functions of aerobreakup fragments with $We = 15$ and various $Oh$ values.}
	\label{fig:size-vel-pdf-Oh-sweep}
\end{figure}

Figure~\ref{fig:size-vel-pdf-Oh-sweep} further shows the evolution of the number of fragments satisfying $d \geq \Delta_{\rm sig}$, the dependence of bag length and width at the onset of bag breakup on the $Oh$ values, and the time- and ensemble-averaged size and speed distribution functions for the three specific $Oh$ values selected in fig.~\ref{fig:Oh_sweep_snapshots}. Figure~\ref{fig:N-frag-Oh-sweep} shows that as the $Oh$ value increases, the number of fragments $N_{\rm frag}$ reached when the bag fully disintegrates decreases, which is because the total area of the bag film decreases, leaving less amount of liquid that feeds the film breakup process. While $N_{\rm frag}$ generally increases over time despite small-scale local oscillations, which most likely arise from relatively rare coalescing events, long periods of time where $N_{\rm frag}$ remains nearly constant can be clearly seen for $Oh = 5 \times 10^{-2}$, as the highly-viscous liquid ligaments can now be sustained for much longer under stretching, and it is their intermittent breakup events that contribute to the isolated sharp growth events in $N_{\rm frag}$. The length $L_{\rm bag}$ and width $d_{\rm bag}$ of the bags just before their breakup for different $Oh$ values are measured and shown in fig.~\ref{fig:Lbag-dbag-Oh-sweep} (which were also presented in fig.~\ref{fig:rm_We_15_comp} as scattered points for validation of our numerical results), where it can be seen that our bags are ‘flattened’ in shape (satisfying $d_{\rm bag} > L_{\rm bag}$), and that for $Oh \geq 5 \times 10^{-4}$, both $L_{\rm bag}$ and $d_{\rm bag}$ decrease with increasing $Oh$, suggesting that the bag area indeed becomes smaller as $Oh$ increases.

Figure~\ref{fig:size-pdf-Oh-sweep} shows that the fragment size distribution functions for $10^{-4} \leq Oh \leq 10^{-2}$ remain very close to each other. When $Oh$ is further increased to $5 \times 10^{-2}$, it is observed that the size distribution function for $d \geq \Delta_{\rm sig}$ becomes much more convex-shaped, with more `intermediate' fragments produced within the range of $\Delta_{\rm sig} \leq d \leq 8\Delta_{\rm sig}$, and fewer large fragments 
with $d \geq 8 \Delta_{\rm sig}$. This arises from the coupled effects of reduction of bag film area (hence smaller chance for formation of `node' fragments, as observed in fig.~\ref{fig:Oh_0_05_snap_3}) and breakup of long viscous ligaments into smaller fragments, and implies a reduction in the average fragment size which will be discussed in more details further below. Finally, fig.~\ref{fig:vel-pdf-Oh-sweep} suggests that increasing the $Oh$ value causes the fragment speed to become more evenly-distributed across different fragment sizes, which is probably due to the combined effects of viscous damping of the internal axial velocity distributions of pre-breakup ligaments \citep{hu2021deformation}, and the ambient airflow becoming much less turbulent as its viscosity also increases under a fixed viscosity ratio $\mu^*$.

\begin{figure}
	\centering
	\subfloat[]{
	    \label{fig:tb-Oh-sweep}
	    \includegraphics[width=.48\textwidth]{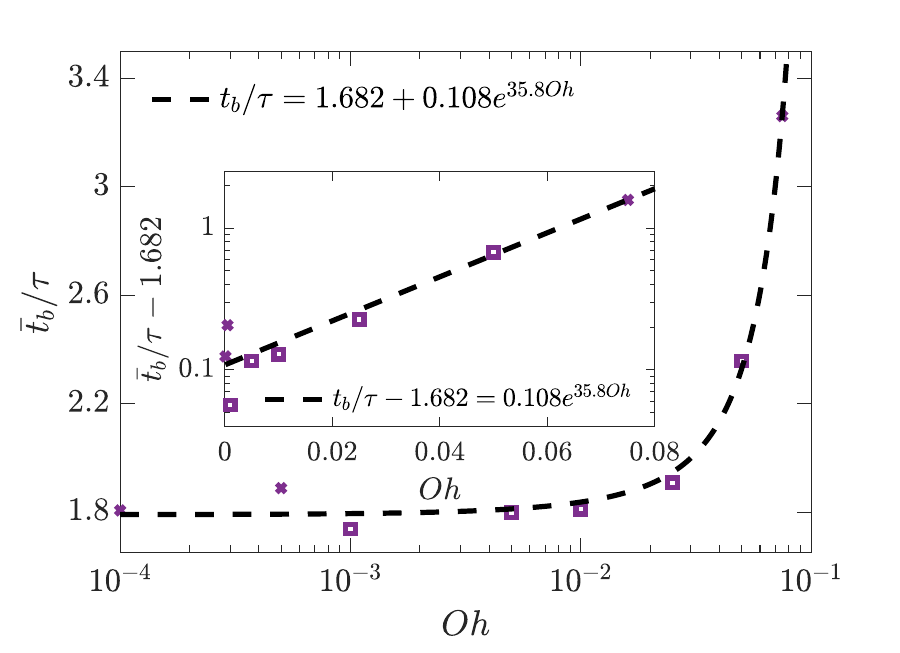}}
	\centering
	\subfloat[]{
	    \label{fig:dave-Oh-sweep}
	    \includegraphics[width=.48\textwidth]{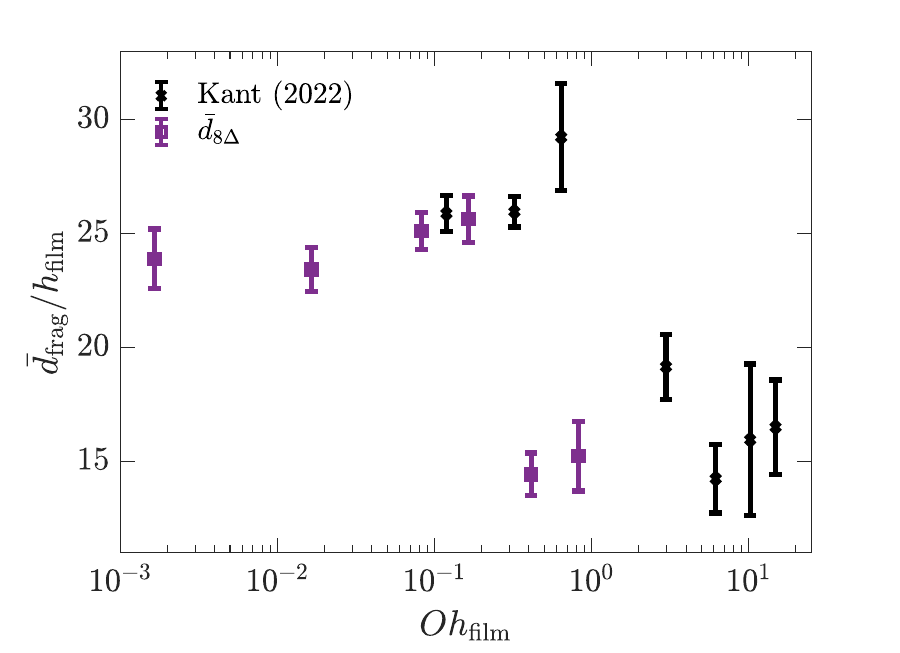}}
	    
	\caption{The breakup onset time $t_b$ (a) and the instantaneous average diameter $\bar{d}$ of fragments satisfying $d \geq 8\Delta_{\rm sig}$ (b) as functions of the $Oh$ values, with an exponential model fit for (a). The non-dimensionalised experimental data of \cite{kant2022bags} are included in (b) for comparison. Squares mean the results have been ensemble-averaged over three individual realisations, and crosses mean data from only one realisation is available. For all cases $We = 15$.}
	\label{fig:tb_dave_Oh_sweep}
\end{figure}

We now analyse the influence of $Oh$ on the ensemble-averaged breakup time $\Bar{t}_b$ (defined as the time when the first hole is generated on the film by the MD algorithm) and the ensemble-averaged instantaneous diameter of fragments $\Bar{d}_{\rm frag}$ in fig.~\ref{fig:tb_dave_Oh_sweep}. It should be noted that due to the significant runtime required on supercomputers, the results at a few $Oh$ values in fig.~\ref{fig:tb_dave_Oh_sweep} are obtained from only one realisation instead of being ensemble-averaged, and these results are differentiated from the others by a cross mark. Nevertheless, from fig.~\ref{fig:tb-Oh-sweep}, it is seen that $\Bar{t}_b$ first remains almost independent of $Oh$ as the latter increases from the low value of $10^{-4}$ to the moderate value of $10^{-2}$, which is likely because the wake flow remains separated from the drop surface, and the thinning process of the bag before the onset of its breakup is determined by capillary and inertial effects, as discussed in \S\ref{subsubsec:drainage}. When $Oh$ exceeds the moderate value of $10^{-2}$, $\Bar{t}_b$ is found to increase exponentially with $Oh$ as shown by the fitted model, which may be the consequence of both the transition of the wake region from a turbulent to a laminar status (hence smaller fore-aft pressure difference on the deforming drop that pushes out the bag), and the bag thinning process coming under the domination of a capillary-viscous balance. The hypothesis of the influence of wake region on the growth of $t_b$ is further supported by examining the freestream Reynolds number $Re$:
\begin{equation}
    Re \equiv \frac{\rho_a U_0 d_0}{\mu_a} = \frac{\mu^* \sqrt{We}}{\sqrt{\rho^*} Oh} = \frac{7.381}{Oh},
\end{equation}
where the critical $Oh$ value of $10^{-2}$ corresponds to $Re = 7.38 \times 10^2$, which agrees with the order of magnitude of previously-reported $Re$ values at which the wake regions behind a sphere transits to turbulence and vortex shedding is initiated \citep{rodriguez2011direct}. Overall, this sharp increase of the breakup time $\Bar{t}_b$ with increasing $Oh$ beyond $10^{-2}$ agrees with the early findings that the $Oh$ values do not have significant influence over breakup regimes when they are below 0.1 \citep{Hsiang1992}. 

Finally, we compute the instantaneous average diameter $\Bar{d}_{\rm frag}$ at the end of film breakup. The averaging is completed for fragments satisfying $d \geq 8\Delta_{\rm sig}$ over different individual realisations with the same $Oh$ value, which enables us to acquire sufficient amounts of numerically converged statistics to produce meaningful results. In fig.~\ref{fig:dave-Oh-sweep}, $\Bar{d}_{\rm frag}$ shows a non-monotonic dependence on the film Ohnesorge number, defined as $Oh_{\rm film} \equiv \mu_l / \sqrt{\rho_l \sigma h_f}$. While the bag films continues thinning as they undergo fragmentation, we select a constant characteristic film thickness value $h_f = D/2^{L_{\rm sig}}$ so that our computed average fragment diameters, non-dimensionalised by $h_f$, match the order of magnitude of the thin-film fragment statistics studied by \cite{kant2022bags}, which are also included in fig.~\ref{fig:dave-Oh-sweep} for comparison. Namely, as $Oh_{\rm film}$ increases from $1.65 \times 10^{-3}$ to 0.826, $\Bar{d}_{\rm frag}$ first remains close to $25h_f$, followed by an abrupt decrease as it approaches $Oh_{\rm film} = 0.4$, which corresponds to the drop $Oh$ value exceeding the moderate value of 0.01. This non-monotonic dependency on $Oh$ is also observed in the results of \cite{kant2022bags}, where an initial increase of $\Bar{d}_{\rm frag}$ to $29.2h_f$ is followed by an abrupt decrease to $14.2h_f$ when $Oh_{\rm film}$ increases beyond unity. Based on our analysis of fig.~\ref{fig:Oh_sweep_snapshots} and fig.~\ref{fig:size-vel-pdf-Oh-sweep}, we ascribe the abrupt decrease of $\Bar{d}_{\rm frag}$ with increasing $Oh$ values to the formation of much fewer large `node' fragments due to the decrease of bag area, and the breakup of ligaments that are stretched much thinner under high viscosity.

\section{Summary of numerical convergence considerations}
\label{sec:num-summary}
Here we provide a brief summary for the influence of the MD algorithm \citep{chirco2021manifold} on the numerical convergence behaviour of bag film fragment statistics, which is of reference value for future works on two-phase flows involving topological changes. It is noted that in the aerobreakup simulations, fragments are produced following a sequence of film perforation, hole expansion, rim collision and ligament breakup, regardless of whether the MD algorithm is applied. However, the MD algorithm controls hole formation through a signature level $L_{\rm sig}$, thereby perforating thin films at a controlled thickness independent of the mesh resolution. Typically, holes are initially isolated and grow for some time before their bordering rims collide with each other. When they do, ligaments form and break up to produce droplets, including primary and satellite drops directly formed out of breaking ligaments (fig.~\ref{fig:lig-breakup-snapshots}) and liquid nodes when their neighbouring ligaments break down completely (fig.~\ref{fig:node-breakup-snapshots}). The independence of the critical film thickness from mesh resolution opens up the possible formation of fragments whose sizes are also grid-independent. In the absence of the MD algorithm, the films undergo `VOF breakup', where they are perforated spontaneously upon reaching the mesh resolution, leading to many adjacent holes which immediately collide, forming small ligaments which in turn rapidly break up into tiny fragments. With only the mesh size as the governing length scale, none of the resulting droplets are grid converged.

It is observed that when the MD algorithm is applied, droplets greater than $8\Delta_{\rm sig}$ show grid-converged size statistics; while those smaller than $8\Delta_{\rm sig}$ do not. Apart from the present results and \cite{chirco2021manifold}, empirical lower bounds for grid convergence in the form of $8\Delta$ have also been proposed in other works involving two-phase breakup phenomena (see e.g., \cite{riviere2021sub}). However, to the knowledge of the authors, there has been no underlying physical mechanism proposed for this lower bound. We suggest a possible explanation as follows. As can be seen in figs.~\ref{fig:lig-breakup-snapshots} and \ref{fig:node-breakup-snapshots}, bag fragments originate from the breakup of liquid ligaments formed from colliding hole rims. When the MD algorithm \citep{chirco2021manifold} is applied, regions on the bag film with thickness around the critical value of $3\Delta_{\rm sig}$ are perforated, and the diameters of the hole rims gradually increase as they recede over the bag film \citep{agbaglah2021breakup}. Consequently, the diameters of colliding rims should satisfy $d_{\rm rim} \geq 3\Delta_{\rm sig}$. Conservation of liquid volume then yields $d_{\rm lig} = \sqrt{2} d_{\rm rim}$, where $d_{\rm lig}$ is the diameter of the fused ligament produced from colliding liquid rims. Further assuming that the fused ligament does not generate transverse liquid lamellae which pinch off into `fine' drops (as seen in \cite{neel2020fines}), but instead break up under the Rayleigh-Plateau (RP) Instability, the size of the primary fragments should then satisfy \citep{pal2021statistics},
\begin{equation}
    d_{\rm RP} = 1.9 d_{\rm lig} \geq 8.0 \Delta_{\rm sig},
\end{equation}
which leads to the lower bound of $8\Delta_{\rm sig}$ observed in Figs.~\ref{fig:size_pdf_single} and \ref{fig:size_pdf}.

On the other hand, the statistics of the smallest droplets are still grid dependent when the MD algorithm is applied, which is also noted by \cite{chirco2021manifold}. These small droplets are most likely satellite drops produced from ligament breakup and not directly controlled by the MD algorithm, whose typical size and number have a strong dependence on the initial perturbations present on the ligament \citep{pal2021statistics} which are under mesh-regularized effects. However, even though their production mechanism is not well-resolved, these droplets themselves are sufficiently large to have well-resolved dynamics captured by the numerical mesh (as discussed in \S\ref{subsec:frag-behaviour}, see especially fig.~\ref{fig:frag-oscillations}).

\section{Conclusions}
\label{sec:conclusions}

We have presented in this study the results of both axisymmetric and three-dimensional numerical simulations of droplet aerobreakup. For the axisymmetric simulations, our results were validated by a good agreement with the experimental results of \cite{Jackiw2021}, and we were able to explain deviation from their theoretical model \eqref{for:jackiw-r-dot} based on the interaction between the drop surface and the wake vortices. We were also able to look into the thinning of bag films before the onset of bag breakup, and found that at small $Oh$ values capillary effects will cause the thinning rate to exceed that predicted by \cite{Villermaux2009}. 

For the three-dimensional aerobreakup simulations, we utilised the MD algorithm \citep{chirco2021manifold} for artificial perforation of thin films, which enabled us to minimise pollution of fragment statistics by spurious numerical breakup, and establish grid convergence of fragment statistics for aerobreakup studies for the first time. Afterwards, we analysed the output fragment statistics, and were able to reconstruct the breakup lineage and evolution of individual fragment properties using the postprocessing toolbox proposed by \cite{chan2021identifying}. It is found that smaller fragments with their diameters satisfying $d \leq 8\Delta_{\rm sig}$ are most likely satellite drops produced from ligament breakup and tend to undergo decaying surface oscillations dominated by the second Rayleigh mode, with their ejection velocity set by the colliding liquid rims receding at the Taylor-Culick velocity; while larger fragments satisfying $8\Delta_{\rm sig} \leq d \leq 0.05d_0$ are most likely primary drops produced from ligament breakup or detached liquid `nodes' bordering three or more holes, and tend to experience secondary local breakup events due to large-amplitude nonlinear oscillations. Destabilisation of receding rims is also found in some individual realisations, although they do not contribute significantly to fragment production under current simulation configurations. 

We find in particular that the bag-breakup problems feature subtle numerical convergence properties:

1. Without the MD algorithm, numerical convergence cannot be achieved for thin film fragmentation owing to the VOF-breakup phenomenon.

2. With the MD algorithm, the production of fragments through thin film fragmentation shows grid convergence for droplet children with diameter $d > 8\Delta_{\rm sig}$.

3. With or without the MD algorithm, the production of small droplets close to the resolution limit resulting from ligament fragmentation occurs independent of VOF- or MD-induced breakup, and numerical convergence for these production mechanisms is yet to be established in the present study.

4. However, once they are produced, the subsequent evolution of small fragments is  well-resolved in the present simulations, even for small fragments approaching the grid resolution.

Finally, we investigated the influence of drop $Oh$ on bag film breakup, and it is found that increasing $Oh$ within the moderate range of $0.005 \leq Oh \leq 0.05$ causes the bag area to decrease and the liquid ligaments to be stretched much thinner, which generally lead to production of fewer fragments with smaller average diameters.

Overall, these results show the utility of the MD algorithm in improving the grid convergence behaviour of fragment statistics and helping to recover previously unresolved fluid physics in two-phase numerical simulations involving breakup of thin films \citep{kant2022bags}, and also shed light on the effect of moderate viscosity on bag breakup which has not been discussed in detail in previous aerobreakup studies \citep{jackiw2022prediction}. They also pave the way for future studies investigating the later development of the remnant rim and the effects of airphase turbulence and initial perturbations on the deformation and breakup of droplets, while also serving as a stepping stone towards a full-scale numerical study of spume drop generation on the air-sea interface under high wind conditions.

\section{Declaration of Interests}
The authors report no conflict of interest.

\section{Acknowledgments}
\label{sec:acknowledge}
The authors would like to thank EPSRC for the computational time made available on the UK supercomputing facility ARCHER2 via the UK Turbulence Consortium (EP/R029326/1). Use of the University of Oxford Advanced Research Computing (ARC) facility is also acknowledged. K. Tang is supported by a Research Studentship at the University of Oxford.

\bibliographystyle{jfm}
\bibliography{references}

\end{document}